\documentclass[11pt]{article}
\usepackage{times}

\usepackage[margin=1in]{geometry}
\usepackage{graphicx}
\usepackage{amsmath, amsthm, amssymb}
\usepackage{braket}
\usepackage{subcaption}
\usepackage{comment}
\usepackage{url}
\usepackage[ruled,lined,linesnumbered]{algorithm2e}
\usepackage{tikz}
\usepackage{diagbox}
\usepackage[colorlinks=true,citecolor=green!70!black,linkcolor=red,urlcolor=blue,bookmarks,breaklinks]{hyperref}
\usepackage{natbib}
\setcitestyle{numbers}

\allowdisplaybreaks

\makeatletter
\newtheorem*{rep@theorem}{\rep@title}
\newcommand{\newreptheorem}[2]{%
\newenvironment{rep#1}[1]{%
 \def\rep@title{#2 \ref{##1} (restated)}%
 \begin{rep@theorem}}%
 {\end{rep@theorem}}}
\makeatother

\makeatletter
\newcommand{\subalign}[1]{%
  \vcenter{%
    \Let@ \restore@math@cr \default@tag
    \baselineskip\fontdimen10 \scriptfont\tw@
    \advance\baselineskip\fontdimen12 \scriptfont\tw@
    \lineskip\thr@@\fontdimen8 \scriptfont\thr@@
    \lineskiplimit\lineskip
    \ialign{\hfil$\m@th\scriptstyle##$&$\m@th\scriptstyle{}##$\hfil\crcr
      #1\crcr
    }%
  }%
}
\makeatother

\newtheorem{thm}{Theorem}
\newtheorem*{thm*}{Theorem}
\newtheorem{cor}[thm]{Corollary}
\newtheorem{con}[thm]{Conjecture}
\newtheorem{lem}[thm]{Lemma}
\newtheorem*{lem*}{Lemma}

\newtheorem{prop}[thm]{Proposition}
\newtheorem{defn}[thm]{Definition}

\newreptheorem{thm}{Theorem}
\newreptheorem{lem}{Lemma}
\newreptheorem{prop}{Proposition}
\newreptheorem{cor}{Corollary}

\DeclareMathOperator*{\argmax}{arg\,max}

\DeclareMathOperator*{\cut}{\textbf{cut}}

\newcommand{\FIXME}[1]{}

\begin{document}

%\scalebox{10}{$\varphi$} \hspace{3cm} \scalebox{10}{$\Psi$}
%\end{document}

\title{Improving the Quantum Approximate Optimization Algorithm with postselection}
\author{Sami Boulebnane\thanks{University College London and PhaseCraft Ltd; {\tt sami.boulebnane@ucl.ac.uk}.}}
\maketitle

\begin{abstract}
Combinatorial optimization is among the main applications envisioned for near-term and fault-tolerant quantum computers. In this work, we consider a well-studied quantum algorithm for combinatorial optimization: the Quantum Approximate Optimization Algorithm (QAOA) applied to the MaxCut problem on 3-regular graphs. We explore the idea of improving the solutions returned by the simplest version of the algorithm (depth-1 QAOA) using a form of postselection that can be efficiently simulated by state preparation. We derive theoretical upper and lower bounds showing that a constant (though small) increase of the fraction of satisfied edges is indeed achievable. Numerical experiments on large problem instances (beyond classical simulatability) complement and support our bounds. We also consider a distinct technique: local updates, which can be applied not only to QAOA but any optimization algorithm. In the case of QAOA, the resulting improvement can be sharply quantified theoretically for large problem instances and in absence of postselection. Combining postselection and local updates, the theory is no longer tractable but numerical evidence suggests that improvements from both methods can be combined.
\end{abstract}

% ------------------------------------------------------------------------------

\section{Introduction}
Quantum-enhanced combinatorial optimization is a possibly promising application of Near-Term Intermediate-Scale Quantum (NISQ) computers \cite{preskill_2018}. The best-known quantum algorithm for this task, the Quantum Approximate Optimization Algorithm (QAOA) was introduced by Farhi et al. in 2014 \cite{1411.4028}. QAOA is a variational quantum algorithm \cite{1509.04279,babbush_wiebe_mcclean_mcclain_neven_chan_2018,moll_barkoutsos_bishop_chow_cross_egger_filipp_fuhrer_gambetta_ganzhorn_et_al_2018,Peruzzo2014} drawing inspiration from the quantum adiabatic algorithm \cite{quant-ph/0001106}; more precisely, the latter may be simulated by QAOA as the depth of the variational ansatz goes to infinity. However, for relevance to near-term quantum computing, the ansatz needs to be kept shallow, in which case QAOA can no longer be analyzed through the lens of the quantum adiabatic algorithm.

QAOA is generally hard to simulate classically under plausible conjectures from complexity theory \cite{1602.07674}. Therefore, until large-scale quantum computers become available, QAOA and similar variational quantum algorithms may only be benchmarked on small problem instances on a classical computer, see e.g. \cite{nannicini_2019}. For a few specific problems, it is also possible to classically evaluate the performance on large instance sizes while full sampling of the variational state remains untractable. An example is the MaxCut problem on sparse graphs (e.g. random regular graphs), where the performance of QAOA (at least for moderate ansatz depth) can be efficiently obtained by tensor network simulations \cite{1909.02559}. The performance is quantified by the cost function of the optimization problem under consideration; for the MaxCut problem, this is the number of satisfied edges. The performance of QAOA is then measured by the cost achieved by bitstrings sampled from the variational state; an example is to consider the expected cost of such a bitstring. For certain problems (including Ising models on sparse graphs \cite{2004.09002} and the Sherrington-Kirkpatrick model in the infinite size limit \cite{1910.08187}), QAOA exhibits a phenomenon called \textit{concentration}, whereby costs of sampled bitstrings oncentrate around this expected value ---in fact, with exponential tails for Ising models on sparse graphs \cite{2004.09002}. In this case, the expected cost function completely summarizes the performance of QAOA indeed. As an example, for MaxCut on 3-regular graphs, the expected size of a cut sampled from depth-1 QAOA is $\left(\frac{3}{4} + \frac{1}{2\sqrt{3}}\right)(n + o(n)) \approx 1.0387(n + o(n))$ for a ``typical"\footnote{See section \ref{subsec:random_graphs} for a precise definition of a ``typical" random regular graph.} 3-regular graph with $n$ vertices and sampling a cut deviating from this expectation by $\Omega(n)$ is exponentially unlikely in $n$.

Recently, a series of rigorous results \cite{1910.08980,1905.07047,1910.08187,2005.08747} shed light on some shortcomings of low-depth QAOA. The limited performance of the latter prompted several attempts to improve the algorithm. A family of proposals consists to modify the cost function used to optimize the variational ansatz classically; examples include the conditional-value-at-risk \cite{barkoutsos_nannicini_robert_tavernelli_woerner_2020} and the Gibbs objective function \cite{li_fan_coram_riley_leichenauer_2020} rather than the usual expectation. To address some limitations of QAOA previously derived in their work, Koenig et al. \cite{1910.08980} propose a variant of QAOA known as Recursive QAOA (RQAOA), finding better performance than standard QAOA in numerical experiments on randomly generated regular graphs. A very recent proposal by Woerner et al. \cite{2009.10095} suggests to initialize the QAOA with an approximate solution of the optimization problem (obtained from a classical approximate algorithm or heuristic). The authors of \cite{2008.08615} attempt to qualitatively described physical mechanisms underlying the performance of quantum combinatorial optimization algorithms and based on this insight, describe a general roadmap to improve the latters. Among the directions proposed, is the possibility of resorting to non-unitary operations. In this work, we explore a variant of this idea, namely, performing a well-chosen projection on the state prepared by the QAOA circuit.

More precisely, we focus on the MaxCut problem on 3-regular graphs and on depth-1 QAOA. For a graph with $n$ vertices, we choose a subset of vertices $V_0$, $|V_0| = \Omega(n)$ that are assigned identically independently distributed values by the QAOA circuit, each vertex taking value $1$ or $-1$ with probability $\frac{1}{2}$. We then consider postselecting the bitstrings sampled from the QAOA conditioned on vertices in $V_0$ taking prescribed values $\left(\sigma_v\right)_{v \in V_0} \in \{-1, 1\}^{V_0}$. This corresponds to applying projection $\prod_{v \in V_0}\frac{1 + \sigma_vZ_v}{2}$ to the QAOA state, before renormalizing. We then ask whether a typical cut sampled after postselection improves on the one without postselection by an amount linear in $n$. The underlying idea is to circumvent the concentration property, which forbids that a cut sampled from QAOA (without postselection) improves over its expected size $1.0387(n + o(n))$ by $\Omega(n)$, except with an exponentially small probability in $n$. In contrast, in the postselected case, for each prescribed set of values $\left(\sigma_v\right)_{v \in V_0} \in \{-1, 1\}^{V_0}$, vertices in $V_0$ take these values with an exponentially small probability $\frac{1}{2^{|V_0|}} = \frac{1}{2^{\Omega(n)}}$ in the QAOA state. Hence, concentration no longer prevents that conditioned on vertices in $V_0$ taking a well-chosen set of prescribed values, the size of the cut sampled from QAOA improves linearly over its (non-postselected) expected value $1.0387(n + o(n))$. This raises the question how to identify an appropriate set of conditioning values $\left(\sigma_v\right)_{v \in V_0}$. Once such values have been identified, it remains to efficiently simulate postselection of bitstrings or equivalently the application of the corresponding projection to the QAOA state. The second problem is in fact straightforward owing to the locality of the QAOA ansatz at constant depth. The first question requires a more involved analysis and constitutes the main point of this paper.

This work is organized as follows. In section \ref{sec:background}, we review some technical preliminaries on the Quantum Approximate Optimization Algorithm together with random graphs and independent sets, which play an important role in the analysis. We then summarize and discuss our main results in section \ref{sec:results}. In section \ref{sec:ring_disagrees}, we apply our modification of QAOA to a ring graph instead of a random 3-regular graph; the analysis is remarkably easier in this case and will serve as future reference. Section \ref{sec:derivation_results} then presents the main line of the derivations, with more technical results postponed to appendix \ref{sec:technical_results_postselect_qaoa}. Finally, in appendix\footnote{These results are disconnected of the main line of this work but of independent interest, hence the choice to include them in appendix.} \ref{sec:bravyi_et_al_generalization}, we prove a few results on the limitations of variational quantum optimization with shallow circuits, generalizing arguments from \cite{1910.08980}.

\section{Our results}

Our main result is that the advantage achieved by postselected QAOA over standard QAOA can be expressed as an Ising model on the prescribed values $\left(\sigma_{v_0}\right)_{v_0 \in V_0}$: $-\sum_{v_0, v_0' \in V_0}J(G, V_0)_{v_0, v_0'}\sigma_{v_0}\sigma_{v_0'}$. The model is on a smaller, but possibly more connected, graph than the original one. The couplings $J(G, V_0)_{v_0, v_0'}$ depend on the 3-regular graph $G$ and the set $V_0$ and are efficiently computable classically. This allows to give upper and lower bounds on the improvement achievable by postselected QAOA.

We start by giving an upper bound holding \textit{in expectation}, which characterizes the advantage for an ``average" random regular graph (in a sense to be precised later). Under a conjecture on the number of 2-independent sets (see section \ref{subsec:independent_sets} for a definition) of prescribed size in a typical random regular graph, this bound can be promoted to one holding \textit{with high probability} on the choice of graph and set $V_0$. The latter bound implies that the postselection strategy just described may achieve a mild advantage over standard QAOA on large graph instances. The improvement, as measured per the increase of the fraction of satisfied edges in the cut, is theoretically upper-bounded by $0.06$ for a typical random 3-regular graph, while numerical experiments on large graph instances (beyond full classical simulation) achieve $0.02$. Finally, an easy theoretical lower bound guarantees $0.0013$.

Our modification to QAOA can be combined with further efficient postprocessing of the sampled bitstrings, relying on local updates of the cut. The advantage of this procedure is rigorously quantifiable in the infinite size limit in absence of postselection: it allows to increase the fraction of satisfied edges by approximately $0.06$. With postselection, we can no longer provide such guarantees, but numerical benchmarks on large random graph instances show an improvement of order $0.08$ in the fraction of satisfied edges when combining postselection with local updates. The postselection improvement adds up with the local update one, suggesting they are of different nature.

To put these figures in context, it has been rigorously shown that a maximum cut in a typical random 3-regular graph includes at most $92.41\%$ \cite{2009.10483} and at least $90.67\%$ \cite{gamarnik_li_2017} of the edges (empirical evidence suggests a lower-bound of $92.13\%$ instead \cite{kardo2012}). In contrast, when applied to a typical random 3-regular graph, depth-1 QAOA outputs a cut comprising $69.2\%$ of edges; this therefore increases to $77\%$ with the modifications of QAOA just described, which lies between the performance of depth-2 and depth-3 QAOA. Therefore, at least in the restricted setting considered, the improvement obtained by our methods effectively increases the depth by 1 or 2. However, it remains modest as compared to high-depth QAOA and classical algorithms \cite{1909.02559}.

Finally, in appendix \ref{sec:bravyi_et_al_generalization}, we derive results generalizing recent work by Bravyi et al. \cite{1910.08980} on the limitations of variational quantum optimization with shallow circuits. In this work, the authors proved upper bounds for the performance of a family of circuits generalizing QAOA: $\mathbf{Z}_2$-symmetric finite-range circuits. In particular, the approximation ratio of such circuits on the (anti)ferromagnetic Ising model on a ring graph was considered. We generalize the arguments to Ising models with arbitrary couplings (drawn from $\{-1, 1\}$) on grid graphs and random regular graphs. We also rigorously show that, maybe surprisingly, the method used by the authors to prove lower bounds matching their upper bounds does not generalize to random regular graphs.

\section{Related work}
The first work on QAOA \cite{1411.4028} investigates the performance of the algorithm on a well-know combinatorial optimization problem: MaxCut on 3-regular graphs for a QAOA ansatz of minimal depth, as the graph size tends to infinity. The quantum algorithm achieved a worse approximation ratio than the best-known classical polynomial-time algorithm: the Goemans-Williamson algorithm \cite{goemans_williamson_1995}. In contrast, it did outperform random assignment, which used to be the reference efficient classical algorithm for decades before the Goemans-Williamson algorithm was discovered. Shortly later, a work by the same authors \cite{1412.6062} applied the QAOA to another constraint satisfaction problem: Max-E3LIN2 and established performance guarantees showing superiority over state-of-the-art efficient classical algorithms at the time. Unfortunately, the excitement was short-lived as a classical algorithm inspired from QAOA but beating it was proposed by Barak et al. \cite{barak_2015} few months later. Since then, it has remained unclear whether the QAOA could achieve any advantage in combinatorial optimization. Indeed, theoretical predictions on the performance of QAOA, whether positive or negative, remain scarce today.

More recently, a series of results \cite{1910.08980,1905.07047,1910.08187,2005.08747} established several limitations to the QAOA and variational quantum optimization in general. Koenig et al. \cite{1910.08980} exihibited the limitations of a family of quantum circuits generalizing the QAOA ansatz: $\mathbf{Z}_2$-symmetric finite range circuits (see e.g. section \ref{subsec:qaoa} for a definition) when applied to the MaxCut problem on certain regular graphs. In \cite{2004.09002}, Farhi et. al demonstrated that for QAOA circuit of constant depth or depth scaling sublogarithmically in the problem size, the approximation ratio on the MaxIndependentSet problem was bounded by $\frac{1}{2} + \frac{1}{2\sqrt{2}} \approx 0.853$, which can likely be taken down to $\frac{1}{2}$. The argument mainly relies on a statistical property of independent sets in random regular graphs (see definition in section \ref{sec:background}), the \textit{clustering property}, which is also the cornerstone of an analogous no-go result for classical algorithms \cite{Rahman2017,Gamarnik2017}. Besides the MaxIndependentSet problem, this property also for generalizations of MaxCut \cite{1707.05386}, but not for MaxCut itself; in other words, the strategy cannot be adapted to prove limitations of QAOA for MaxCut on random regular graphs. The QAOA, when applied to graph problems, bears similarities to an important family of classical algorithms: local algorithms \cite{1409.5214,Bamas2019}. A study by Hastings \cite{1905.07047} compares QAOA and classical local algorithms applied to the Max-3-LIN-2 and MaxCut problems; classical local algorithms are found to outperform QAOA at $p = 1$ and the author gives evidence that this should persist at higher depth. Depsite the restrictivity of these results for the performance of QAOA, they only cover the case where the depth of the ansatz is constant or varies sublogarithmically with the problem size. However, as pointed out in \cite{2004.09002}, a single-digit depth may violate this assumption even for reasonable optimization problems on millions of variables, so these results may not be so restrictive including for near-term quantum computing.
\section{Technical preliminaries}
\label{sec:background}
In this section, we review a few technical preliminaries required for the precise description of postselected QAOA, the statement of our results and their derivation. We start by discussing relevant concepts and results from random graph theory (section \ref{subsec:random_graphs}). We then introduce $k$-independent sets (section \ref{subsec:independent_sets}) which are a generalization of independent sets. Finally, we recap we properties of the Quantum Approximate Optimization algorithm relevant to this work (section \ref{subsec:qaoa}).

\subsection{Random graphs}
\label{subsec:random_graphs}
The analysis of postselected QAOA will require to consider random 3-regular graphs. We therefore start with a few definitions and results from (random) graph theory. We will systematically denote by $V$ the vertex set and $E$ the edge set of a graph.

We recall the definition of a $d$-regular graph, an important family of graphs in the analysis of classical and quantum optimization algorithms:
\begin{defn}[$d$-regular graph]
	A graph $G$ is called $d$-regular if each vertex of $G$ has exactly $d$ neighbours. 
\end{defn}
When trying to establish lower or upper bounds on the performance of an algorithm on $d$-regular graphs with $n$ vertices, it is frequently easier to show that the bounds hold not for all graphs but for a fraction of graphs which approaches $1$ as $n$ goes to infinity. The latter is equivalent to stating that the bound holds \textit{with high probability}, meaning with probability approaching $1$ as $n \to \infty$, for a random graph uniformly sampled from $d$-regular graphs. This motivates the notion of \textit{random $d$-regular graph}:
\begin{defn}[Random $d$-regular graph]
	\label{def:random_d_regular_graph}
	A random $d$-regular graph is a random variable which is a graph, sampled with uniform probability from all $d$-regular graphs.
\end{defn}
There exist efficient algorithms to generate random $d$-regular graphs (see e.g. \cite{steger_wormald_1999}). To analyze the performance of algorithms on random regular graphs, it is often convenient to describe the latters through the so-called \textit{configuration model}:
\begin{defn}[Configuration model]
\label{def:configuration_model}
	The configuration model on $n$ vertices of degree $d$ is a probability distribution on multigraphs\footnote{A multigraph differs from a graph in that it may have loops (edge from a vertex to itself) and multi-edges (repeated edge between two vertices).} of $n$ vertices and degree $d$ such that a multigraph is sampled according to the following process. First, consider the set $\left\{(1, 1), (1, 2), \ldots, (1, d), (2, 1), \ldots, (2, d), \ldots, (n, 1), \ldots, (n, d)\right\}$ and uniformly sample a perfect matching $\Delta$ of this set. Then, associate a multigraph $G$ with vertex set $\{1, \ldots, n\}$ to this perfect matching according to the relation:
	\begin{align}
		\{i, i'\}\textrm{ is a edge of } G \iff \exists 1 \leq j, j' \leq d\, \leq \{(i, j), (i', j')\} \in \Delta
	\end{align}
\end{defn}
Intuitively, element $(i, j)$ from the set\\ $\left\{(1, 1), (1, 2), \ldots, (1, d), (2, 1), \ldots, (2, d), \ldots, (n, 1), \ldots, (n, d)\right\}$ can be regarded as a \textit{half-edge} attached to vertex $i$, which, when matched to another half-edge (either belonging to $i$ or another vertex), gives an edge in the multigraph. We will often make use of this informal terminology in the following proofs. We will denote by $\mathbf{P}_G[\ldots]$ and $\mathbf{E}_G[\ldots]$ the probabilities and expectations calculated over the configuration model, where the $\ldots$ involve a random variable $G$ sampled from the configuration model.

The following proposition now specifies the connection between the configuration model and random regular graphs:
\begin{prop}[Configuration model and random regular graphs {\cite[section 3]{1503.03923}}]
	For fixed $d$, the probability of an $n$-vertices degree-$d$ multigraph sampled from the configuration model being a graph (i.e., having no loop or multiple edge) is lower-bounded by a constant as $n \to \infty$. Besides, conditioned on being a graph, a multigraph sampled from the configuration model is distributed as a random $d$-regular graph.
\end{prop}
This implies that if a property holds with high probability for a multigraph sampled from the configuration model, it also holds with high probability for a random regular graph. This fact proves useful since analyzing (multi)graphs sampled from the configuration model is usually more tractable than analyzing random regular graphs.

We now state a useful result on the local structure of random $d$-regular graphs, saying that locally, such a graph looks like a tree. For that purpose, we need to define the neighbourhood of a vertex in a graph.
\begin{defn}[Neighbourhood of a vertex in a graph]
\label{def:neighbourhood_vertex_graph}
	Let $G = (V, E)$ denote an arbitrary graph (or multigraph). For $v \in V$ and $r \geq 0$, the $r$-neighbourhood of $v$ in $G$, denoted by $B_G(v, r)$, is the set of vertices $w \in V$ such that there exists a path of length $\leq r$ from $v$ to $w$. Furthermore, we define $\partial B_G(v, 0) := B_G(v, 0)$ and for all $r \geq 1$, $\partial B_G(v, r) := B_G(v, r) - B_G(v, r - 1)$ so that $\partial B_G(v, r)$ contains the vertices which can be connected to $v$ by a length-$r$ path but by no shorter path.
\end{defn}
For convenience, we also introduce the distance between two vertices in a graph:
\begin{defn}
    Given a graph or multigraph $G = (V, E)$ and two vertices $v, v' \in V$, the distance between $v$ and $v'$ in $G$, denoted by $d_G(v, v')$, is the length of the shortest path between $v$ and $v'$ in $G$.
\end{defn}
We can now state the following classical result on vertex neighbourhoods in random regular graphs:
\begin{prop}[Neighbourhoods of random regular graphs are trees {\cite[proposition 2.2]{gamarnik_sudan_2014}}]
	\label{prop:tree_neighbourhood}
	Let $d \geq 2$ and $r \geq 0$ be fixed. As $n \to \infty$, with high probability almost all neighbourhoods of an $n$-vertices random $d$-regular graph are trees. More precisely, for any $\varepsilon > 0$, for large enough $n$,
	\begin{align}
		\mathbf{P}\left[\left|\left\{v \in V : B_G(v, r) \textnormal{ not tree}\right\}\right| \geq \varepsilon n\right] & \leq \varepsilon,
	\end{align}
	where the probability distribution is on random $d$-regular graphs.
\end{prop}

\subsection{$k$-independent sets}
\label{subsec:independent_sets}
The description and analysis of postselected QAOA relies on the construction of $2$-independent sets, a generalization of independent sets. The general definition of a $k$-independent set is given hereafter:
\begin{defn}[$k$-independent set {\cite{duckworth_zito_2003}}]
    Let $G = (V, E)$ a graph. A set $V_0 \subset V$ is called $k$-independent if the distance between any two vertices of $V_0$ is at least $k + 1$.
\end{defn}
This case $k = 1$ corresponds to the familiar notion of independent set. We will sometimes abbreviate ``$k$-independent" as ``$k$-id" for space reasons. It is NP-hard to find an independent set of maximum cardinality in graph and the same holds for $k$-independent set when $k \geq 2$ \cite{duckworth_zito_2003}. \cite{wormald_2003} proposes a greedy algorithm that constructs a $k$-independent set in a random $d$-regular graph and gives an explicit lower bound on the size of the latter for $k \in \{2, 3, 4, 5\}$ and $d \in \{3, 4, 5, 6, 7\}$. It also states an upper bound for the maximum size of a $k$-independent set, holding for all but an exponentially small subset of regular graphs; these bounds are consistent with the estimates derived in section \ref{subsec:2_independent_sets_estimates} of this paper. For convenience, we restate the results from \cite{wormald_2003} that apply to 2-independent sets of 3-regular graphs:
\begin{prop}[Lower bound on size of 2-independent set in typical 3-regular graph {\cite{wormald_2003}}]
\label{prop:lower_bound_2_independent_set}
    There exists an efficient algorithm which, given a random 3-regular graph of $n$ vertices, produces asymptotically almost surely a 2-independent set of size $\geq 0.204n$.
\end{prop}

\begin{prop}[Upper bound on size of 2-independent set in typical 3-regular graph {\cite{wormald_2003}}]
\label{prop:upper_bound_k_independent_set}
    There exists a constant\footnote{The notation is consistent with $\alpha^*_d$ introduced in section \ref{subsec:2_independent_sets_estimates}.} $\alpha^*_3 \approx 0.236$ such that all 3-regular graphs of size $n$, except for an exponentially small fraction (in $n$) of them, contain no independent set of size $\geq \alpha_3n$.
\end{prop}

\subsection{The Quantum Approximate Optimization Algorithm (QAOA)}
\label{subsec:qaoa}
The Quantum Approximate Optimization Algorithm (QAOA), originally introduced in \cite{1411.4028}, is a variational quantum algorithm which aims at finding an approximate ground state of a Hamiltonian on $n$ qubits $\hat{H}_C$. The variational ansatz is given by:
\begin{align}
\label{eq:qaoa_ansatz}
    \ket{\psi} = \ket{\psi(\beta_1, \gamma_1, \ldots, \beta_p, \gamma_p)} & := \overrightarrow{\prod_{1 \leq k \leq p}}e^{-\frac{i}{2}\beta_k\hat{H}_B}e^{-\frac{i}{2}\gamma_k\hat{H}_C}\ket{+}^{\otimes n},
\end{align}
where
\begin{align}
    \hat{H}_B & := \sum_{0 \leq k < n}\hat{X}_k.
\end{align}
The $2p$ parameters $\beta_1, \gamma_1, \ldots, \beta_p, \gamma_p$ are optimized to obtain a state which is suitably close to the ground state of $\hat{H}_C$ This is usually done by minimizing the expected energy
\begin{align}
    \braket{\psi(\beta_1, \gamma_1, \ldots, \beta_p, \gamma_p)|\hat{H}_C|\psi(\beta_1, \gamma_1, \ldots, \beta_p, \gamma_p)},
\end{align}
though using different cost functions has occasionally been proposed \cite{barkoutsos_nannicini_robert_tavernelli_woerner_2020,li_fan_coram_riley_leichenauer_2020}. The $p$ parameter will sometimes be referred to as the \textit{depth parameter} of the QAOA (the depth of the quantum circuit implementing ansatz \ref{eq:qaoa_ansatz} is proportional to $p$ indeed). QAOA with depth parameter $p$ will also be referred to as \textit{depth-$p$ QAOA}.

The case where $\hat{H}_C$ is the Hamiltonian of a classical Ising Hamiltonian on some graph is well-studied. Given a graph $G = (V, E)$ ($V = \{0, 1, \ldots, n - 1\}$: vertex set; $E$: edge set), an Ising Hamitonian on $G$ has the form
\begin{align}
    \hat{H}_C & = \sum_{e = \{e_0, e_1\} \in E}J_e\hat{Z}_{e_0}\hat{Z}_{e_1},
\end{align}
where the $\left(J_e\right)_{e \in E}$ are arbitrary real numbers and we indexed the qubits by vertices of the graph. The MaxCut Hamiltonian on $G$ is an important special case:
\begin{align}
    \hat{H}_{\textnormal{MaxCut}}(G) & := -\sum_{e = \{e_0, e_1\} \in E}\frac{1 - \hat{Z}_{e_0}\hat{Z}_{e_1}}{2}.
\end{align}
A computational basis state $\ket{b_{n - 1}\ldots b_0}$ is an eigenstate of $\hat{H}_{\textnormal{MaxCut}}(G)$ with eigenvalue minus the number of satisfied edges in the cut defined by vertex sets $\{v \in V\,:\,b_v = 0\}$ and $\{v \in V\,:\,b_v = 1\}$.

When qubits are labelled by the vertices of a graph, as is natural when considering an Ising Hamiltonian on a graph, the notion of range-$R$ quantum circuit can be defined, following \cite{1910.08980}:
\begin{defn}
Let $G = (V, E)$ a graph. A circuit $U$ acting on qubits labelled by the vertices of $V$ is said to have range $R$ if for all $v \in V$, all single-qubit observable $\mathcal{O}_v$ supported on $v$, $U^{\dagger}\mathcal{O}_vU$ is supported on the $R$-neighbourhood of $v$ in $G$.
\end{defn}
The following easy proposition states that for a $\mathbf{Z}_2$-symmetric range-$R$ circuit, measuring $n_0$ qubits belonging to an $2R$-independent set is equivalent to performing $n_0$ independent coin flips:
\begin{prop}
	\label{prop:range_r_circuit_sparse_set}
	Let $G = (V, E)$ a graph and let $\ket{\psi}$ a state on qubits labelled by $V$ that is prepared by applying a range-$R$ (with respect to $G$) $\mathbf{Z}_2$-symmetric circuit to a product state. Let $V_0$ be a $2R$-independent set of $G$. Consider the measurement on $\ket{\psi}$ of qubits labelled by vertices from $V_0$ in the computational basis. Then the measurement outcomes are i.i.d. and for each qubit, the outcome is $-1$ with probability $\frac{1}{2}$ and $1$ with probability $\frac{1}{2}$. In other words, for all $\left(\sigma_{v_0}\right)_{v_0 \in V_0}$,
	\begin{align}
	\left\lVert\prod_{v_0 \in V_0}\frac{1 + \sigma_{v_0}Z_{v_0}}{2}\ket{\psi}\right\rVert^2 = \frac{1}{2^{|V_0|}}.
	\end{align}
\end{prop}
Following the same reference, one can also define $\mathbf{Z}_2$-symmetric circuits:
\begin{defn}[$\mathbf{Z}_2$-symmetric states and circuits]
    A state $\ket{\psi} \in \mathbf{C}^{2^n}$ on $n$ qubits is called $\mathbf{Z}_2$-symmetric if $X^{\otimes n}\ket{\psi} = \ket{\psi}$. A quantum circuit $U \in \mathbf{U}(2^n)$ acting on $n$ qubits is called $\mathbf{Z}_2$-symmetric if $X^{\otimes n}UX^{\otimes n} = U$. A state will be called ``prepared by a $\mathbf{Z}_2$-symmetric circuit" if it can be obtained by applying a $\mathbf{Z}_2$-symmetric circuit to a $\mathbf{Z}_2$-symmetric product state.
\end{defn}
The following is an easy consequence of $\mathbf{Z}_2$-symmetry:
\begin{prop}
For a $\mathbf{Z}_2$-symmetric state $\ket{\psi}$ on $n$ qubits: $\braket{\psi|Z_k|\psi} = 0$ for all $0 \leq k < n$.
\end{prop}
It is straightforward to establish that depth-$p$ QAOA is a range-$p$ $\mathbf{Z}_2$-symmetric quantum circuit.

\section{Results}
\label{sec:results}

In this section, we review the main results of the paper on the performance of postselected QAOA, with proofs deferred to section \ref{sec:derivation_results}. We first precisely recall the principle of postselected QAOA. Then, we present theoretical upper and lower bounds on its performance. Finally, we comment on related numerical experiments.

\subsection{Postselected QAOA}
\label{subsec:postselected_qaoa}
Given a 3-regular graph $G = (V, E)$ ($n := |V|$), postselected QAOA for MaxCut on $G$ proceeds as follows:
\begin{itemize}
    \item Find the optimal parameters $\beta, \gamma$ for depth-1 QAOA on $G$. With high probability, these parameters can be taken as $\beta^* = -\frac{\pi}{4}(1 + o(1)), \gamma^* = \arctan\left(\frac{1}{\sqrt{2}}\right)\left(1 + o(1)\right)$ as $n \to \infty$ \cite{1411.4028}.
    \item Select a 2-independent set $V_0$ of $G$.
    \item Find postselected values $\left(\sigma_{v_0}\right)_{v_0 \in V_0} \in \{-1, 1\}^{V_0}$ for vertices in $V_0$ so as to maximize 
    \begin{align}
        \frac{\braket{\psi_G|\prod_{v_0 \in V_0}\frac{1 + \sigma_{v_0}Z_{v_0}}{2}H_{\textnormal{MaxCut}}(G)\prod_{v_0 \in V_0}\frac{1 + \sigma_{v_0}Z_{v_0}}{2}|\psi_G}}{\left\lVert\prod_{v_0 \in V_0}\frac{1 + \sigma_{v_0}Z_{v_0}}{2}\ket{\psi_G}\right\rVert^2}.
    \end{align}
    (alternatively, one may approximately maximize, using a heuristic or approximate algorithm instead of an exact one to find $\left(\sigma_{v_0}\right)_{v_0 \in V_0}$).
    \item Sample bitstrings from the state $\frac{\prod_{v_0 \in V_0}\frac{1 + \sigma_{v_0}Z_{v_0}}{2}\ket{\psi_G}}{\left\lVert \prod_{v_0 \in V_0}\frac{1 + \sigma_{v_0}Z_{v_0}}{2}\ket{\psi_G}\right\rVert}$ (which can be obtained by applying the QAOA circuit to a well-chosen initial state as implied by proposition \ref{prop:simulate_postselection}.
\end{itemize}
The last step poses the challenge of simulating sampling from the state $\frac{\prod_{v_0 \in V_0}\frac{1 + \sigma_{v_0}Z_{v_0}}{2}\ket{\psi_G}}{\left\lVert \prod_{v_0 \in V_0}\frac{1 + \sigma_{v_0}Z_{v_0}}{2}\ket{\psi_G}\right\rVert}$. A possibility would be to sample from $\ket{\psi_G}$ and postselect on spins from $V_0$ having values $\left(\sigma_{v_0}\right)_{v_0 \in V_0}$, but since $\left\lVert\prod_{v_0 \in V_0}\frac{1 + \sigma_{v_0}Z_{v_0}}{2}\ket{\psi_G}\right\rVert^2 = \frac{1}{2^{|V_0|}}$ (proposition \ref{prop:range_r_circuit_sparse_set}) and $|V_0| = \Omega(n)$, time $2^{\Omega(n)}$ would be required before seeing one postselected sample. Fortunately, there is a more efficient alternative: preparing $\frac{\prod_{v_0 \in V_0}\frac{1 + \sigma_{v_0}Z_{v_0}}{2}\ket{\psi_G}}{\left\lVert \prod_{v_0 \in V_0}\frac{1 + \sigma_{v_0}Z_{v_0}}{2}\ket{\psi_G}\right\rVert}$ directly thanks to the finite-range of QAOA. The result is contained in the following proposition:
\begin{prop}
\label{prop:simulate_postselection}
	Let $G = (V, E)$ a graph and let $\ket{\psi} = U\ket{\varphi}^{\otimes |V|}$ a state of qubits labelled by $V$ be prepared by a range-$R$ circuit $U$ applied to a product state $\ket{\varphi}^{\otimes |V|}$. Let $V_0 \subset V$ a $2R$-independent set in $G$. Then postselection on the measurement outcomes of qubits labelled by $V_0$ can be simulated by applying $U$ to a state different from $\ket{\varphi}^{\otimes |V|}$. Precisely, given $\left(\sigma_{v_0}\right)_{v_0 \in V_0} \in \{-1, 1\}^{V_0}$, there exists an explicitly constructible quantum circuit $\widetilde{U}\left(\left(\sigma_{v_0}\right)_{v_0 \in V_0}\right)$ depending on the postselected measurement outcomes of qubits labelled by $V_0$, such that
	\begin{align*}
		\frac{\prod_{v_0 \in V_0}\frac{1 + \sigma_{v_0}Z_{v_0}}{2}\ket{\psi}}{\left\lVert\prod_{v_0 \in V_0}\frac{1 + \sigma_{v_0}Z_{v_0}}{2}\ket{\psi}\right\rVert} & = U\widetilde{U}\left(\left(\sigma_{v_0}\right)_{v_0 \in V_0}\right)\ket{\varphi}^{\otimes |V|}.
	\end{align*}
	$\widetilde{U}$ can be implemented by $|V_0|$ parallel unitaries, each of which corresponds to a vertex of $V_0$; the unitary associated to $v_0$ acts on the qubits in $B_G(v_0, R)$.
	\begin{proof}
		The result follows easily from commuting the projectors through the circuit unitary $U$:
		\begin{align*}
			& \frac{\prod_{v_0 \in V_0}\frac{1 + \sigma_{v_0}Z_{v_0}}{2}\ket{\psi}}{\left\lVert\prod_{v_0 \in V_0}\frac{1 + \sigma_{v_0}Z_{v_0}}{2}\ket{\psi}\right\rVert}\\%
			& \propto U\left(\prod_{v_0 \in V_0}U^{\dagger}\frac{1 + \sigma_{v_0}Z_{v_0}}{2}U\right)\ket{\varphi}^{\otimes |V|}
		\end{align*}
		Now, recalling the range-$R$ assumption on $U$, $U^{\dagger}\frac{1 + \sigma_{v_0}Z_{v_0}}{2}U$ is supported on $B_G(v_0, R)$. Since $V_0$ is $2R$-independent, the $B_G(v_0, R)$ are pairwise disjoint (when iterating over $v_0 \in V_0$) and therefore the  $U^{\dagger}\frac{1 + \sigma_{v_0}Z_{v_0}}{2}U$ have disjoint supports. This completes the proof.
	\end{proof}
\end{prop}
Applied to depth-1 QAOA for MaxCut of 3-regular graphs, this proposition implies that postselection on the values of vertices belonging to a 2-independent set can be simulated by state preparation.

\subsection{Theoretical upper and lower bounds}
We now describe upper and lower bounds on the performance of the algorithm just described. The latter is measured by the increase in the expected number of satisfied edges in cuts sampled from the postselected state $\frac{\prod_{v_0 \in V_0}\frac{1 + \sigma_{v_0}Z_{v_0}}{2}\ket{\psi_G}}{\left\lVert \prod_{v_0 \in V_0}\frac{1 + \sigma_{v_0}Z_{v_0}}{2}\ket{\psi_G}\right\rVert}$ as compared to the non-postselected state $\ket{\psi_G}$; that is,
\begin{align}
\label{eq:cost_postselected_qaoa}
    \max_{\left(\sigma_{v_0}\right)_{v_0 \in V_0} \in \{-1, 1\}^{V_0}}\frac{\braket{\psi_G|\prod_{v_0 \in V_0}\frac{1 + \sigma_{v_0}Z_{v_0}}{2}H_{\textnormal{MaxCut}}(G)\prod_{v_0 \in V_0}\frac{1 + \sigma_{v_0}Z_{v_0}}{2}|\psi_G}}{\left\lVert\prod_{v_0 \in V_0}\frac{1 + \sigma_{v_0}Z_{v_0}}{2}\ket{\psi_G}\right\rVert^2} - \braket{\psi_G|H_{\textnormal{MaxCut}(G)}|\psi_G}
\end{align}
The following proposition is a partial result bounding this quantity:
\begin{prop}
\label{prop:upper_bound_conditioning_qaoa_expectation}
    Let $V_0 \subset V = [n]$ a set of vertices with $\alpha := \frac{|V_0|}{n} \in (0, \alpha_3^*)$ (where $\alpha_d^*$ is defined in paragraph \ref{subsec:independent_sets}). Then the expected improvement of postselected QAOA over QAOA conditioned on $V_0$ being 2-independent is bounded as follows (where the expectation is taken over random regular graphs $G$ on vertex set $V$ in the configuration model):
    \begin{align}
    \label{eq:upper_bound_expectation}
        & \mathbf{E}_G\left[\max_{\left(\sigma_{v_0}\right)_{v_0 \in V_0} \in \{-1, 1\}^{V_0}}\frac{\braket{\psi_G|\prod_{v_0 \in V_0}\frac{1 + \sigma_{v_0}Z_{v_0}}{2}H_{\textnormal{MaxCut}}(G)\prod_{v_0 \in V_0}\frac{1 + \sigma_{v_0}Z_{v_0}}{2}|\psi_G}}{\left\lVert\prod_{v_0 \in V_0}\frac{1 + \sigma_{v_0}Z_{v_0}}{2}\ket{\psi_G}\right\rVert^2}\right.\nonumber\\
        & \hspace{0.1\textwidth} - \braket{\psi_G|H_{\textnormal{MaxCut}(G)}|\psi_G}\,\bigg|\,V_0\textrm{ 2-id in } G\Bigg] \leq \frac{\alpha^2(0.18125 - 0.83875\alpha + 0.99\alpha^2)}{(0.5 - \alpha)^3}n + \mathcal{O}(1).
    \end{align}
\end{prop}
This is proven in section \ref{sec:derivation_results}. Unfortunately, the result is not very natural as stated. Indeed, it describes the situation where one chooses a set $V_0 \subset [n]$ (independent of any graph) before drawing a graph $G$ at random with vertex set $[n]$; then, conditioned on $V_0$ being 2-independent in $G$, the improvement achievable by postselected QAOA is bounded in expectation by the right-hand-side of equation \ref{eq:upper_bound_expectation}. However, in the algorithm described in section \ref{subsec:postselected_qaoa}, $V_0$ is allowed to depend on $G$. Besides, it would be desirable to obtain bounds on the improvement that hold with high probability (on the random graph and choice of 2-independent set) and not only in expectation. Such results can be derived from the previous one at the cost of assuming a conjecture on the number of 2-independent sets of prescribed size in random regular graphs. To introduce the conjecture, we first need the following definition:
\begin{defn}[2-independent-set-typical graph]
	\label{def:sparse_set_typical_graph}
	Let $d \geq 3$ and $\alpha \in \left(0, \alpha^*_d\right)$. A $d$-regular graph $G_0$ with $n$ vertices is called 2-independent-set-$(\alpha, \varepsilon)$-typical if the following holds:
	\begin{align}
	\sum_{\substack{V_0 \subset V\\|V_0| = \alpha n}}\mathbf{1}_{V_0\textrm{ 2-id in } G_0} & \geq \exp\left(-\varepsilon n\right)\mathbf{E}_{G}\left[\sum_{\substack{V_0 \subset V\\|V_0| = \alpha n}}\mathbf{1}_{V_0\textrm{ 2-id in } G}\right].
	\end{align}
\end{defn}
The conjecture states that a random regular graph is typical with high probability:
\begin{con}
	\label{conj:independent_set_typical_graph_frequent}
	For every $\alpha \in \left(0, \alpha^*_3\right)$ and $\varepsilon > 0$, there exists $\overline{n} = \overline{n}(\alpha, \varepsilon)$ such that for $n \geq \overline{n}(\varepsilon)$, a $n$-vertices 3-regular graph is $(\alpha, \varepsilon)$-typical with high probability.
\end{con}
Loosely speaking, the conjecture means that for most regular graphs, the number of 2-independent sets of a prescribed size cannot be much smaller than its expected value on random regular graphs. Though the question of determining the maximum size of an independent set has been extensively studied in the literature (see e.g. \cite{Ding2016} for a recent sharp result on random regular graphs), much less work has been dedicated to counting independent sets. Examples include \cite{Engbers2013,Lehner2017,GAN2014,2002.03189}, which consider extremal values of the number of independent sets of fixed size for several classes of graphs. However, we are not aware of works counting the number of ($k$-)independent sets of prescribed cardinality in typical graphs and typical random regular graphs in particular. For $\alpha = 0.204$ (a value motivated by proposition \ref{prop:lower_bound_2_independent_set}), numerical evidence can be collected for the validity of the conjecture up to graphs of size $n \approx 40$. Assuming conjecture \ref{prop:upper_bound_conditioning_qaoa_expectation}, the following ``high-probability variant" of proposition \ref{prop:upper_bound_conditioning_qaoa_expectation} holds
\begin{prop}
	\label{prop:upper_bound_conditioning_qaoa}
    Let $\alpha \in (0, \alpha_3^*)$. Let $G = (V, E)$ a random 3-regular graph with $n$ vertices sampled from the configuration model and $V_0$ a 2-independent set of $G$ of size $\alpha n$. Then under conjecture \ref{conj:independent_set_typical_graph_frequent}, for all $\varepsilon > 0$, there exists $\overline{n} = \overline{n}(\varepsilon)$ such that for all $n \geq \overline{n}$,
	\begin{align}
		& \max_{\left(\sigma_{v_0}\right)_{v_0 \in V_0} \in \{-1, 1\}^{V_0}}\frac{\braket{\psi_G|\prod_{v_0 \in V_0}\frac{1 + \sigma_{v_0}Z_{v_0}}{2}H_{\textnormal{MaxCut}}(G)\prod_{v_0 \in V_0}\frac{1 + \sigma_{v_0}Z_{v_0}}{2}|\psi_G}}{\left\lVert\prod_{v_0 \in V_0}\frac{1 + \sigma_{v_0}Z_{v_0}}{2}\ket{\psi_G}\right\rVert^2} - \braket{\psi_G|H_{\textnormal{MaxCut}(G)}|\psi_G}\nonumber\\
		& \hspace{0.1\textwidth} \leq \left(\frac{\alpha^2(0.18125 - 0.83875\alpha + 0.99\alpha^2)}{(0.5 - \alpha)^3} + 0.59\varepsilon\right)n + O(1).
	\end{align}
	with high probability on $G$ and the choice of 2-independent set $V_0$ of $G$.
\end{prop}
For instance, taking $\varepsilon = \frac{1}{100}$ and $\alpha = 0.204$ (following proposition \ref{prop:lower_bound_2_independent_set}) in the latter proposition, the improvement is upper-bounded by $0.06|E|$ for sufficiently large $n$. The proof is given in section \ref{sec:derivation_results}.

This upper bound can be compared to a lower bound (not relying on any conjecture). The latter rests on algorithm \ref{prop:lower_bound} to select a 2-independent set.
\begin{algorithm}
	\caption{2-independent vertex set}
	\label{alg:systematic_sparse_vertex_set}
	$allowedVertices \leftarrow V$\;
	$chosenVertices \leftarrow \varnothing$\;
	$candidatePairs \leftarrow \{ \textrm{pairs of vertices with distance 4} \}$\;
	\While{$candidatePairs$ not empty}{
	    $pair \leftarrow \textrm{any element from } candidatePairs$\;
	    $candidatePairs \leftarrow candidatePairs - \{\textrm{pairs having at least one vertex }$
	\hspace{185px}$\textrm{distant from any vertex of } pair \textrm{ by at most } 5\}$
	}
	\Return $chosenVertices$
\end{algorithm}
Choosing $V_0$ with this algorithm and postselecting on all vertices from $V_0$ taking the same value yields a very modest, though linear improvement in $|E|$ (which would be exponentially unlikely without postselection):
\begin{prop}
\label{prop:lower_bound}
	Let $G$ a random 3-regular graph and $V_0$ a 2-independent set of $G$ selected by algorithm \ref{alg:systematic_sparse_vertex_set}. Then, with high probability (on the choice of $G$), conditioned on the vertices from $V_0$ being measured to $1$, the cut sampled from the QAOA state is $0.0013|E|$ above the cut sampled from the unconditioned QAOA.
\end{prop}

Finally, we derive general negative results on the performance of $\mathbf{Z}_2$-symmetric finite range circuits (which generalize QAOA, see section \ref{subsec:qaoa}) on the MaxCut problem. Since these results are disconnected from the main line of this work but possibly of independent interest, we chose to defer them to appendix \ref{sec:bravyi_et_al_generalization}. They generalize negative results obtained in \cite{1910.08980} for the variational optimization of the (anti)ferromagnetic Ising model on ring graphs. Our main contribution a modification to the argument of the authors, generalizing their upper bounds to other Ising models and graphs. Also, we establish that the technique used in \cite{1910.08980} to construct lower bounds matching their upper bounds on ring graphs is inoperative in the case of random regular graphs. This suggests that emulating the performance of QAOA with ``simpler" $\mathbf{Z}_2$-symmetric constant-range circuits is significantly more challenging for random regular graphs.

\subsection{Numerical experiments}
We supported and complemented the results described above with numerical experiments. We carried out postselected QAOA on many large random 3-regular graph instances as described in section \ref{subsec:postselected_qaoa} (except for the sampling phase, which is classically hard) and evaluated the improvement obtained for these instances exactly. Besides, motivated by the simpler example of MaxCut on the ring graph discussed in section \ref{sec:ring_disagrees}, we evaluated to what extent the optimal values $\left(\sigma^*_{v_0}\right)_{v_0 \in V_0}$ for the postselected vertices gave a ``good" cut suggestion. These results are described in paragraph \ref{subsubsec:improvement_postselection}. Finally, we considered combining postselection with another simple postprocessing strategy (local updates) on the cuts sampled from the QAOA circuit. We discuss the ideas in paragraph \ref{subsubsec:improvement_local_updates}.

\subsubsection{Improvement from postselection}
\label{subsubsec:improvement_postselection}
\begin{figure}[!htbp]
    \centering
    \begin{subfigure}{0.49\columnwidth}
        \centering
        \includegraphics[width=\columnwidth]{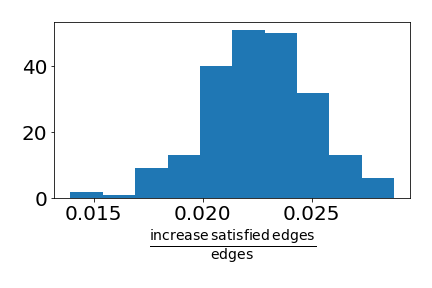}
        \caption{With exact optimization}
    \end{subfigure}
    \begin{subfigure}{0.49\columnwidth}
        \centering
        \includegraphics[width=\columnwidth]{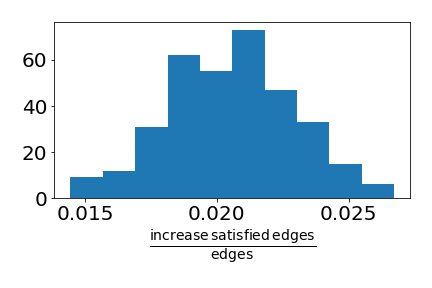}
        \caption{With approximate optimization (Goemans-Williamson)}
    \end{subfigure}
    \caption{Increase in fraction of satisfied edges}
    \label{fig:increase_satisfied_edges}
\end{figure}

We evaluated the improvement obtained from postselection on uniformly randomly generated 3-regular instances of size $200$. We therefore required the corresponding 2-independent sets to have size 40 (see proposition \ref{prop:lower_bound_2_independent_set}). Finding optimal (or sufficiently good) values $\left(\sigma_{v_0}\right)_{v_0 \in V_0}$ for the postselected vertices requires to optimize an Ising model on a smaller yet possibly denser graph than the original one (see proposition \ref{prop:sparse_vertex_set_ising_model}). For the graph sizes considered, optimizing exactly (using the branch-and-bound solver \verb|BiqCrunch| \cite{krislock_malick_roupin_2017}) and approximately (Goemans-Williamson) was feasible and we compare both methods on figure \ref{fig:increase_satisfied_edges}. Here, we plot the increase in the fraction of satisfied edges resulting from postselection (as compared to QAOA without postselection). The typical increase is around $0.02$, consistent with the $0.06$ upper bound we derived by invoking proposition \ref{prop:upper_bound_conditioning_qaoa}.

\begin{figure}[!htbp]
    \centering
    \begin{subfigure}{0.49\columnwidth}
        \centering
        \includegraphics[width=\columnwidth]{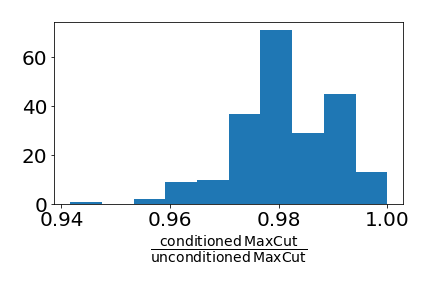}
        \caption{With exact optimization}
    \end{subfigure}
    \begin{subfigure}{0.49\columnwidth}
        \centering
        \includegraphics[width=\columnwidth]{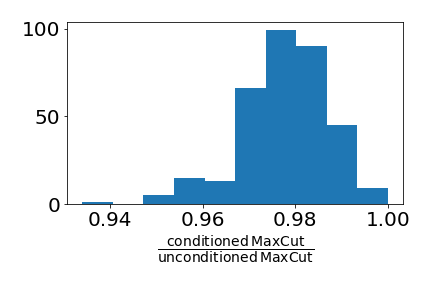}
        \caption{With approximate optimization}
    \end{subfigure}
    \caption{Conditioned vs. unconditioned MaxCut}
    \label{fig:conditioned_vs_unconditioned_max_cut}
\end{figure}

We then tried to empirically assess whether the optimal values $\left(\sigma_{v_0}^*\right)_{v_0 \in V_0}$ gave a relevant suggestion for a ``good" cut. More precisely, for each graph instance, we exactly solved the usual MaxCut problem on the graph (``unconditioned MaxCut") and the MaxCut problem with vertices from $V_0$ constrained to take values $\left(\sigma_{v_0}^*\right)_{v_0 \in V_0}$ (``conditioned MaxCut"). We then computed the ratio between the optimal cut sizes in the latter case and former cases. The results, displayed on figure \ref{fig:conditioned_vs_unconditioned_max_cut}, show that conditioning on optimal values $\left(\sigma_{v_0}^*\right)_{v_0 \in V_0}$ reduces the max cut by up to $5\,\%$.

\subsubsection{Improvement from local updates}
\label{subsubsec:improvement_local_updates}
We also explored a different strategy to improve the cut returned by the QAOA. The latter is based on making local improvements to the cut. More precisely, given a cut of a graph $G$ and a $k$-independent set $V_0$ of $G$, for each $v_0 \in V_0$, one performs locally optimal updates on the vertices lying in $B_G(v_0, k - 1)$. The idea is formalized in the following algorithm:
\begin{algorithm}
	\caption{Local MaxCut improvement on 3-regular graph}
	\label{alg:local_maxcut_improvement}
	\KwData{Graph $G = (V, E)$; Cut $\left(\sigma_v\right)_{v \in V} \in \{-1, 1\}^V$; depth $d$.}
	\KwResult{Improved cut $\left(\sigma_v'\right)_{v \in V}$.
	$V_0 \leftarrow \textrm{a } (2d - 1)-\textrm{-independent set of } G$}
	\ForAll {$v_0 \in V_0$}{
	    $\left(\sigma'_{v}\right)_{v \in B_G\left(v_0, d - 1\right)}$
	    \hspace{0.08\textwidth} $\leftarrow \argmax_{\left(\widetilde{\sigma}_{v}\right)_{v \in B_G\left(v_0, d - 1\right)}}\cut\left(G(B_G(v_0, d)); \left(\widetilde{\sigma}_v\right)_{v \in B_G(v_0, d - 1)}, \left(\sigma_v\right)_{v \in \partial B_G(v_0, d)}\right)$
	}
	\ForAll {$v \in V - \bigcup_{v_0 \in V_0}B_G(v_0, d - 1)$}{
	    $\sigma'_v \leftarrow \sigma_v$
	}
	\Return $\left(\sigma_v'\right)_{v \in V}$
\end{algorithm}
Here, we denote by $G(V')$ the graph induced by $G = (V, E)$ and a vertex set $V' \subset V$ and; besides for all graph $G' = (V', E')$, $\cut\left(G', \left(\sigma_{v'}\right)_{v' \in V'}\right)$ is the number of satisfied edges in the cut of $G'$ where vertices are assigned values $\left(\sigma_{v'}\right)_{v' \in V'} \in \{-1, 1\}^{V'}$.

\begin{figure}[!htbp]
    \centering
    \includegraphics[width=0.6\textwidth]{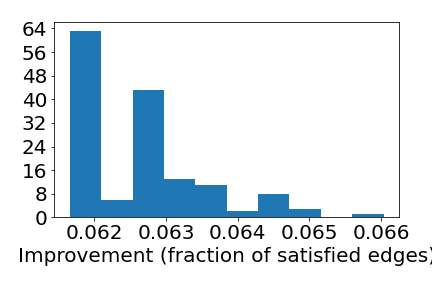}
    \caption{Improvement on graph instances from local updates}
    \label{fig:qaoa_improve_instance_local_updates}
\end{figure}

\begin{figure}[!htbp]
    \centering
    \includegraphics[width=0.6\textwidth]{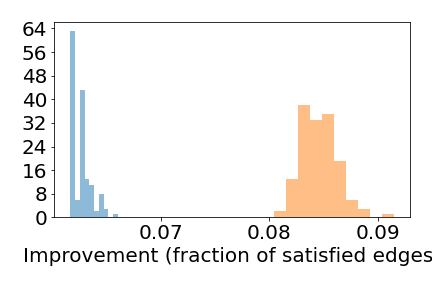}
    \caption{Expected increase (after postselection and local updates) of fraction of satisfied edges obtainable for $p = 1$ QAOA (blue: local update before postselection; orange: local update after postselection). 150 randomly generated instances.}
    \label{fig:qaoa_postselection_local_update_improvement}
\end{figure}

\begin{figure}[!htbp]
    \centering
    \includegraphics[width=0.6\textwidth]{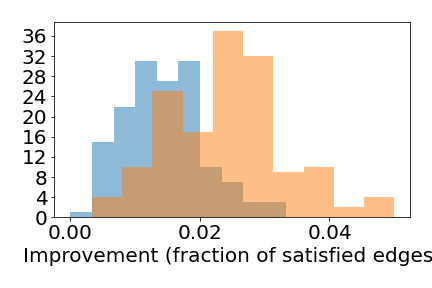}
    \caption{Expected increase (after local updates) of fraction of satisfied edges in a cut sampled from Goemans-Williamson (blue: single round of local updates; orange: 20 rounds of local updates).}
    \label{fig:gw_local_improve}
\end{figure}

This algorithm can be applied to improve cuts sampled from QAOA but also from any classical algorithm. However, in the case of QAOA, the performance of the algorithm can be rigorously and sharply quantified for moderate depth parameters $p$, at least in the infinite size limit; details are in appendix \ref{sec:local_improvement_infinite_size}. Here, we consider applying the algorithm to graphs of finite size, similar to the previous paragraphs. Besides, the depth parameter in algorithm \ref{alg:local_maxcut_improvement} is set to $1$; therefore, a 2-independent set is required. Random 3-regular graphs instances have 200 vertices and we construct 2-independent sets of size 40. Figure \ref{fig:qaoa_improve_instance_local_updates} shows the improvement (increase in fraction of satisfied edges) resulting from applying local updates to cuts sampled from depth-1 QAOA.
Finally, we consider combining postselection and local updates. After carrying out postselection as described in paragraph \ref{subsec:postselected_qaoa}, we compute the advantage resulting from local updates. Figure \ref{fig:qaoa_postselection_local_update_improvement} compares the improvement resulting from local updates only and the one resulting from postselection followed by local updates. The advantages obtain from both methods appear to add up.
Further improvements could possibly be obtained by performing more iterations of local updates, as is the case when applying algorithm \ref{alg:local_maxcut_improvement} to cuts sampled from the Goemans-Williamson algorithm (see figure \ref{fig:gw_local_improve}). Unfortunately, in the case of QAOA, our methods would not allow to efficiently compute classically the advantage of performing multiple local update iterations.
\section{The ring graph case}
\label{sec:ring_disagrees}
In this section, we analyze the postselected QAOA described in section \ref{subsec:postselected_qaoa} on the $n$-vertices ring graph. The analysis is considerably simpler than for 3-regular graphs, as there is a canonical way of choosing a 2-independent set $V_0$ in a ring graph (choose 1 vertex out of 3) and the Ising model in the postselected values $\left(\sigma_{v_0}\right)_{v_0 \in V_0}$ to optimize also lives on a ring graph. However, the model still proves relevant to introduce the general idea and can serve as a useful reference for comparison with the 3-regular case.

Given a $n$-vertices ring graph with vertex set $[n]$, the following proposition quantifies the improvement in the cut after postselecting on vertices $3k$, $k \in \left[0, \frac{n}{3}\right)$.
\begin{prop}
\label{prop:conditioned_qaoa_ring_disagrees}
	Let $n$ be a multiple of $3$ (for simplicity) and consider the MaxCut $p = 1$ QAOA on the $n$-vertices ring. Let $\ket{\psi}$ be the state prepared by the QAOA circuit with parameters $(\beta, \gamma)$ and let $\ket{\psi'}$ be the (non-normalized) state conditioned on measuring qubits $0, 3, \ldots, 3n  -3$ to $(\sigma_{3j})_{0 \leq j < \frac{n}{3}} \in \{-1, 1\}^{n/3}$, i.e.
	\begin{align*}
		\ket{\psi'} & = \prod_{0 \leq j < \frac{n}{3}}\frac{1 + \sigma_{3j}Z_{3j}}{2}\ket{\psi}.
	\end{align*}
	Then the expected cut obtained by sampling from $\ket{\psi'}$ is:
	\begin{align}
	\label{eq:expected_cut_conditioned_ring}
		\frac{\braket{\psi'|\sum_{0 \leq j < n}\frac{1 - Z_jZ_{j + 1}}{2}|\psi'}}{\left\lVert\ket{\psi'}\right\rVert} & = \left(\frac{1}{2} - \frac{\sin(2\beta)\sin(2\gamma)}{4}\right)n\nonumber\\
		& \hspace*{0.05\textwidth} - \sum_{0 \leq j < \frac{n}{3}}\left(\frac{(5 + 3\cos(2\gamma))\sin^2(2\beta)\sin^2\gamma}{16} + \frac{\sin^2\beta\sin^2(2\gamma)}{8}\right)\sigma_{3j}\sigma_{3(j + 1)}.
	\end{align}
\begin{proof}
	First, recalling proposition \ref{prop:range_r_circuit_sparse_set},
	\begin{align*}
		\left\lVert\ket{\psi'}\right\rVert^2 & = \braket{\psi|\prod_{0 \leq k < \frac{n}{3}}\frac{1 + \sigma_{3k}Z_{3k}}{2}|\psi}\\
		& = \frac{1}{2^{n/3}}.
	\end{align*}
	Next, consider $j \in \left[0, \frac{n}{3}\right)$.
	\begin{align*}
		\braket{\psi'|Z_{3j}Z_{3j + 1}|\psi'} & = \braket{\psi|Z_{3j}Z_{3j + 1}\prod_{0 \leq k < \frac{n}{3}}\frac{1 + \sigma_{3k}Z_{3k}}{2}|\psi}\\
		& = \bra{+}^{\otimes n}U^{\dagger}Z_{3j}Z_{3j + 1}\prod_{0 \leq k < \frac{n}{3}}\frac{1 + \sigma_{3k}Z_{3k}}{2}U\ket{+}^{\otimes n}.
	\end{align*}
	Now, by the range-1 property of $U$, it is easy to see that only the factors $k \in \{j, j + 1\}$ contribute to the expectation. Besides, using the $\mathbf{Z}_2$-symmetry of $U$, the above becomes.
	\begin{align*}
		\braket{\psi'|Z_{3j}Z_{3j + 1}|\psi'} & = \frac{1}{2^{n/3}}\bra{+}^{\otimes n}U^{\dagger}Z_{3j}Z_{3j + 1}U\ket{+}^{\otimes n}\\
		& \hspace{0.05\textwidth} + \sigma_{3j}\sigma_{3(j + 1)}\frac{1}{2^{n/3}}\bra{+}^{\otimes n}U^{\dagger}Z_{3j}Z_{3j + 1}Z_{3j}Z_{3j + 3}U\ket{+}^{\otimes n}\\
		& = \frac{1}{2^{n/3}}\bra{+}^{\otimes n}U^{\dagger}Z_{3j}Z_{3j + 1}U\ket{+}^{\otimes n}\\
		& \hspace{0.05\textwidth} + \sigma_{3j}\sigma_{3(j + 1)}\frac{1}{2^{n/3}}\bra{+}^{\otimes n}U^{\dagger}Z_{3j + 1}Z_{3j + 3}U\ket{+}^{\otimes n}.
	\end{align*}
	This can easily be computed from the formulae in appendix \ref{appendix:p1_qaoa}. We obtain:
	\begin{align*}
		\braket{\psi'|Z_{3j}Z_{3j + 1}|\psi'} & = \frac{1}{2^{n/3}}\left(\frac{\sin(2\beta)\sin(2\gamma)}{2}  + \sigma_{3j}\sigma_{3(j + 1)}\frac{\sin^2\beta\sin^2(2\gamma)}{4}\right)
	\end{align*}
	One can likewise establish:
	\begin{align*}
		\braket{\psi'|Z_{3j + 1}Z_{3j + 2}|\psi'} & = \frac{1}{2^{n/3}}\left(\frac{\sin(2\beta)\sin(2\gamma)}{2} + \sigma_{3j}\sigma_{3(j + 1)}\frac{\left(5 + 3\cos(2\gamma)\right)\sin^2(2\beta)\sin^2\gamma}{8}\right)\\
		\braket{\psi'|Z_{3j + 2}Z_{3j + 3}|\psi'} & = \frac{1}{2^{n/3}}\left(\frac{\sin(2\beta)\sin(2\gamma)}{2} + \sigma_{3j}\sigma_{3(j + 1)}\frac{\sin^2\beta\sin^2(2\gamma)}{4}\right)
	\end{align*}
	This establishes:
	\begin{align*}
		& \frac{\braket{\psi'|\sum_{0 \leq j < n}\frac{1 - Z_jZ_{j + 1}}{2}|\psi'}}{\left\lVert\ket{\psi'}\right\rVert}\\
		& = \left(\frac{1}{2} - \frac{\sin(2\beta)\sin(2\gamma)}{4}\right)n\\
		& \hspace*{0.05\textwidth} - \sum_{0 \leq j < \frac{n}{3}}\left(\frac{(5 + 3\cos(2\gamma))\sin^2(2\beta)\sin^2\gamma}{16} + \frac{\sin^2\beta\sin^2(2\gamma)}{8}\right)\sigma_{3j}\sigma_{3(j + 1)}.
	\end{align*}
\end{proof}
\end{prop}
The expected cut on the ring for the non-postselected state corresponds to the first term in equation \ref{eq:expected_cut_conditioned_ring}. The effect of postselecting is expressed in the sum, which is an Ising model on the values $(\sigma_{3j})_{0 \leq j < \frac{n}{3}}$ of the conditioning vertices. In particular, if $\frac{n}{3}$ is even, letting $\sigma_{3(j + 1)} = -\sigma_{3j}$ (which is possible since the model is not frustrated) gives an expected cut size
\begin{align*}
	\left(\frac{1}{2} - \frac{\sin(2\beta)\sin(2\gamma)}{4} + \frac{(5 + 3\cos(2\gamma))\sin^2(2\beta)\sin^2\gamma}{48} + \frac{\sin^2(\beta)\sin^2(2\gamma)}{24}\right)n
\end{align*}
which is approximately $0.823n$ for the optimal QAOA parameters\footnote{Note that these parameters are the optimal ones for QAOA \textit{without} postselection; they maximize $\frac{1}{2} - \frac{\sin(2\beta)\sin(2\gamma)}{4}$ instead of the function with two extra terms above. Optimizing the latter would yield a slightly better cut, namely $0.824n$.} $(\beta, \gamma) = \left(-\frac{\pi}{4}, \frac{\pi}{4}\right)$. This corresponds to a $0.073n$ improvement over the standard QAOA.

In this case, the values assigned to the postselected vertices: $\sigma_{3(j + 1)} = -\sigma_{3j}$ can be extended to a maximum cut. This suggests the possibility of using postselected QAOA not as a cut sampler (similar to QAOA), but to identify ``good assignments" of a restricted vertex set $V_0$. It is then natural to ask whether postselected values for vertices in $V_0$ determined by optimizing equation \ref{fig:conditioned_vs_unconditioned_max_cut} can still be extended to a maximum or ``good" cut for a 3-regular graph. This motivated the numerical experiment reported on figure \ref{fig:conditioned_vs_unconditioned_max_cut} and commented in section \ref{subsubsec:improvement_postselection}.
\section{Derivation of results}
\label{sec:derivation_results}

\subsection{Improvement for fixed graph and 2-independent set}
\label{subsec:improvement_fixed_graph_set}
In this section, we derive an expression for the expected energy of postselected QAOA \begin{align}
    \frac{\braket{\psi_G|\prod_{v_0 \in V_0}\frac{1 + \sigma_{v_0}Z_{v_0}}{2}H_{\textnormal{MaxCut}}(G)\prod_{v_0 \in V_0}\frac{1 + \sigma_{v_0}Z_{v_0}}{2}|\psi_G}}{\left\lVert \prod_{v_0 \in V_0}\frac{1 + \sigma_{v_0}Z_{v_0}}{2}\ket{\psi_G}\right\rVert^2}.
\end{align}
From proposition \ref{prop:range_r_circuit_sparse_set}, this simplifies to
\begin{align}
    \braket{\psi_G|\prod_{v_0 \in V_0}\left(1 + \sigma_{v_0}Z_{v_0}\right)H_{\textnormal{MaxCut}}(G)\prod_{v_0 \in V_0}\left(1 + \sigma_{v_0}Z_{v_0}\right)|\psi_G}.
\end{align}
We now show that the expression above reduces to an Ising model in the variables $\left(\sigma_{v_0}\right)_{v_0 \in V_0}$ whose coupling coefficients are efficiently computable classically.
\begin{prop}
\label{prop:sparse_vertex_set_ising_model}
	Let $G = (V, E)$ be an arbitrary 3-regular graph and $V_0$ a 2-independent vertex set in $G$. Then there exist (classically efficiently computable) constants $\left(J(G, V_0)_{v_0v_0'}\right)_{v_0, v_0' \in V_0}$, depending in $G$ and $V_0$, such that for any $\left(\sigma_{v_0}\right)_{v_0 \in V_0} \in \{-1, 1\}^{V_0}$,
	\begin{align}
	    & \braket{\psi_G|\prod_{v_0 \in V_0}\left(1 + \sigma_{v_0}Z_{v_0}\right)H_{\textnormal{MaxCut}}(G)\prod_{v_0 \in V_0}\left(1 + \sigma_{v_0}Z_{v_0}\right)|\psi_G}\nonumber\\
	    & = \braket{\psi_G|H_{\textnormal{MaxCut}}(G)|\psi_G} + \sum_{\{v_0, v_0'\} \subset V_0}J(G, V_0)_{v_0, v_0'}\sigma_{v_0}\sigma_{v_0'}
	\end{align}
	where $\ket{\psi_G}$ is the state prepared by the $p = 1$ QAOA.
	The couplings $J(G, V_0)_{v_0v_0'}$ can be expressed:
	\begin{align}
		\label{eq:conditioned_qaoa_ising_couplings}
		J(G, V_0)_{v_0v_0'} & = \sum_{\substack{e = \{e_0, e_1\} \in E\\d(v_0, e) \leq 2\\d(v_0', e) \leq 2}}\braket{\psi|Z_{e_0}Z_{e_1}Z_{v_0}Z_{v_0'}|\psi},
	\end{align}
	where we denoted by $d(v_0, e)$ the distance between the vertex $v_0$ and the edge $e$, i.e. $d(v_0, \{e_0, e_1\}) := d_G(v_0, e_0) \wedge d_G(v_0, e_1)$.
	\begin{proof}
        The problem is to evaluate
        \begin{align*}
            & \braket{\psi_G|\prod_{v_0 \in V_0}\left(1 + \sigma_{v_0}Z_{v_0}\right)H_{\textnormal{MaxCut}}(G)\prod_{v_0 \in V_0}\left(1 + \sigma_{v_0}Z_{v_0}\right)|\psi_G}\\
            & = \braket{\psi_G|\prod_{v_0 \in V_0}\left(1 + \sigma_{v_0}Z_{v_0}\right)\sum_{e = \{e_0, e_1\} \in E}\frac{1 - Z_{e_0}Z_{e_1}}{2}\prod_{v_0 \in V_0}\left(1 + \sigma_{v_0}Z_{v_0}\right)|\psi_G}\\
            & = \frac{|E|}{2} - \frac{1}{2}\sum_{e = \{e_0, e_1\} \in E}\braket{\psi_G|Z_{e_0}Z_{e_1}\prod_{v_0 \in V_0}\left(1 + \sigma_{v_0}Z_{v_0}\right)|\psi_G}.
        \end{align*}
		Let us start by expanding the product $\prod_{v_0 \in V_0}\left(1 + \sigma_{v_0}Z_{v_0}\right)$:
		\begin{align*}
			& \prod_{v_0 \in V_0}\left(1 + \sigma_{v_0}Z_{v_0}\right)\\
			& = \sum_{0 \leq k \leq |V_0|}\sum_{\substack{W_0 \subset V_0\\|W_0| = k}}\left(\prod_{v_0 \in W_0}\sigma_{v_0}\right)\left(\prod_{v_0 \in W_0}Z_{v_0}\right)
		\end{align*}
		$\mathbf{Z}_2$-symmetry of QAOA immediately implies that only even $k$ may give nonzero contributions. We now show that the terms with $k > 4$ do not contribute to the expectation \begin{align*}
			\braket{\psi_G|Z_{e_0}Z_{e_1}\prod_{v_0 \in V_0}\left(1 + \sigma_{v_0}Z_{v_0}\right)|\psi_G}
		\end{align*}
		for any edge $e$. Indeed, it is easily checked that since $V_0$ is 2-independent, $e$ has at most 4 vertices of $V_0$ in its 2-neighbourhood ---denote them temporarily by $V_0'$. All other vertices of $V_0$ have distance at least 3 from $e$ (because they do not lie in the 2-neighbourhood) and distance at least 3 from $V_0'$ (because $V_0$ is 2-independent). Therefore, for $W_0 \subset V_0, |W_0| \geq 5$,
		\begin{align*}
			\braket{\psi|Z_{e_0}Z_{e_1}\prod_{v_0 \in W_0}Z_{v_0}|\psi} & = \braket{\psi|Z_{e_0}Z_{e_1}\prod_{v_0 \in W_0 \cap V_0'}Z_{v_0}\prod_{v_0 \in W_0 - V_0'}Z_{v_0}|\psi}\\
			& = \braket{\psi|Z_{e_0}Z_{e_1}\prod_{v_0 \in W_0 \cap V_0'}Z_{v_0}|\psi}\braket{\psi|\prod_{v_0 \in W_0 - V_0'}Z_{v_0}|\psi}\\
			& = \braket{\psi|Z_{e_0}Z_{e_1}\prod_{v_0 \in W_0 \cap V_0'}Z_{v_0}|\psi}\prod_{v_0 \in W_0 - V_0'}\underbrace{\braket{\psi|Z_{v_0}|\psi}}_{= 0}\\
			& = 0
		\end{align*}
		($W_0 - V_0'$ nonempty since $|W_0| \geq 5$ and $V_0' \leq 4$). The calculation above also shows that only the $v_0 \in V_0$ lying in the 2-neighbourhood of $e$ may give nonzero contributions. Finally, to handle the possibility $k = 4$, we may start with the case where the 2-neighbourhood is a tree. It is then easy to write down all possible configurations of 4 vertices from $V_0$ in the 2-neighbourhood of $e$ that are compatible with the 2-independence assumption; however, an explicit evaluation of the expectation of such configurations gives $0$. In case the 2-neighbourhood is not a tree, it is in fact impossible to fit 4 vertices from $V_0$ in it. Therefore, only the $k = 2$ terms contribute; consequently, the expectation $-\frac{1}{2}\braket{\psi|\sum_{\{e_0, e_1\} \in E}Z_{e_0}Z_{e_1}\prod_{v_0 \in V_0}\left(1 + \sigma_{v_0}Z_{v_0}\right)|\psi}$ does reduce to an Ising model.
		
		The latter can be explicitly expressed as:
		\begin{align}
			& -\frac{1}{2}\braket{\psi|\sum_{\{e_0, e_1\} \in E}Z_{e_0}Z_{e_1}\prod_{v_0 \in V_0}(1 + \sigma_{v_0}Z_{v_0})|\psi}\\
			& = -\frac{1}{2}\braket{\psi|\sum_{\{e_0, e_1\} \in E}Z_{e_0}Z_{e_1}|\psi} - \frac{1}{2}\sum_{\{e_0, e_1\} \in E}\sum_{\substack{v_0, v_0' \in V_0\\d(v_0, e) \leq 2\\d(v_0', e) \leq 2}}\braket{\psi|Z_{e_0}Z_{e_1}Z_{v_0}Z_{v_0'}|\psi}\sigma_{v_0}\sigma_{v_0'}.
		\end{align}
	\end{proof}
\end{prop}

\begin{figure}[!htbp]
	\centering
	\begin{subfigure}{0.3\textwidth}
		\centering
		\includegraphics[width=0.8\columnwidth]{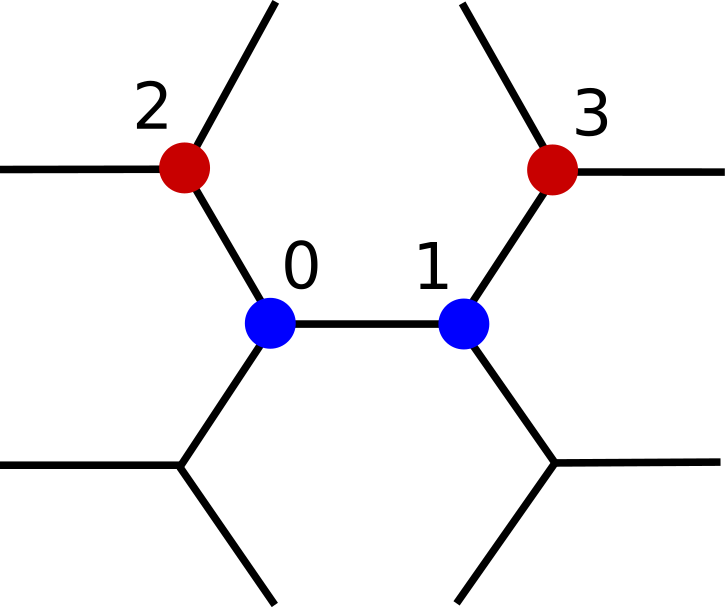}
		\caption{Configuration 1}
	\end{subfigure}
	\begin{subfigure}{0.3\textwidth}
		\centering
		\includegraphics[width=0.8\columnwidth]{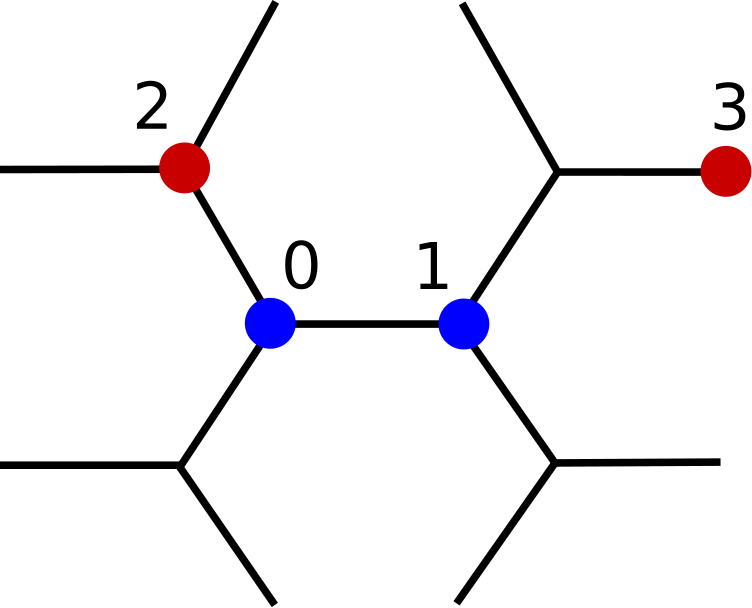}
		\caption{Configuration 2}
	\end{subfigure}
	\begin{subfigure}{0.3\textwidth}
		\centering
		\includegraphics[width=0.8\columnwidth]{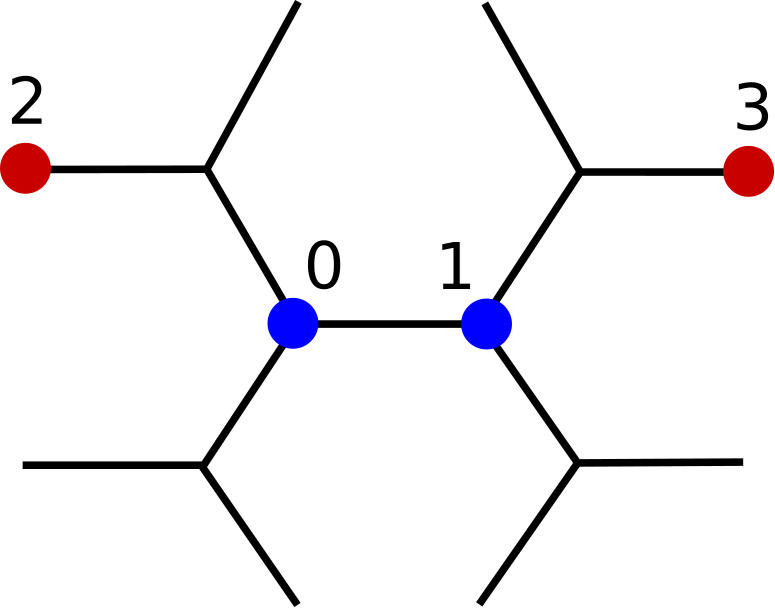}
		\caption{Configuration 3}
	\end{subfigure}
	\begin{subfigure}{0.3\textwidth}
		\centering
		\includegraphics[width=0.8\columnwidth]{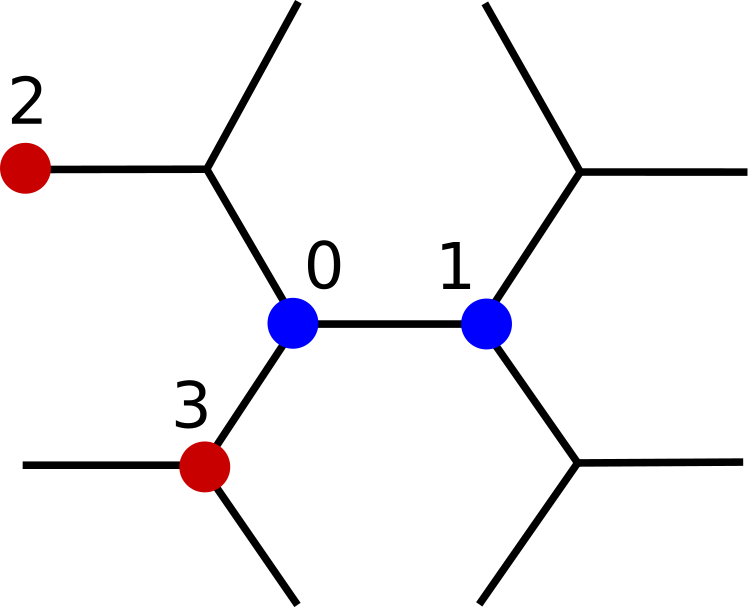}
		\caption{Configuration 4}
	\end{subfigure}
	\begin{subfigure}{0.3\textwidth}
		\centering
		\includegraphics[width=0.8\columnwidth]{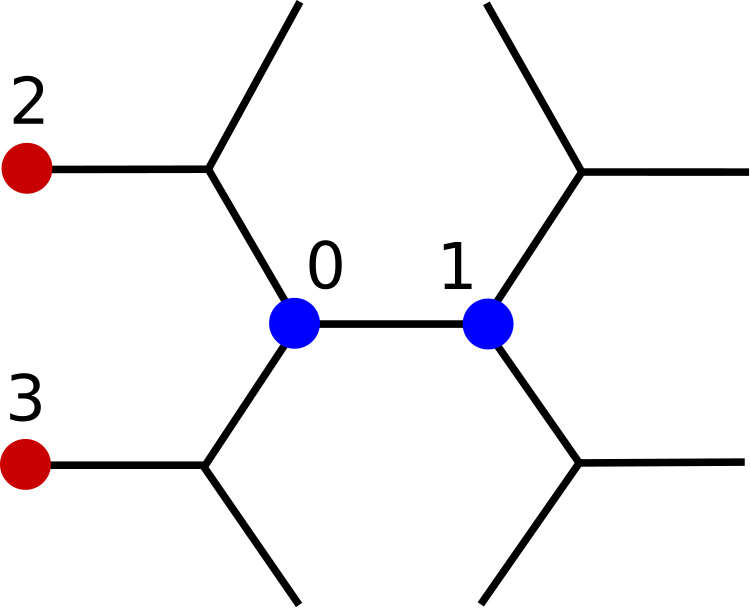}
		\caption{Configuration 5}
	\end{subfigure}
	\begin{subfigure}{0.3\textwidth}
		\centering
		\includegraphics[width=0.8\columnwidth]{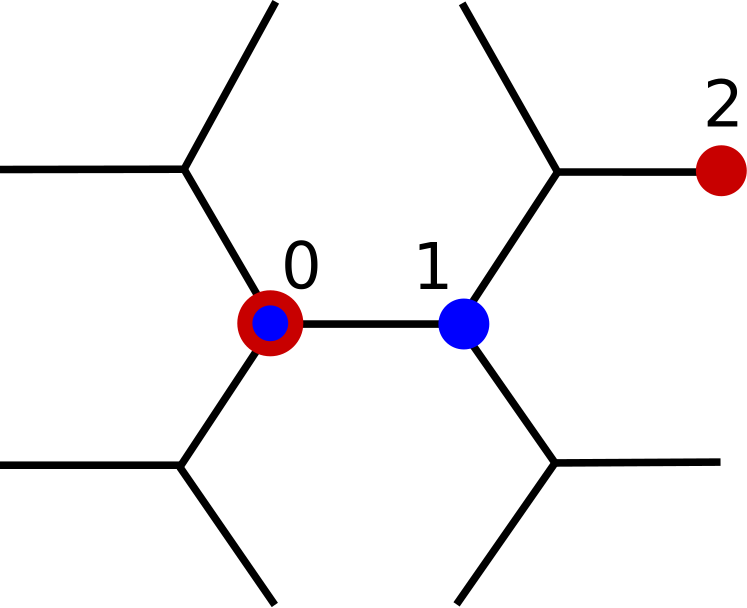}
		\caption{Configuration 6}
	\end{subfigure}
	\caption{Possible configurations of a pair of conditioned vertices (red) with respect to an edge (vertices in blue). Black edges indicate the 2-neighbourhood of the edge.}
	\label{fig:frozen_vertices_configurations}
\end{figure}

According to equation \ref{eq:conditioned_qaoa_ising_couplings}, to compute the Ising coupling between $v_0, v_0' \in V_0$, it suffices to enumerate edges $\{e_0, e_1\}$ such that $v_0, v_0'$ lie in the 2-neighbourhood of $e$. We now focus on the case where the latter is a tree, for reasons that will be formalized later (but the intuition is, random regular graphs locally look like trees by proposition \ref{prop:tree_neighbourhood}). The possible configurations of a pair of vertices from $V_0$ with respect to an edge (up to tree isomorphism\footnote{By the symmetry of QAOA, the value of the couplings are manifestly invariant under tree isomorphism.}) are represented on figure \ref{fig:frozen_vertices_configurations}. The coupling corresponding to each configuration, calculated for the optimal $p = 1$ QAOA parameters for 3-regular graphs $(\beta, \gamma) = \left(-\frac{\pi}{4}, \arctan\left(\frac{1}{\sqrt{2}}\right)\right)$ are reported in table \ref{table:sparse_set_vertices_ising_couplings}.

\begin{table}[!htbp]
	\centering{
		\begin{tabular}{c|c}
			\textbf{Configuration} & \textbf{Ising coupling} \\
			\hline
			1 & -0.0833\\
			\hline
			2 & 0.0178\\
			\hline
			3 & -0.00412\\
			\hline
			4 & 0.00926\\
			\hline
			5 & -0.00356\\
			\hline
			6 & -0.0370
	\end{tabular}}
	\caption{Ising coefficients between conditioned vertices for $p = 1$ QAOA on random 3-regular graphs}
	\label{table:sparse_set_vertices_ising_couplings}
\end{table}

\subsection{Neighbourhoods of 2-independent sets and expectation upper bound}
In section \ref{subsec:improvement_fixed_graph_set}, we established proposition \ref{prop:sparse_vertex_set_ising_model}, which rephrases the problem of choosing the best prescribed values for the qubits in the $2$-independent set to that of optimizing an Ising model over these values. Precisely, using the notation of the latter proposition, the cut sampled from the QAOA circuit conditioned on the vertices from the $2$-independent set having spins $\left(\sigma_{v_0}\right)_{v_0 \in V_0}$ increases by \begin{align}
    \sum_{\{v_0, v_0'\} \in E}J(G, V_0)_{v_0v_0'}\sigma_{v_0}\sigma_{v_0'}
\end{align}
compared to the standard QAOA. In particular, it is upper-bounded by $\sum_{\{v_0, v_0'\} \in E}\left|J(G, V_0)_{v_0v_0'}\right|$. The purpose of this section is to analyze the latter upper bound for a fixed set $V_0$ and random regular graph $G$, \textit{conditioned on $V_0$ being 2-independent in $G$}. This will lead to a proof of proposition \ref{prop:upper_bound_conditioning_qaoa}.

As discussed in section \ref{subsec:improvement_fixed_graph_set}, the coupling $J(G, V_0)_{v_0v_0'}$ between two vertices $v_0, v_0' \in V_0$ depends on the distance between $v_0$ and $v_0'$ in the graph; indeed, this distance determines the configurations, among those represented on figure \ref{fig:frozen_vertices_configurations}, the pair appears in relative to edges. This motivates, given a vertex $v_0 \in V_0$, to count the vertices $V_0 \cap \partial B_G(v_0, r)$ at a distance $r$ from $V_0$ which also lie in $V_0$. Given $V_0$ is 2-independent and depth-1 QAOA has range 1, it will suffice to restrict to $r \in \{3, 4, 5\}$.

We then start with following lemma, which gives a complicated expression for the probability distribution of the number of vertices from $V_0$ at a distance $3, 4$ or $5$ from $v_0 \in V_0$. (Fortunately, this will be simplified shortly.) The proof, relying on some tedious counting, is deferred to appendix \ref{sec:technical_results_postselect_qaoa}.
\begin{lem}
	\label{lemma:sparse_set_distance_3_4_5_successors}
	Let $V_0 \subset [n]$ a set of vertices of size $n_0 \geq cn$ ($c > 0$) and $v_0 \in V_0$. For an $n$-vertices multigraph $G$ drawn from the configuration model, conditional on $V_0$ being 2-independent, the probability that the $5$-neighbourhood of $v_0$ in $G$ is a tree and that $v_0$ contains $k_i$ vertices from $V_0$ among its distance $i$ successors for all $i \in \{3, 4, 5\}$ is given by:
	\begin{align}
	\label{eq:p_3_4_5}
	& p_{V_0}(k_3, k_4, k_5)\\
	& := \sum_{\subalign{0 \leq k_6 \leq & d(d - 1)^4\\& - (d - 1)^2k_3\\& - (d - 1)k_4\\& - k_5}}\binom{d(d - 1)}{k_3}\binom{d(d - 1)^2 - k_3}{k_4}\binom{d(d - 1)^3 - (d - 1)k_3 - k_4}{k_5}\nonumber\\
	& \hspace{0.05\textwidth} \times \binom{d(d - 1)^4 - (d - 1)^2k_3 - (d - 1)k_4 - k_5}{k_6}\left(\frac{d(d - 1)n_0}{nd - 2n_0d}\right)^{k_3 + k_4 + k_5 + k_6}\nonumber\\
	& \hspace*{0.05\textwidth} \times \left(\frac{nd - d(d + 1)n_0}{nd - 2n_0d}\right)^{d(d - 1)\frac{(d - 1)^4 - 1}{d - 2} - (d^2 - d + 2)k_3 - (d + 1)k_4 - 2k_5 - k_6}\nonumber\\
	& \hspace{0.05\textwidth} \times \left(1 + \mathcal{O}\left(\frac{1}{n}\right)\right)
	\end{align}
	where the implicit constant in the $\mathcal{O}\left(\frac{1}{n}\right)$ depends only on $c$ such that $n_0 \geq cn$ and $d$.
\end{lem}
We now apply this lemma to derive two propositions which characterize the neighbourhood of every vertex $v_0 \in V_0$ in $G$ conditioned $V_0$ being 2-independent in $G$. The first of them states that for each $v_0 \in V_0$, the 5-neighbourhood of $V_0$ is a tree with high probability:
\begin{prop}
	\label{prop:5_neighbourhood_whp}
	Let $v_0$ a set of vertices on size $n_0$, with $n_0 = \alpha n$, $\alpha \in \left(0, \alpha_d^*\right)$ ($\alpha_d^*$ is defined in proposition \ref{prop:upper_bound_k_independent_set}). Then, conditioned on $V_0$ being $2$-independent, for fixed $v_0 \in V_0$, the $5$-neighbourhood of $v_0$ is a tree with probability $1 - \mathcal{O}\left(\frac{1}{n}\right)$ (the constant hidden in the $\mathcal{O}$ depends on $d$). In other words,
	\begin{align}
	\sum_{k_3, k_4, k_5 \geq 0}p_{V_0}(k_3, k_4, k_5) = 1 - \mathcal{O}\left(\frac{1}{n}\right)
	\end{align}
	\begin{proof}
		The result follows from applying the binomial expansion formula to evaluate the sum on $k_6$ in equation \ref{eq:p_3_4_5} ($k_3, k_4, k_5$ fixed) without the $\mathcal{O}\left(\frac{1}{n}\right)$, followed by the same strategy for the sum over $k_5$ and finally $k_4, k_3$.
	\end{proof}
\end{prop}
The next proposition estimates the expected number of distance-3, 4 and 5 neighbours of each vertex $v_0 \in V_0$ conditioned on $V_0$ being 2-independent in $G$.
\begin{prop}
	\label{prop:sparse_set_distance_3_4_5_successors_estimate}
	Let $v_0 \in V_0$ a set of vertices on size $n_0$, with $n_0 = \alpha n$, $\alpha \in \left(0, \alpha_d^*\right)$ ($\alpha_d^*$ is defined in proposition \ref{prop:upper_bound_k_independent_set}). Let $v_0 \in V_0$. Then, conditioned on $V_0$ being $2$-independent, the expected number of distance 3 successors of $v_0$ lying in $V_0$ is:
	\begin{align}
	& d(d - 1)^2\frac{n_0}{n - 2n_0} + \mathcal{O}\left(\frac{1}{n}\right)
	\end{align}
	The expected number of distance 4 successors of $v_0$ lying in $V_0$ is:
	\begin{align}
	& d(d - 1)^3\frac{n_0(n - 3n_0)}{(n - 2n_0)^2} + \mathcal{O}\left(\frac{1}{n}\right)
	\end{align}
	The expected number of distance 5 successors of $v_0$ lying in $V_0$ is:
	\begin{align}
	& d(d - 1)^4\frac{n_0(n - 3n_0)^2}{(n - 2n_0)^3} + \mathcal{O}\left(\frac{1}{n}\right)
	\end{align}
	\begin{proof}
		The expected number of distance 3 successors of $v_0$ is
		\begin{align*}
		& \sum_{k_3, k_4, k_5\geq 0}p(k_3, k_4, k_5)k_3 + \left(1 - \sum_{k_3, k_4, k_5}p(k_3, k_4, k_5)\right)\mathcal{O}\left(d(d - 1)^2\right)
		\end{align*}
		since any vertex has at most $d(d - 1)^2$ distance 3 successors. By proposition \ref{prop:5_neighbourhood_whp}, this becomes
		\begin{align*}
		& \sum_{k_3, k_4, k_5\geq 0}p(k_3, k_4, k_5)k_3 + \mathcal{O}\left(\frac{1}{n}\right).
		\end{align*}
		By a calculation similar to the one in the proof of proposition \ref{prop:5_neighbourhood_whp}, the first term can, up to an error $\mathcal{O}\left(\frac{1}{n}\right)$, be estimated to
		\begin{align*}
		& \sum_{k_3, k_4, k_5, k_6 \geq 0}\binom{d(d - 1)}{k_3}\left(\frac{d(d - 1)n_0}{nd - 2n_0d}\right)^{k_3}\left(\frac{nd - d(d + 1)n_0}{nd - 2n_0d}\right)^{d(d - 1) - k_3}k_3\\%
		& = d(d - 1)^2\frac{n_0}{n - 2n_0}
		\end{align*}
		The estimates for the distance 4 and 5 successors proceed similarly.
	\end{proof}
\end{prop}
We are now in position, to prove proposition \ref{prop:upper_bound_conditioning_qaoa_expectation}, the main result of this section. 
\begin{repprop}{prop:upper_bound_conditioning_qaoa_expectation}
    Let $V_0 \subset V = [n]$ a set of vertices with $\alpha := \frac{|V_0|}{n} \in (0, \alpha_3^*)$ (where $\alpha_d^*$ is defined in section \ref{subsec:independent_sets}). Then the expected improvement of postselected QAOA over QAOA conditioned on $V_0$ being 2-independent is bounded as follows (where the expectation is taken over random regular graphs $G$ on vertex set $V$ in the configuration model):
    \begin{align}
        & \mathbf{E}_G\left[\max_{\left(\sigma_{v_0}\right)_{v_0 \in V_0} \in \{-1, 1\}^{V_0}}\frac{\braket{\psi_G|\prod_{v_0 \in V_0}\frac{1 + \sigma_{v_0}Z_{v_0}}{2}H_{\textnormal{MaxCut}}(G)\prod_{v_0 \in V_0}\frac{1 + \sigma_{v_0}Z_{v_0}}{2}|\psi_G}}{\left\lVert\prod_{v_0 \in V_0}\frac{1 + \sigma_{v_0}Z_{v_0}}{2}\ket{\psi_G}\right\rVert^2}\right.\nonumber\\
        & \hspace{0.1\textwidth} - \braket{\psi_G|H_{\textnormal{MaxCut}(G)}|\psi_G}\,\bigg|\,V_0\textrm{ 2-id in } G\Bigg] \leq \frac{\alpha^2(0.18125 - 0.83875\alpha + 0.99\alpha^2)}{(0.5 - \alpha)^3}n + \mathcal{O}(1).
    \end{align}
    \begin{proof}
		Recalling the notation in proposition \ref{prop:sparse_vertex_set_ising_model}, the expression in the expectation can be rewritten as $\sum_{v_0, v_0' \in V_0}J(G, V_0)_{v_0v_0'}\sigma_{v_0}\sigma_{v_0'}$, hence bounded by $\sum_{v_0, v_0' \in V_0}\left|J(G, V_0)_{v_0v_0'}\sigma_{v_0}\sigma_{v_0'}\right|$. In all the following, we may assume that $V_0$ is 2-independent in $G$ by conditioning. Now, fix $v_0 \in V_0$ and let us estimate $J(G, V_0)_{v_0, v_0'}$ for all $v_0' \in V_0 - \{v_0\}$. First consider the case where $B_G(v_0, 5)$ is a tree. We will systematically refer to figure \ref{fig:frozen_vertices_configurations}, detailing the possible configurations of a pair of vertices from a 2-independent set relative to an edge.
		\begin{itemize}
			\item For $v_0' \in B_G(v_0, 3)$, the pair $\{v_0, v_0'\}$ appears once in configuration 1, twice in configuration 4 and twice in configuration 6. In this case, $J(G, V_0)_{v_0v_0'} = 1 \times (-0.0833) + 2 \times (0.00926) + 2 \times (-0.0370)$ and $|J(G, V_0)_{v_0, v_0'}| \leq 0.140$.
			\item For $v_0' \in B_G(v_0, 4)$, the pair $\{v_0, v_0'\}$ appears twice in configuration 2 and thrice in configuration 5. Therefore, $J(G, V_0)_{v_0, v_0'} = 2 \times (0.0178) + 3 \times (-0.00356)$ and $|J(G, V_0)_{v_0, v_0'}| \leq 0.0252$.
			\item For $v_0' \in B_G(v_0, 5)$, the pair $\{v_0, v_0'\}$ appears thrice in configuration 3. Therefore $J(G, V_0)_{v_0, v_0'} = 3 \times (-0.00412)$ and $|J(G, V_0)_{v_0v_0'}| \leq 0.0124$.
		\end{itemize}
		In case $B_G(v_0, 5)$ is not a tree, we crudely bound $|J(G, V_0)_{v_0, v_0'}| \leq 1$ for all $v_0' \in V_0 - \{v_0\}$.
		
		Therefore,
		\begin{align*}
		& \sum_{\{v_0, v_0'\} \subset V_0}|J(G, V_0)_{v_0v_0'}|\\
		& = \frac{1}{2}\sum_{v_0 \in V_0}\sum_{v_0' \in V_0 - \{v_0\}}|J(G, V_0)_{v_0v_0'}|\\
		& = \frac{1}{2}\sum_{v_0 \in V_0}\mathbf{1}_{B_G(v_0, 5)\textrm{ tree}}\sum_{v_0' \in V_0 - \{v_0\}}|J(G, V_0)_{v_0v_0'}| + \frac{1}{2}\sum_{v_0 \in V_0}\mathbf{1}_{B_G(v_0, 5)\textrm{ not tree}}\sum_{v_0' \in V_0 - \{v_0\}}|J(G, V_0)_{v_0v_0'}|\\
		& \leq \frac{1}{2}\sum_{v_0 \in V_0\,:\,B_G(v_0, 5)\textrm{ tree}}\left(0.140|B_G(v_0, 3)| + 0.0252|B_G(v_0, 4)| + 0.0124|B_G(v_0, 5)|\right)\\
		& \hspace*{0.05\textwidth} + \frac{1}{2}\sum_{v_0 \in V_0\,:\,B_G(v_0, 5)\textrm{ not tree}}1
		\end{align*}
		Now, taking the expectation $\mathbf{E}\left[\cdot\,|\,V_0\,\textrm{2-id}\right]$ on both sides of the inequality, we may use proposition \ref{prop:sparse_set_distance_3_4_5_successors_estimate} to bound the first line by \begin{align*}
			& \left(0.84\frac{\alpha^2}{1 - 2\alpha} + 0.31\frac{\alpha^2(1 - 3\alpha)}{(1 - 2\alpha)^2} + 0.30\frac{\alpha^2(1 - 3\alpha)^2}{(1 - 2\alpha)^3}\right)n + \mathcal{O}(1)
		\end{align*}
		and proposition \ref{prop:5_neighbourhood_whp} the second line by $\mathcal{O}(1)$. This gives the wanted result.
	\end{proof}
\end{repprop}

\subsection{High-probability upper bound}
Now we established an upper bound has been holding in expectation, we discuss an upper bound holding with high probability assuming conjecture \ref{conj:independent_set_typical_graph_frequent}. The proof is very similar to that of proposition \ref{prop:upper_bound_conditioning_qaoa_expectation}; the main change is to replace the estimates on neighbourhoods of vertices from $V_0$ that held in expectation in proposition \ref{prop:sparse_set_distance_3_4_5_successors_estimate} by similar estimates holding with high probability. This is done in the following proposition, whose proof is deferred to appendix \ref{sec:technical_results_postselect_qaoa}.
\begin{prop}
\label{prop:concentration_tree_neighbourhoods}
	Let $\varepsilon > 0$ and $\alpha \in \left(0, \alpha^*_d\right)$ and set $n_0 := \alpha n$. Under conjecture \ref{conj:independent_set_typical_graph_frequent}, the following holds: for all but an exponentially small fraction (in $n$) of 2-independent-set-typical graphs $G_0$ and all but an exponentially small fraction of 2-independent sets $V_0$ of $G_0$,
	\begin{align}
	\left|\sum_{v_0 \in V_0}\left|\partial B_{G_0}(v_0, 3)\right| - d(d - 1)^2\frac{n_0^2}{n - 2n_0}\right| & \leq \varepsilon n\\
	\left|\sum_{v_0 \in V_0}\left|\partial B_{G_0}(v_0, 4)\right| - d(d - 1)^3\frac{n_0^2(n - 3n_0)}{(n - 2n_0)^2}\right| & \leq \varepsilon n\\
	\left|\sum_{v_0 \in V_0}\left|\partial B_{G_0}(v_0, 5)\right| - d(d - 1)^4\frac{n_0^2(n - 3n_0)^2}{(n - 2n_0)^3}\right| & \leq \varepsilon n\\
	\left|\sum_{v_0 \in V_0}\mathbf{1}_{B(v_0, 5)\textrm{ is a tree}} - n_0\right| & \leq \varepsilon n
	\end{align}
	for sufficiently large $n$.
\end{prop}
We can now prove proposition \ref{prop:upper_bound_conditioning_qaoa} along similar lines as proposition \ref{prop:upper_bound_conditioning_qaoa_expectation} from the previous subsection:
\begin{repprop}{prop:upper_bound_conditioning_qaoa}
    Let $\alpha \in (0, \alpha_3^*)$. Let $G = (V, E)$ a random 3-regular graph with $n$ vertices sampled from the configuration model and $V_0$ a 2-independent set of $G$ of size $\alpha n$. Then under conjecture \ref{conj:independent_set_typical_graph_frequent}, for all $\varepsilon > 0$, there exists $\overline{n} = \overline{n}(\varepsilon)$ such that for all $n \geq \overline{n}$,
	\begin{align}
		& \max_{\left(\sigma_{v_0}\right)_{v_0 \in V_0} \in \{-1, 1\}^{V_0}}\frac{\braket{\psi_G|\prod_{v_0 \in V_0}\frac{1 + \sigma_{v_0}Z_{v_0}}{2}H_{\textnormal{MaxCut}}(G)\prod_{v_0 \in V_0}\frac{1 + \sigma_{v_0}Z_{v_0}}{2}|\psi_G}}{\left\lVert\prod_{v_0 \in V_0}\frac{1 + \sigma_{v_0}Z_{v_0}}{2}\ket{\psi_G}\right\rVert^2} - \braket{\psi_G|H_{\textnormal{MaxCut}(G)}|\psi_G}\nonumber\\
		& \hspace{0.1\textwidth} \leq \left(\frac{\alpha^2(0.18125 - 0.83875\alpha + 0.99\alpha^2)}{(0.5 - \alpha)^3} + 0.59\varepsilon\right)n.
	\end{align}
	with high probability on $G$ and the choice of 2-independent set $V_0$ of $G$.
\end{repprop}
\begin{proof}
    Fix a graph $G$ and a 2-independent set $V_0$ of $G$ satisfying proposition \ref{prop:concentration_tree_neighbourhoods}. Recalling the notation in proposition \ref{prop:sparse_vertex_set_ising_model}, the advantage achieved by postselection can be recast as\\$\max_{\left(\sigma_{v_0 \in V_0}\right)_{v_0 \in V_0} \in \{-1, 1\}^{V_0}}\sum_{v_0, v_0' \in V_0}J(G, V_0)_{v_0v_0'}\sigma_{v_0}\sigma_{v_0'}$, hence bounded by $\sum_{v_0, v_0' \in V_0}\left|J(G, V_0)_{v_0v_0'}\right|$. Let us now fix $v_0 \in V_0$ and estimate $J(G, V_0)_{v_0, v_0'}$ for all $v_0'$. First consider the case where $B_G(v_0, 5)$ is a tree. We will systematically refer to figure \ref{fig:frozen_vertices_configurations}, detailing the possible configurations of a pair of vertices from a 2-independent set relative to an edge.
	\begin{itemize}
		\item For $v_0' \in B_G(v_0, 3)$, the pair $\{v_0, v_0'\}$ appears once in configuration 1, twice in configuration 4 and twice in configuration 6. In this case, $J(G, V_0)_{v_0v_0'} = 1 \times (-0.0833) + 2 \times (0.00926) + 2 \times (-0.0370)$ and $|J(G, V_0)_{v_0, v_0'}| \leq 0.140$.
		\item For $v_0' \in B_G(v_0, 4)$, the pair $\{v_0, v_0'\}$ appears twice in configuration 2 and thrice in configuration 5. Therefore, $J(G, V_0)_{v_0, v_0'} = 2 \times (0.0178) + 3 \times (-0.00356)$ and $|J(G, V_0)_{v_0, v_0'}| \leq 0.0252$.
		\item For $v_0' \in B_G(v_0, 5)$, the pair $\{v_0, v_0'\}$ appears thrice in configuration 3. Therefore $J(G, V_0)_{v_0, v_0'} = 3 \times (-0.00412)$ and $|J(G, V_0)_{v_0v_0'}| \leq 0.0124$.
	\end{itemize}
	Now, in case $B_G(v_0, 5)$ is not a tree, we crudely bound $|J(G, V_0)_{v_0, v_0'}| \leq 1$.
	
	Therefore,
	\begin{align*}
	& \sum_{\{v_0, v_0'\} \subset V_0}|J(G, V_0)_{v_0v_0'}|\\
	& = \frac{1}{2}\sum_{v_0 \in V_0}\sum_{v_0' \in V_0 - \{v_0\}}|J(G, V_0)_{v_0v_0'}|\\
	& = \frac{1}{2}\sum_{v_0 \in V_0}\mathbf{1}_{B_G(v_0, 5)\textrm{ tree}}\sum_{v_0' \in V_0 - \{v_0\}}|J(G, V_0)_{v_0v_0'}| + \frac{1}{2}\sum_{v_0 \in V_0}\mathbf{1}_{B_G(v_0, 5)\textrm{ not tree}}\sum_{v_0' \in V_0 - \{v_0\}}|J(G, V_0)_{v_0v_0'}|\\
	& \leq \frac{1}{2}\sum_{v_0 \in V_0\,:\,B(v_0, 5)\textrm{ tree}}\left(0.140|B(v_0, 3)| + 0.0252|B(v_0, 4)| + 0.0124|B(v_0, 5)|\right)\\
	& \hspace*{0.05\textwidth} + \frac{1}{2}\sum_{v_0 \in V_0\,:\,B(v_0, 5)\textrm{ not tree}}1
	\end{align*}
	Now, we may use proposition \ref{prop:concentration_tree_neighbourhoods} to bound the first line by \begin{align*}
		& \left(0.84\frac{\alpha^2}{1 - 2\alpha} + 0.31\frac{\alpha^2(1 - 3\alpha)}{(1 - 2\alpha)^2} + 0.30\frac{\alpha^2(1 - 3\alpha)^2}{(1 - 2\alpha)^3} + 0.09\varepsilon\right)n
	\end{align*}
	and the second line by
	\begin{align*}
		\frac{\varepsilon}{2}n
	\end{align*}
	with high probability on the choice of $G$ and 2-independent set $V_0$ in $G$.
\end{proof}

\subsection{Lower bound}
Having completed the derivation of upper bounds, let us consider the (weaker) lower bound stated in proposition \ref{prop:lower_bound}:
\begin{repprop}{prop:lower_bound}
	Let $G$ a random 3-regular graph and $V_0$ a 2-independent set of $G$ selected by algorithm \ref{alg:systematic_sparse_vertex_set}. Then, with high probability (on the choice of $G$), conditioned on the vertices from $V_0$ being measured to $1$, the cut sampled from the QAOA state is $0.0013|E|$ above the cut sampled from the unconditioned QAOA.
\end{repprop}
\begin{proof}
	We refer to figure \ref{fig:frozen_vertices_configurations} to calculate the contributions of the conditioned vertices to the expect cut size, similar to the proof of proposition \ref{prop:upper_bound_conditioning_qaoa}. For any $\delta > 0$, initially, w.h.p the graph contains $\frac{1}{2}3 \times 2^{4 - 1}n(1 - \delta) = 12(1 - \delta)n$ pairs of vertices with distance 4 (this results from the tree structure of constant-depth neighbourhoods in random regular graphs). Each round of the loop removes at most $2 \times 3 \times (2^{5 - 1} + \ldots + 2^0) = 186$ pairs from $candidatePairs$. Therefore, w.h.p the algorithm returns more than $\frac{2}{31}(1 - \delta)n$ pairs. This corresponds to an enhancement of the expected energy of $0.0013|E|$. Now, since by construction of the algorithm, any pair of vertices in the 2-independent set either have a distance 4 or a distance $\geq 6$, configurations 2 and 5 are the only possible configurations. A pair of vertices at distance 4 from each other contributes two configuration 2 and one configuration 5. Therefore, the cut sampled from the conditioned QAOA state is
	\begin{align*}
		& \frac{2}{31}(1 - \delta)n\left(2 \times \underbrace{0.0177}_{\textrm{configuration 2 coupling}} + 1 \times \underbrace{(-0.00357)}_{\textrm{configuration 5 coupling}}\right)\\%
		& \geq 0.0013|E|
	\end{align*}
	(for small enough $\delta$).
\end{proof}
\section{Conclusion}
In this work, we considered an approach to improve the solutions to MaxCut on 3-regular graph obtained by depth-1 QAOA. We explored the idea of postselecting sampled bitstrings conditioned on a subset of bits taking well-chosen values. We established an upper bound (assuming a conjecture on the number of 2-independent sets of random regular graphs), leaving open the possibility of a mild advantage using this method. Numerical experiments on large graph instances (beyond full classical simulation) achieved an improvement matching the upper bound. We then combined postselection of sampled bitstrings with local updates. Though the advantage offered by the latter could be rigorously quantified in the infinite size limit in absence of postselection, we could merely rely on numerical experiments with postselection. The latters suggest that improvements from both methods combine.

In this paper, we considered postselecting on vertices from a 2-independent set. Though this is the simplest scenario to analyze, this is not the only classically tractable one and more insight on the performance of our approach could be gained by analyzing other configurations. Finally, given that postselection can be simulated by state preparation in depth-1 QAOA, our approach amounts to changing the initial state of QAOA, in line with other recent proposals to improve the algorithm. It would be interesting to explore other changes to the initial state more systematically for the same problem.

\subsection*{Acknowledgements}

This work was supported by the EPSRC Centre for Doctoral Training in Delivering Quantum Technologies, grant ref. EP/S021582/1.

% ------------------------------------------------------------------------------

\bibliographystyle{hapalike}
\bibliography{bibliography}

% ------------------------------------------------------------------------------

\appendix

\section{Technical results for postselected QAOA}
\label{sec:technical_results_postselect_qaoa}

\subsection{2-independents sets and their neighbourhood: expectation estimates}
\label{subsec:2_independent_sets_estimates}
In this section, we derive an estimate for the expected number of 2-independent set in a random $d$-regular graph (proposition \ref{prop:expected_number_2_independent_sets}). Then, given a set of vertices $V_0$ and $v_0 \in V_0$, we establish a lemma (lemma \ref{lemma:sparse_set_distance_3_4_5_successors}) describing how the neighbourhoods of $v_0$ intersect $V_0$ in expectation, conditioned on $V_0$ being 2-independent.
The following proposition gives the expected number of 2-independent sets embeddable in a multigraph sampled from the configuration model:
\begin{prop}
\label{prop:expected_number_2_independent_sets}
	Let $G$ a random $d$-regular multigraph on vertex set $\left\{1, \ldots, n\right\}$ sampled uniformly from the configuration model. Then for all integer $n_0 \geq 1$, the expected number of $2$-independent vertex sets of size $n_0$ in $G$ is:
	\begin{align}
	\binom{n}{n_0}\frac{d^{n_0d}(n - n_0)!(nd - 1 - 2n_0d)!!}{(n - n_0 - n_0d)!(nd - 1)!!} & = \frac{(2d)^{n_0d}n!\left(\frac{nd}{2}\right)!\left(nd - 2n_0d\right)!}{n_0!(n - n_0 - n_0d)!(nd)!\left(\frac{nd}{2} - n_0d\right)!}
	\end{align}
	\begin{proof}
		Let us consider one of the $\binom{n}{n_0}$ sets of vertices of size $n_0$ and compute the probability that this set be $2$-independent for a multigraph uniformly sampled from the configuration model. Denote by $V_0$ this set.
		
		First, a necessary and sufficient condition for $V_0$ to be $2$-independent is that the half-edges of each vertex in $V_0$ be matched to half-edges belonging to pairwise distinct vertices not in $V_0$. We call a \textit{good matching} a perfect matching satisfying this property.
		
		Then, observe that a uniformly sampled perfect matching between the $nd$ half-edges can be obtained by successively matching the $d$ half-edges of the first vertex of $V_0$, then the remaining half-edges of the second vertex of $V_0$, etc. The matching extends to a good one after the first step iff the first half-edge of the first vertex of $V_0$ is matched to a half-edge of a vertex $v_1$ not in $V_0$, which happens with probability $\frac{(n - n_0)d}{nd - 1}$. Then, given the matching after the first step extends to a good one, the matching after the second step extends to a good one iff the second half-edge of the first vertex of $V_0$ is matched to a half-edge belonging to a vertex in $V - V_0 - \{v_1\}$, which happens with probability $\frac{(n - n_0 - 1)d}{nd - 3}$. Therefore, the matching obtained after the second step extends to a good one with probability $\frac{(n - n_0)(n - n_0 - 1)d^2}{nd - 1}$. Iterating up to step $n_0d$ (so as to match all half-edges belonging to vertices in $V_0$), the probability of obtaining a good matching is
		\begin{align*}
		d^{n_0d}\prod_{0 \leq k < n_0d}\frac{n - n_0 - k}{nd - 1 - 2k} & = \frac{d^{n_0d}(n - n_0)!(nd - 1 - 2n_0d)!!}{(n - n_0 - n_0d)!(nd - 1)!!}\\
		& = \frac{(2d)^{n_0d}(n - n_0)!\left(\frac{nd}{2}\right)!\left(nd - 2n_0d\right)!}{(n - n_0 - n_0d)!(nd)!\left(\frac{nd}{2} - n_0d\right)!}
		\end{align*}
		The result follows.
	\end{proof}
\end{prop}
An asymptotic expansion of the estimate from this proposition yields the following corollary, from which proposition \ref{prop:upper_bound_k_independent_set} can be recovered:
\begin{cor}
	\label{cor:alpha_d}
	There exists $\alpha_d^* \in \left(0, \frac{1}{d + 1}\right)$ such that for all $\alpha \in \left(0, \alpha_d^*\right)$, the expected number of 2-independent sets of size $\alpha n$ in a random $d$-regular multigraph with $n$ vertices sampled from the configuration model increases exponentially with $n$, while it decreases exponentially with $n$ for $\alpha > \alpha_d^*$. In particular, for $d = 3$, one can estimate $\alpha_3^* \in (0.235, 0.236)$.
\end{cor}
The following lemma now characterizes the neighbourhoods of a vertex from a 2-independent set:
\begin{replem}{lemma:sparse_set_distance_3_4_5_successors}
	Let $V_0$ a set of vertices of size $n_0 \geq cn$ ($c > 0$) and $v_0 \in V_0$. For a multigraph $G$ drawn from the configuration model, conditional on $V_0$ being 2-independent, the probability that the $5$-neighbourhood of $v_0$ in $G$ is a tree and that $v_0$ contains $k_i$ vertices from $V_0$ among its distance $i$ successors for all $i \in \{3, 4, 5\}$ is given by:
	\begin{align}
	& p(k_3, k_4, k_5)\\
	& := \sum_{\subalign{0 \leq k_6 \leq & d(d - 1)^4\\& - (d - 1)^2k_3\\& - (d - 1)k_4\\& - k_5}}\binom{d(d - 1)}{k_3}\binom{d(d - 1)^2 - k_3}{k_4}\binom{d(d - 1)^3 - (d - 1)k_3 - k_4}{k_5}\nonumber\\
	& \hspace{0.05\textwidth} \times \binom{d(d - 1)^4 - (d - 1)^2k_3 - (d - 1)k_4 - k_5}{k_6}\left(\frac{d(d - 1)n_0}{nd - 2n_0d}\right)^{k_3 + k_4 + k_5 + k_6}\nonumber\\
	& \hspace*{0.05\textwidth} \times \left(\frac{nd - d(d + 1)n_0}{nd - 2n_0d}\right)^{d(d - 1)\frac{(d - 1)^4 - 1}{d - 2} - (d^2 - d + 2)k_3 - (d + 1)k_4 - 2k_5 - k_6}\nonumber\\
	& \hspace{0.05\textwidth} \times \left(1 + \mathcal{O}\left(\frac{1}{n}\right)\right)
	\end{align}
	where the implicit constant in the $\mathcal{O}\left(\frac{1}{n}\right)$ depends only on $c$ such that $n_0 \geq cn$ and $d$.
\end{replem}
\begin{proof}
	Consider forming the matching by starting to match the half-edges of the $n_0$ vertices of $V_0$. Then the set $V_0$ is 2-independent for the selected matching iff. after $n_0d$ matching steps, the half-edges of the vertices belonging to $V_0$ are matched to half-edges belonging to pairwise distinct vertices not in $V_0$. For each $v_0 \in V_0$, we denote by $v_0^{(i)}$, $1 \leq i \leq d$ the vertices now connected to $v_0$. In this case, the matching after $n_0d$ steps must look like a collection of disjoint $(d + 1)$-vertices star graphs, where the centers of stars are the vertices in $V_0$. Now, fix some $v_0 \in V_0$ and given the initial configuration just described (with $n_0d$ pairs of half-edges already formed), let us count the complete matchings such that the $5$-neighbourhood of $v_0$ is a tree and $v_0$ has respectively $k_3$, $k_4$, $k_5$ distance $3$, $4$, $5$ successors.
	
	We consider complementing the initial matching by first matching the $(d - 1)$ remaining half-edges of each of the descendants $v_0^{(1)}, \ldots, v_0^{(d)}$ of $v_0$. To obtain $k_3$ distance-3 successors of $v_0$ lying in $V_0$ while ensuring that its $5$-neighbourhood be a tree, $k_3$ half-edges must be matched to half-edges belonging to vertices $v_0'^{(i)}$ ($v_0' \notin v_0$), where the $v_0'$ have to be pairwise distinct, and $d(d - 1) - k_3$ half-edges must be matched to half-edges belonging to vertices in $V - V_0 \cup \left\{v^{(i)}\,:\,v \in V_0, 1 \leq i \leq d\right\}$. This gives
	\begin{align*}
	& \binom{d(d - 1)}{k_3}\binom{n_0 - 1}{k_3}k_3![d(d - 1)]^{k_3}\\
	& \hspace*{0.05\textwidth} \times (n - (d + 1)n_0)d \times (n - (d + 1)n_0 - 1)d \times \ldots \times (n - (d + 1)n_0 - (d(d - 1) - k_3 + 1))d\\
	& = \binom{d(d - 1)}{k_3}\binom{n_0 - 1}{k_3}k_3![d(d - 1)]^{k_3}\frac{(n - (d + 1)n_0)!}{(n - (d + 1)n_0 - (d(d - 1) - k_3))!}d^{d(d - 1) - k_3}\\
	& = \binom{d(d - 1)}{k_3}\left[d(d - 1)n_0\right]^{k_3}\left[nd - d(d + 1)n_0\right]^{d(d - 1) - k_3}\left(1 + \mathcal{O}\left(\frac{1}{n}\right)\right)
	\end{align*}
	In the line before the last, the $\binom{d(d - 1)}{k_3}$ is the number of choices for the $k_3$ half-edges belonging to vertices $v_0^{(i)}$ that will be matched to half-edges belonging to vertices $v_0'^{(i)}$ ($v_0' \neq v_0$). The $\binom{n_0 - 1}{k_3}$ is the number of choices for $k_3$ pairwise distinct $v_0'$. $k_3!$ is the number of mappings between the $k_3$ chosen half-edges attached to $v_0^{(i)}$ and the $k_3$ chosen vertices $v_0'$. Given chosen vertices $v_0' \neq v_0$, $[d(d - 1)]^{k_3}$ is the number of choices for the half-edges of the $v_0'{(i)}$ that will be matched to half-edges of the $v_0^{(i)}$. Finally, having done the $k_3$ matchings previously described, one has yet to match the remaining $d(d - 1) - k_3$ half-edges attached to the $v_0^{(i)}$ with half-edges belonging to pairwise-distinct vertices from $V - V_0 \cup \left\{v^{(i)}\,:\,v \in V_0, 1 \leq i \leq d\right\}$. There are therefore $(n - (d + 1)n_0)d$ choices for the first such half-edge; given the first matching, there are $(n - (d + 1)n_0 - 1)d$ choices for the second such half-edge, etc., which justifies $\frac{(n - (d + 1)n_0)!}{(n - (d + 1)n_0 - (d(d - 1) - k_3))!}d^{d(d - 1) - k_3}$. 
	
	Having attributed a tree $2$-neighbourhood and $k_3$ distance-3 successors to $v_0$ in the way described above, let us now consider the number of possibilities of attributing $k_4$ distance-4 successors to $v_0$. First, let us denote by $v_0^{(i, j)}$ ($1 \leq i \leq d$, $1 \leq j \leq d - 1$) the vertices connected to $v_0^{(i)}$ in the previous step. Using this notation, we recall from the previous step that $k_3$ of the $v_0^{(i, j)}$ are also a $v_0'^{(i)}$ for some $v_0' \in V_0 - \{v_0\}$ while $d(d - 1)$ of them are in $V - V_0 \cup \left\{v^{(i)}\,:\,v \in V_0, 1 \leq i \leq n\right\}$. We now continue to construct the matching by matching the still free half-edges of the $v_0^{(i, j)}$. Each of the $v_0^{(i, j)}$ that is also a $v_0'^{(i)}$ has $(d - 2)$ free half edges remaining; the other $v_0^{(i ,j)}$ have $(d - 1)$ such half-edges; therefore, there is a total number of $k_3(d- 2) + (d(d - 1) - k_3)(d - 1) = d(d - 1)^2 - k_3$ half-edges to match. To enforce $k_4$ distance-4 successors on $v_0$ while guaranteeing that the $5$-neighbourhood be a tree, one must match $k_4$ of these half-edges to half-edges belonging to vertices $v_0'^{(i)}$, where $v_0' \neq v_0$ and $v_0'$ was not used in the previous steps of the matching (violating the latter would compromise the tree neighbourhood requirement). The remaining $d(d - 1)^2 - k_3 - k_4$ half-edges must be matched to half-edges belonging pairwise distinct vertices in $V - V_0 \cup \left\{v^{(i)}\,:\,v \in V_0, 1 \leq i \leq d\right\} \cup \left\{v_0^{(i, j)}\,:\,1 \leq i \leq d, 1 \leq j \leq d - 1\right\}$. By arguments similar to the previous one, there are
	\begin{align*}
	&\binom{d(d - 1)^2 - k_3}{k_4}\binom{n_0 - 1 - k_3}{k_4}k_4![d(d - 1)]^{k_4}\\
	& \hspace*{0.05\textwidth} \times \frac{(n - (d + 1)n_0 - (d(d - 1) - k_3))!}{(n - (d + 1)n_0 - (d(d - 1)- k_3) - (d(d - 1)^2 - k_3 - k_4))!}d^{d(d - 1)^2 - k_3 - k_4}\\
	& = \binom{d(d - 1)^2 - k_3}{k_4}\left[d(d - 1)n_0\right]^{k_4}\left[nd - d(d + 1)n_0\right]^{d(d - 1)^2 - k_3 - k_4}\left(1 + \mathcal{O}\left(\frac{1}{n}\right)\right)
	\end{align*}
	ways of extending the matching. We denote by $v_0^{(i, j, k)}$, where $1 \leq k \leq d - 2$ if $v_0^{(i, j)}$ is a $v_0'^{(i)}$ and $1 \leq k \leq d - 1$ otherwise, the vertices connected to $v_0^{(i, j)}$ in this step.
	
	The partial matching currently done fixes the number distance 3 and 4 successors of $v_0$ while ensuring that the $5$-neighbourhood of $v_0$ be a tree. Given this partial matching, the next step matches the free half-edges of vertices $v_0^{(i, j, k)}$, prescribing the number $k_5$ of distance-5 successors for $v_0$ and maintaining the tree neighbourhood invariant. Considerations similar to those of the previous step then reveal there are
	\begin{align*}
	& \binom{d(d - 1)^3 - k_3(d - 1) - k_4}{k_5}\binom{n_0 - 1 - k_3 - k_4}{k_5}k_5![d(d - 1)]^{k_5}\\
	& \hspace*{0.05\textwidth} \times \frac{(n - (d + 1)n_0 + 2k_3 + k_4 - d^2(d - 1))!d^{d(d - 1)^3 - (d - 1)k_3 - k_4 - k_5}}{(n - (d + 1)n_0 + (d + 1)k_3 + 2k_4 + k_5 - d(d - 1)\frac{(d - 1)^3 - 1}{d - 2})!}\\
	& = \binom{d(d - 1)^3 - k_3(d - 1) - k_4}{k_5}\left[d(d - 1)n_0\right]^{k_4}\left[nd - d(d + 1)n_0\right]^{d(d - 1)^3 - (d - 1)k_3 - k_4 - k_5}\\
	& \hspace*{0.05\textwidth} \times \left(1 + \mathcal{O}\left(\frac{1}{n}\right)\right)
	\end{align*}
	ways of extending the matching satisfying the required constraints. Similar to what precedes, we denote by $v_0^{(i, j, k, l)}$ ($1 \leq l \leq d - 2$ or $1 \leq l \leq d - 1$ depending on whether $v_0^{(i, j, k)}$ is a $v_0'^{(i)}$ or not) the vertices that are connected to $v_0^{(i, j, k)}$. $k_5$ of the $v_0^{(i, j, k, l)}$ are also a $v_0'^{(i)}$, $d(d - 1)^3 - k_3(d - 1) - k_4 - k_5$ are vertices not used before.
	
	To finally fulfill the tree 5-neighbourhood requirement, the free half-edges of the $v_0^{(i, j, k, l)}$ must be matched to half-edges belonging to pairwise distinct vertices. These vertices may be chosen from the yet unused ones, but also from the $v_0'^{(i)}$ such that no half-edge was matched to any $v_0'^{(i)}$ in the first two matching steps but a half-edge was possibly matched to some $v_0'^{(i')}$, $i' \neq i$ in the third matching step. The first category contains
	\begin{align*}
	& n - (d + 1)n_0 - (d(d - 1) - k_3) - (d(d - 1)^2 - k_3 - k_4) - (d(d - 1)^3 - (d - 1)k_3 - k_4 - k_5)\\
	& = n - (d + 1)n_0 + (d + 1)k_3 + 2k_4 + k_5 - d(d - 1)\frac{(d - 1)^3 - 1}{d - 2}
	\end{align*}
	vertices, the second one
	\begin{align*}
	(n_0 - k_3 - k_4 - k_5)d + k_5(d - 1) & = n_0d - k_3d - k_4d - k_5
	\end{align*}
	vertices. Now, if we match a free half-edge of some $v_0^{(i, j, k, l)}$ to a half-edge attached to a vertex from the first category, then there are $d$ possible choices for the latter half-edge. However, if we match to a half-edge attached to a vertex from the second category, there are $d - 1$ choices of half-edge. There are $d(d - 1)^4 - (d - 1)^2k_3 - (d - 1)k_4 - k_5$ half-edges from
	$v_0^{(i, j, l, k)}$ to match. Denoting by $k_6$ the number of half-edges matched to half-edges belonging to vertices from the second category, there are then
	\begin{align*}
	& \sum_{\subalign{0 \leq k_6 \leq & d(d - 1)^4\\ & - (d - 1)^2k_3\\ & - (d - 1)k_4\\ &  - k_5}}\binom{n_0d - k_3d - k_4d - k_5}{k_6}\binom{d(d - 1)^4 - (d - 1)^2k_3 - (d - 1)k_4 - k_5}{k_6}\\
	& \hspace*{0.02\textwidth} \times k_6!(d - 1)^{k_6}\\
	& \hspace*{0.02\textwidth} \times \frac{\left(n - (d + 1)n_0 + (d + 1)k_3 + 2k_4 + k_5 - d(d - 1)\frac{(d - 1)^3 - 1}{d - 2}\right)!}{\left(n - (d + 1)n_0 + (d^2 - d + 2)k_3 + (d + 1)k_4 + 2k_5 + k_6 - d(d - 1)\frac{(d - 1)^4 - 1}{d - 2}\right)!}\\
	& \hspace{0.02\textwidth} \times d^{d(d - 1)^4 - (d - 1)^2k_3 - (d - 1)k_4 - k_5 - k_6}\\
	& = \binom{d(d - 1)^4 - (d - 1)^2k_3 - (d - 1)k_4 - k_5}{k_6}\left[n_0d(d - 1)\right]^{k_6}\\
	& \hspace*{0.05\textwidth} \times \left[nd - d(d + 1)n_0\right]^{d(d - 1)^4 - (d - 1)^2k_3 - (d - 1)k_4 - k_5}\left(1 + \mathcal{O}\left(\frac{1}{n}\right)\right)
	\end{align*}
	
	Once this has been done, $v_0$ is guaranteed to have a tree $5$-neighbourhood, and $k_3, k_4, k_5$ distance $3, 4, 5$ successors. The still free half-edges can then be matched completely arbitrarily. There are
	\begin{align*}
	& nd - 2n_0d - 2\left[d(d - 1) + d(d - 1)^2 - k_3 + d(d - 1)^3 - (d - 1)k_3 - k_4\right.\\
	& \left. \hspace*{0.2\textwidth} + d(d - 1)^4 - (d - 1)^2k_3 - (d - 1)k_4 - k_5\right]\\
	& = nd - 2n_0d - 2\left(d(d - 1)\frac{(d - 1)^4 - 1}{d - 2} - (d^2 - d + 1)k_3 - k_4d - k_5\right)
	\end{align*}
	such half-edges. Therefore, there are
	\begin{align*}
	& \left(nd - 2n_0d - 2\left(d(d - 1)\frac{(d - 1)^4 - 1}{d - 2} - (d^2 - d + 1)k_3 - k_4d - k_5\right) - 1\right)!!
	\end{align*}
	of completing the matching.
	
	To obtain $p(k_3, k_4, k_5)$, we multiply together the ways of realizing each step and divide by $(nd - 2n_0d - 1)!!$, the number of possible matchings conditioned on $V_0$ being 2-independent.
\end{proof}

\subsection{2-independent sets and their neighbourhood: high-probability estimates}
In this section, we derive estimates characterizing the neighbourhoods of 2-independent sets ---similar to the previous subsection, except the results hold with high probability and not only in expectation. However these estimates do depend on conjecture \ref{conj:independent_set_typical_graph_frequent}.

Results holding with high probability will be established using concentration inequalities. Since in the configuration model \ref{def:configuration_model}, random regular graphs can be described as perfect matchings, we then start with a few concentration results applying to functions of matchings.
\begin{prop}[Concentration inequality for functions of matching {\cite[theorem 2.19]{wormald_1999}}, see also {\cite[corollary 3.27]{Bordenave16}}]
	
	\label{prop:uniform_matchings_concentration_inequality}
	
	Let $\Delta$ be a set with an even number of elements and $F$ a real function on the matchings of $\Delta$. Assume that whenever matchings $m, m'$ differ only by a switch,
	\begin{align}
	\left|F(m) - F(m')\right| & \leq c.
	\end{align}
	Then, for $m$ drawn uniformly from the matchings on $\Delta$,
	
	\begin{align}
	\mathbf{P}\left[F(m) - \mathbf{E}F(m) \geq t\right] & \leq \exp\left(-\frac{t^2}{|\Delta|c^2}\right).
	\end{align}
\end{prop}
To demonstrate proposition \ref{prop:concentration_tree_neighbourhoods}, leading to the main result \ref{prop:concentration_tree_neighbourhoods}, we will need a slightly stronger version of this result, which will be proved thanks to the following variant of the Azuma-Hoeffding inequality:
\begin{lem}[Variant of Azuma-Hoeffding inequality]
	\label{lemma:variation_azuma_hoeffding}
	Let $\{X_0, X_1, \ldots\}$ be a supermartingale with respect to filtration $\{\mathcal{F}_0, \mathcal{F}_1, \ldots\}$. Assume that there exists a constant $c$ such that for all $k \geq 0$, $|X_{k + 1} - X_k| \leq c$. Then for all $n \geq 1$ and all event $A_0 \in \mathcal{F}_0$,
	\begin{align}
	\mathbf{P}\left[X_n - X_0 \geq t, A_0\right] & \leq \exp\left(-\frac{t^2}{2nc^2}\right)\mathbf{P}[A_0].
	\end{align}
\end{lem}
The following proposition is then the promised strengthening of the concentration inequality on matchings:
\begin{prop}[Modified concentration inequality for functions of matching]
	\label{prop:uniform_matchings_modified_concentration_inequality}
	Let $\Delta$ a set with $2n$ elements ($n$ positive integer). Let $(m_k)_{0 \leq k \leq n}$ a random process taking values in the partial matchings of $\Delta$, such that $|m_0| = 0$ and for all $k \geq 0$, $m_{k + 1}$ extends $m_k$ by one pair (therefore, $m_n$ is a complete matching). Assume that for $k \geq k_0$, $m_{k + 1}$ is uniformly distributed among (partial) matchings extending $m_k$ by one pair (in particular, this implies that for any $k \geq k_0$, $m_n$ is uniformly distributed among matchings completing $m_k$). Denote by $\left(\mathcal{F}_{k}\right)_{0 \leq k \leq n}$ the $\sigma$-algebra generated by $(m_k)_{0 \leq k \leq n}$. Then, for any event $A_{k_0} \in \mathcal{F}_{k_0}$ and any function $F$ of the matchings of $\Delta$ whose variation is bounded by $c$ under a switching,
	\begin{align}
	\mathbf{P}\left[F(m_n) - \mathbf{E}\left[F(m_n)\,|\,\mathcal{F}_{k_0}\right] \geq t, A_{k_0}\right] & \leq \exp\left(-\frac{t^2}{(|\Delta| - 2k_0)c^2}\right)\mathbf{P}\left[A_{k_0}\right].
	\end{align}
	\begin{proof}
		The proof is a repetition of \cite[corollary 3.27]{Bordenave16} up to a few details. We will apply the variation of the Azuma-Hoeffding inequality to the martingale $\left(\mathbf{E}\left[F(m_n)\,|\,\mathcal{F}_k\right]\right)_{0 \leq k \leq n}$. Given any set $E$, we temporarily denote by $\mathcal{M}(E)$ the set of matchings of $E$. Given a set $E$ and a partial matching $m$ on $E$, we also denote, in a slight abuse of notation, by $E \setminus m$ the set $E$ minus the elements of $E$ appearing in the pairs of $m$. We need to bound the difference between 2 consecutive terms of the martingale. For $k \geq k_0$
		\begin{align*}
		& \mathbf{E}\left[F(m_n)\,|\,\mathcal{F}_{k + 1}\right] - \mathbf{E}\left[F(m_n)\,|\,\mathcal{F}_k\right]\\
		& = \frac{1}{(2n - 2k - 3)!!}\sum_{m \in \mathcal{M}\left(\Delta \setminus m_{k + 1}\right)}F(m_{k + 1}; m) - \frac{1}{(2n - 2k - 1)!!}\sum_{m \in \mathcal{M}(\Delta \setminus m_k)}F(m_k; m),
		\end{align*}
		where, following the notation of \cite[corollary 3.27]{Bordenave16}, we denoted by $m; m'$ the disjoint union of partial matchings $m$ and $m'$.
		We now transform the two sums. To achieve that, let us denote, following the idea of \cite[corollary 3.27]{Bordenave16}, by $v_k$ the smallest element of $\Delta \setminus m_k$ and by $w_k$ the element to which $v_k$ is matched in $m_{k + 1}$. Besides, denote, for any $w \in \Delta \setminus (m_k \cup \{w\})$, by $\mathcal{M}_{w}\left(\Delta \setminus m_k\right)$ the matchings of $\Delta \setminus m_k$ where $v_k$ is matched to $w$. Using these notations,
		\begin{align*}
		\sum_{m \in \mathcal{M}(\Delta \setminus m_k)}F(m_k; m) & = \sum_{w \in \Delta \setminus (m_k \cup \{v_k\})}\sum_{m \in \mathcal{M}_w\left(\Delta \setminus m_k\right)}F(m_k; m)\\
		\sum_{m \in \mathcal{M}\left(\Delta \setminus m_{k + 1}\right)}F(m_{k + 1}; m) & = \sum_{m \in \mathcal{M}_{w_k}(\Delta \setminus m_k)}F(m_k; m)\\
		& = \frac{1}{2n - 2k - 1}\sum_{w \in \Delta \setminus (m_k \cup \{v_k\})}\sum_{m \in \mathcal{M}_{w_k}(\Delta \setminus m_k)}F(m_k; m)
		\end{align*}
		This gives
		\begin{align*}
		& \mathbf{E}\left[F(m_n)\,|\,\mathcal{F}_{k + 1}\right] - \mathbf{E}\left[F(m_n)\,|\,\mathcal{F}_k\right]\\
		& = \frac{1}{(2n - 2k - 1)!!}\sum_{w \in \Delta \setminus (m_k \cup \{v_k\})}\left(\sum_{m \in \mathcal{M}_{w_k}(\Delta \setminus m_k)}F(m_k; m) - \sum_{m \in \mathcal{M}_w(\Delta \setminus m_k)}F(m_k; m)\right)
		\end{align*}
		Now, as detailed in \cite[corollary 3.27]{Bordenave16}, matchings from $\mathcal{M}_w(\Delta \setminus m_k)$ can be put in one-to-one correspondence with matchings from $\mathcal{M}_{w_k}(\Delta \setminus m_k)$, such that two corresponding matchings differ by a switching. The former expression can then be bounded in absolute value by
		\begin{align*}
		\frac{1}{(2n - 2k - 1)!!} \times (2n - 2k - 1) \times (2n - 2k - 3)!! \times c & = c
		\end{align*}
		The result then follows from the variation of the Azuma-Hoeffding inequality in lemma \ref{lemma:variation_azuma_hoeffding}.
	\end{proof}
\end{prop}
Combining the latter concentration inequality and conjecture \ref{conj:independent_set_typical_graph_frequent} leads then leads to:
\begin{repprop}{prop:concentration_tree_neighbourhoods}
	Let $\varepsilon > 0$ and $\alpha \in \left(0, \alpha^*_d\right)$ and set $n_0 := \alpha n$. Under conjecture \ref{conj:independent_set_typical_graph_frequent}, the following holds: for all but an exponentially small fraction (in $n$) of 2-independent-set-typical graphs $G_0$ and all but an exponentially small fraction of 2-independent sets $V_0$ of $G_0$,
	\begin{align}
	\left|\sum_{v_0 \in V_0}\left|\partial B_{G_0}(v_0, 3)\right| - d(d - 1)^2\frac{n_0^2}{n - 2n_0}\right| & \leq \varepsilon n\\
	\left|\sum_{v_0 \in V_0}\left|\partial B_{G_0}(v_0, 4)\right| - d(d - 1)^3\frac{n_0^2(n - 3n_0)}{(n - 2n_0)^2}\right| & \leq \varepsilon n\\
	\left|\sum_{v_0 \in V_0}\left|\partial B_{G_0}(v_0, 5)\right| - d(d - 1)^4\frac{n_0^2(n - 3n_0)^2}{(n - 2n_0)^3}\right| & \leq \varepsilon n\\
	\left|\sum_{v_0 \in V_0}\mathbf{1}_{B(v_0, 5)\textrm{ is a tree}} - n_0\right| & \leq \varepsilon n
	\end{align}
	for sufficiently large $n$.
\end{repprop}
\begin{proof}
	Given a $d$-regular multigraph $G_0$ sampled from the configuration model, let us call a vertex set $V_0 \subset V$, $|V_0| = n_0$ ``good" if
	\begin{align*}
	\left|\sum_{v_0 \in V_0}\left|\partial B_{G_0}(v_0, 3)\right| - \mathbf{E}_G\left[\sum_{v_0 \in V_0}\left|\partial B_G(v_0, 3)\right|\,\bigg|\,V_0\textrm{ 2-id}\right]\right| & \leq \varepsilon n
	\end{align*}
	Let us now evaluate 
	\begin{align*}
	& \mathbf{E}\left[\sum_{\substack{V_0 \subset V\\|V_0| = n_0}}\mathbf{1}_{V_0\textrm{ 2-id}}\mathbf{1}_{V_0\textrm{ is not good}}\right]\\
	& = \sum_{\substack{V_0 \subset V\\|V_0| = n_0}}\mathbf{E}\left[\mathbf{E}\left[\mathbf{1}_{V_0\textrm{ is not good}}\,|\,V_0\textrm{ 2-id}\right]\mathbf{1}_{V_0\textrm{ 2-id}}\right]\\
	& = \sum_{\substack{V_0 \subset V\\|V_0| = n_0}}\mathbf{E}\left[\mathbf{P}\left[\left|\sum_{v_0 \in V_0}\left|\partial B_{G_0}(v_0, 3)\right| - \mathbf{E}_G\left[\sum_{v_0 \in V_0}\left|\partial B_G(v_0, 3)\right|\,\bigg|\,V_0\textrm{ 2-id}\right]\right| > \varepsilon n\,\bigg|\,V_0\textrm{ 2-id}\right]\right.\\
	& \hspace{0.15\textwidth} \mathbf{1}_{V_0\textrm{ 2-id}}\bigg]
	\end{align*}
	Now, fix $V_0 \subset V$ and rewrite a $\mathbf{E}\left[\sum_{v_0 \in V_0}\left|\partial B_G(v_0, 3)\right|\,\bigg|\,V_0\textrm{ 2-id}\right]$ in a form that will lend itself to applying a concentration inequality. For that purpose, consider the uniform sampling of a matching as a random process where a pair of half-edges is matched at each step. Besides, we start by pairing the half-edges belonging to vertices from $V_0$. Denote by $\mathcal{F}_t$ the $\sigma$-algebra generated by the first $t$ pairings. It is then easy to see that on $V_0$ 2-independent, for each $v_0 \in V_0$, $\mathbf{E}\left[|\partial B_G(v_0, 3)|\,\bigg|\,\mathcal{F}_{n_0d}\right]$ is in fact a constant (in other words, it is independent of the first $n_0d$ pairings, the set $V_0$ and $v_0$). This constant evaluates to:
	\begin{align*}
	\frac{\mathbf{E}_G\left[\mathbf{E}_G\left[|\partial B_G(v_0, 3)|\,\bigg|\,\mathcal{F}_{n_0d}\right]\mathbf{1}_{V_0\textrm{ 2-id}}\right]}{\mathbf{P}(V_0\textrm{ 2-id})} & = \frac{\mathbf{E}_G\left[\mathbf{E}_G\left[|\partial B_G(v_0, 3)|\mathbf{1}_{V_0\textrm{ 2-id}}\,\bigg|\,\mathcal{F}_{n_0d}\right]\right]}{\mathbf{P}[V_0\textrm{ 2-id}]}\\
	& = \frac{\mathbf{E}\left[|\partial B_G(v_0, 3)|\mathbf{1}_{V_0\textrm{ 2-id}}\right]}{\mathbf{P}[V_0\textrm{ 2-id}]}\\
	& = \mathbf{E}\left[|\partial B_G(v_0, 3)|\,\bigg|\,V_0\textrm{ 2-id}\right]
	\end{align*}
	Therefore, we can rewrite
	\begin{align*}
	   & \mathbf{E}\left[\mathbf{1}_{V_0\textrm{ 2-id}}\mathbf{1}_{V_0\textrm{ not good}}\right]\\
    	& = \mathbf{E}\left[\mathbf{P}\left[\left|\sum_{v_0 \in V_0}\left|\partial B_{G_0}(v_0, 3)\right| - \mathbf{E}_G\left[\sum_{v_0 \in V_0}\left|\partial B_G(v_0, 3)\right|\,\bigg|\,\mathcal{F}_{n_0d}\right]\right| > \varepsilon n\,\bigg|\,V_0\textrm{ 2-id}\right.\right]\\
    	& \hspace{0.08\textwidth} \mathbf{1}_{V_0\textrm{ 2-id}}\bigg]
	\end{align*}
	We now want to apply the concentration inequality in proposition \ref{prop:uniform_matchings_modified_concentration_inequality} to the function $G \longmapsto \sum_{v_0 \in V_0}|\partial B_G(v_0, 3)|$ (where the multigraph $G$, expressed in the configuration model, is regarded as a matching). Since one may assume $V_0$ 2-independent because of the $\mathbf{1}_{V_0\textrm{ 2-id}}$ in the expectation, it is easy to see that the function varies by at most 8 when perfoming a switch on the matching. It follows
	\begin{align*}
	& \mathbf{P}\left[\left|\sum_{v_0 \in V_0}\left|\partial B_{G_0}(v_0, 3)\right| - \mathbf{E}_G\left[\sum_{v_0 \in V_0}\left|\partial B_G(v_0, 3)\right|\,\bigg|\,\mathcal{F}_{n_0d}\right]\right| > \varepsilon n\,\bigg|\,V_0\textrm{ 2-id}\right]\\
	& \leq 2\exp\left(-\frac{\varepsilon^2n^2}{128(n - n_0)d}\right),
	\end{align*}
	hence
	\begin{align*}
	\mathbf{E}\left[\mathbf{1}_{V_0\textrm{ 2-id}}\mathbf{1}_{V_0\textrm{ is not good}}\right] & \leq 2\exp\left(-\frac{\varepsilon^2n^2}{128(n - n_0)d}\right)\mathbf{E}\left[\mathbf{1}_{V_0\textrm{ 2-id}}\right],
	\end{align*}
	hence, by summing over $V_0$,
	\begin{align*}
	\sum_{\substack{V_0 \subset V\\|V_0| = n_0}}\mathbf{E}\left[\mathbf{1}_{V_0\textrm{ 2-id}}\mathbf{1}_{V_0\textrm{ is not good}}\right] & \leq 2\exp\left(-\frac{\varepsilon^2n^2}{128(n - n_0)d}\right)\sum_{\substack{V_0 \subset V\\|V_0| = n_0}}\mathbf{E}\left[\mathbf{1}_{V_0\textrm{ 2-id}}\right]
	\end{align*}
	Now, recalling the definition of a 2-independent-set-typical graph (definition \ref{def:sparse_set_typical_graph}),
	\begin{align*}
	& \mathbf{P}\left[\sum_{\substack{V_0 \subset V\\|V_0| = n_0}}\mathbf{1}_{V_0\textrm{ 2-id}}\mathbf{1}_{V_0 \textrm{ is not good}} > \exp\left(-\frac{\varepsilon^2n^2}{512(n -  n_0)d}\right)\sum_{\substack{V_0 \subset V\\|V_0| = n_0}}\mathbf{1}_{V_0\textrm{ 2-id}},\right.\\
	& \hspace{0.1\textwidth}\,G\textrm{ 2-independent-set typical}\Bigg]\\
	& \leq \mathbf{P}\left[\sum_{\substack{V_0 \subset V\\|V_0| = n_0}}\mathbf{1}_{V_0\textrm{ 2-id}}\mathbf{1}_{V_0\textrm{ is not good}} > \exp\left(-\frac{\varepsilon^2n^2}{256(n - n_0)d}\right)\mathbf{E}\left[\sum_{\substack{V_0 \subset V\\|V_0| = n_0}}\mathbf{1}_{V_0\textrm{ 2-id}}\right],\right.\\
	& \hspace*{40px}G\textrm{ 2-independent-set typical}\Bigg] \qquad \textrm{(2-independent-set-typicality)}\\
	& \leq \mathbf{P}\left[\sum_{\substack{V_0 \subset V\\|V_0| = n_0}}\mathbf{1}_{V_0\textrm{ 2-id}}\mathbf{1}_{V_0\textrm{ is not good}} > \exp\left(-\frac{\varepsilon^2n^2}{256(n - n_0)d}\right)\mathbf{E}\left[\sum_{\substack{V_0 \subset V\\|V_0| = n_0}}\mathbf{1}_{V_0\textrm{ 2-id}}\right]\right]\\
	& \leq 2\exp\left(\frac{\varepsilon^2n^2}{256(n - n_0)d}\right)\frac{\mathbf{E}\left[\sum_{\substack{V_0 \subset V\\|V_0| = n_0}}\mathbf{1}_{V_0\textrm{ 2-id}}\mathbf{1}_{V_0\textrm{ is not good}}\right]}{\mathbf{E}\left[\sum_{\substack{V_0 \subset V\\|V_0| = n_0}}\mathbf{1}_{V_0\textrm{ 2-id}}\right]} \qquad \textrm{(Markov inequality)}\\
	& \leq 2\exp\left(-\frac{\varepsilon^2n^2}{256(n - n_0)d}\right)
	\end{align*}
	Plugging-in conjecture \ref{conj:independent_set_typical_graph_frequent} on the frequency of 2-independent-set-typical graphs, this implies
	\begin{align*}
	& \mathbf{P}\left[\sum_{\substack{V_0 \subset V\\|V_0| = n_0}}\mathbf{1}_{V_0\textrm{ 2-id}}\mathbf{1}_{V_0 \textrm{ is not good}} > \exp\left(-\frac{\varepsilon^2n^2}{512(n -  n_0)d}\right)\sum_{\substack{V_0 \subset V\\|V_0| = n_0}}\mathbf{1}_{V_0\textrm{ 2-id}}\right.\\
	& \left.\hspace{0.1\textwidth}\,\bigg|\,G\textrm{ 2-independent-set-typical}\right]\\
	& \leq \mathcal{O}(1)\exp\left(-\frac{\varepsilon^2n^2}{256(n - n_0)d}\right)
	\end{align*}
	In other words, supposing the conjecture to be true, for all but an exponentially small fraction of 2-independent-set-typical graphs, all but an exponentially small fraction of 2-independent sets are ``good", meaning
	\begin{align*}
	\left|\sum_{v_0 \in V_0}\left|\partial B_{G_0}(v_0, 3)\right| - \mathbf{E}_G\left[\sum_{v_0 \in V_0}\left|\partial B_G(v_0, 3)\right|\,\bigg|\,V_0\textrm{ 2-id}\right]\right| & \leq \varepsilon n
	\end{align*}
	for all but exceptional 2-independent-set-typical $G_0$ and all but exceptional 2-independent sets $V_0$ of $G_0$. By invoking proposition \ref{prop:sparse_set_distance_3_4_5_successors_estimate}, the expectation above evaluates to $d(d - 1)^2\frac{n_0^2}{n - 2n_0} + \mathcal{O}(1)$.
	
	Similarly we demonstrate that for all but a fraction $\mathcal{O}(1)\exp\left(-\frac{\varepsilon^2n^2}{1296(n - n_0)d}\right)$ of 2-independent-set-typical graphs $G_0$ and all but a fraction $\exp\left(-\frac{\varepsilon^2n^2}{5192(n - n_0)d}\right)$ of the 2-independent sets of $G_0$,
	\begin{align*}
	\left|\sum_{v_0 \in V_0}\left|\partial B_{G_0}(v_0, 4)\right| - \left(d(d - 1)^3\frac{n_0^2(n - 3n_0)}{(n - 2n_0)^2} + \mathcal{O}(1)\right)\right| & \leq \varepsilon n
	\end{align*}
	and that for all but a fraction $\mathcal{O}(1)\exp\left(-\frac{\varepsilon^2n^2}{5184(n - n_0)d}\right)$ of 2-independent-set-typical graph $G_0$ and all but a fraction $\exp\left(-\frac{\varepsilon^2n^2}{10368(n - n_0)d}\right)$ of the 2-independent sets of $G_0$,
	\begin{align*}
	\left|\sum_{v_0 \in V_0}\left|\partial B_{G_0}(v_0, 5)\right| - \left(d(d - 1)^4\frac{n_0^2(n - 3n_0)^2}{(n - 2n_0)^3} + \mathcal{O}(1)\right)\right| & \leq \varepsilon n.
	\end{align*}
	The very last bound on $\left|\sum_{v_0 \in V_0}\mathbf{1}_{B(v_0, 5)\textrm{ is a tree}} - n_0\right|$ is established similarly.
\end{proof}
\section{Performance of local updates in infinite size limit}
\label{sec:local_improvement_infinite_size}
In this section, we consider the performance of local update algorithm \ref{alg:local_maxcut_improvement}, introduced in section \ref{subsubsec:improvement_local_updates}, when applied to a large graph.

\begin{figure}[!htbp]
    \centering
    \includegraphics[width=0.2\textwidth]{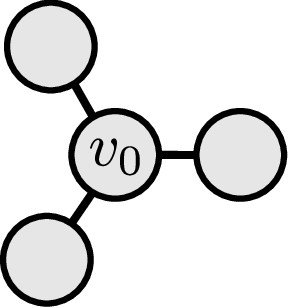}
    \caption{1-neighbourhood of $v_0$}
    \label{fig:1_neighbourhood}
\end{figure}

\begin{figure}[!htbp]
    \centering
    \includegraphics[width=0.8\textwidth]{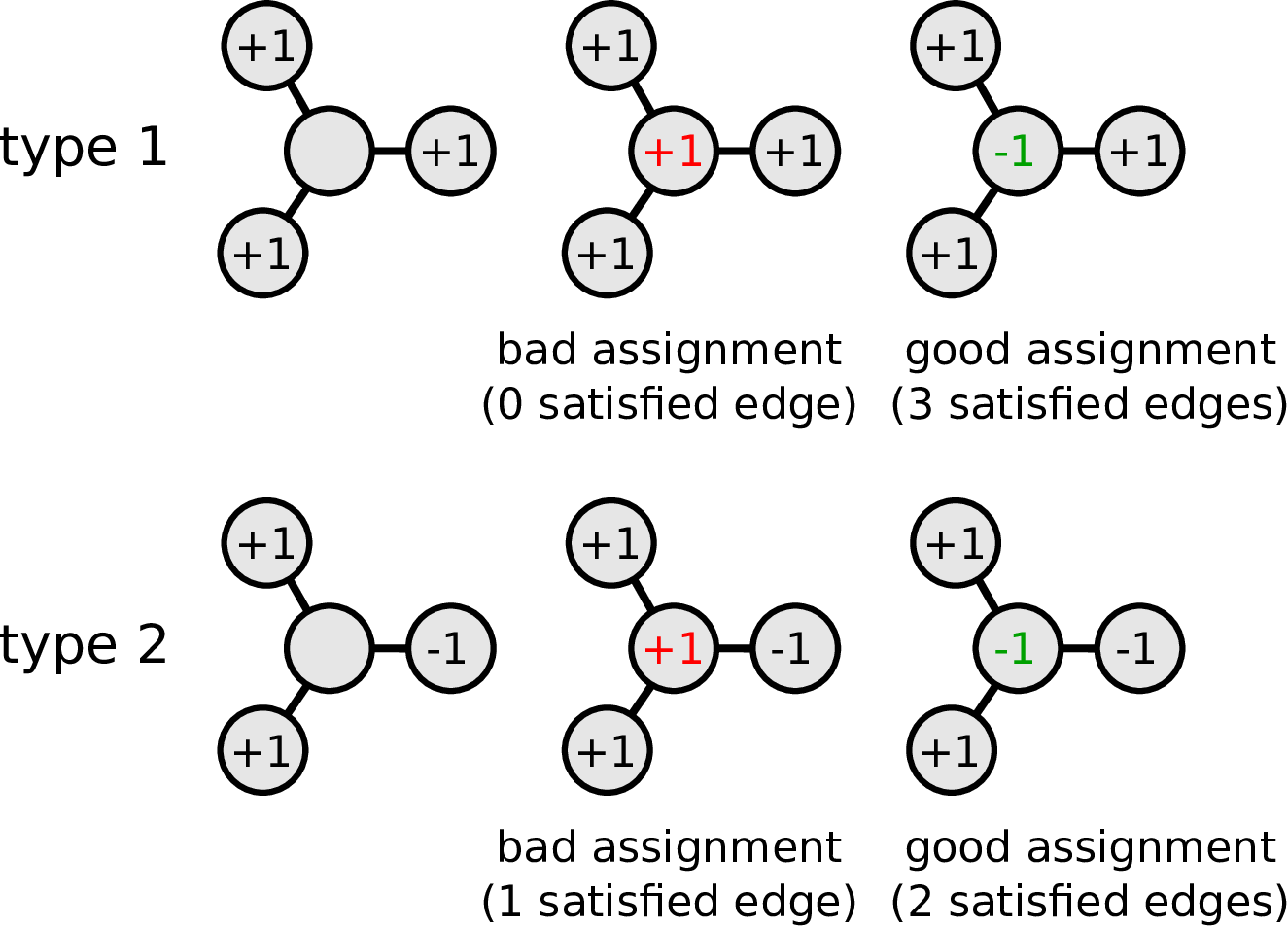}
    \caption{Configuration types for the neighbours of $v_0$.}
    \label{fig:tree_classes}
\end{figure}

Let us start by the simple case $d = 1$. In this case, the neighbourhood $B_G(v_0, d) = B_G(v_0, 1)$ will look like the star graph on figure \ref{fig:1_neighbourhood}.
The step of the for loop which looks at $v_0$ will then assign $v_0$ the value that maximizes that maximizes the number of satisfied edges in this star graph given the values of the neighbours of $v_0$; in this case, this value is simply the minority of the neighbours of $v_0$. The possible cases for the values of the neighbours of $v_0$ (up to permutation of neighbours and $\mathbf{Z}_2$ symmetry) are represented on figure \ref{fig:tree_classes}, with the corresponding ``good" and ``bad" assignment for $v_0$. Therefore, for each occurrence of a badly assigned $v_0$ with type 1 (resp. type 2) neighbourhood, our algorithm improves the number of satisfied edges by 3 (resp. 1).
Given a $1$-independent set $V_0$, the expected improvement on a cut sampled from a QAOA is then (denoting by $\ket{\psi}$ the state prepared by the circuit and by $v^{(0)}, v^{(1)}, v^{(2)}$ the neighbours of $v$ in $G$):
\begin{align}
\label{eq:qaoa_local_improvement_full}
    \textrm{expected improvement} & = \bra{\psi}\sum_{v_0 \in V_0}\left[3.\frac{1 - Z_{v_0}}{2}\frac{1 - Z_{v_0^{(1)}}}{2}\frac{1 - Z_{v_0^{(2)}}}{2}\frac{1 - Z_{v_0^{(3)}}}{2}\right.\nonumber\\
    & + 3.\frac{1 + Z_{v_0}}{2}\frac{1 + Z_{v_0^{(1)}}}{2}\frac{1 + Z_{v_0^{(2)}}}{2}\frac{1 + Z_{v_0^{(3)}}}{2}\nonumber\\
    & + 1.\frac{1 - Z_{v_0}}{2}\frac{1 - Z_{v_0^{(1)}}}{2}\frac{1 - Z_{v_0^{(2)}}}{2}\frac{1 + Z_{v_0^{(3)}}}{2}\nonumber\\
    & + 1.\frac{1 - Z_{v_0}}{2}\frac{1 - Z_{v_0^{(1)}}}{2}\frac{1 + Z_{v_0^{(2)}}}{2}\frac{1 - Z_{v_0^{(3)}}}{2}\nonumber\\
    & + 1.\frac{1 - Z_{v_0}}{2}\frac{1 + Z_{v_0^{(1)}}}{2}\frac{1 - Z_{v_0^{(2)}}}{2}\frac{1 - Z_{v_0^{(3)}}}{2}\nonumber\\
    & + 1.\frac{1 + Z_{v_0}}{2}\frac{1 + Z_{v_0^{(1)}}}{2}\frac{1 + Z_{v_0^{(2)}}}{2}\frac{1 - Z_{v_0^{(3)}}}{2}\nonumber\\
    & + 1.\frac{1 + Z_{v_0}}{2}\frac{1 + Z_{v_0^{(1)}}}{2}\frac{1 - Z_{v_0^{(2)}}}{2}\frac{1 + Z_{v_0^{(3)}}}{2}\nonumber\\
    & + 1.\frac{1 + Z_{v_0}}{2}\frac{1 - Z_{v_0^{(1)}}}{2}\frac{1 + Z_{v_0^{(2)}}}{2}\frac{1 + Z_{v_0^{(3)}}}{2}\Bigg]\ket{\psi}
\end{align}
For a fixed $v_0$, assuming $B_G(v_0, 2)$ is a tree (which holds if $v_0$ is any vertex except $o(n)$ of them), the expectation of the operator above simplifies considerably by symmetry, giving
\begin{align}
    \textrm{expected improvement from } v_0 & = \frac{3}{4}\left(1 + 2\braket{\psi|Z_{v_0}Z_{v_0^{(1)}}|\psi} + \braket{\psi|Z_{v_0^{(1)}}Z_{v_0^{(2)}}|\psi}\right)
\end{align}
This expression does not depend on $v_0$ (provided the neighbourhood assumption is satisfied) but only on the $\beta, \gamma$ parameters of the QAOA. For instance, in the following proposition, we explicitly compute the improvement for $p = 1$ QAOA.
\begin{prop}
    Let $\alpha > 0$ a fixed constant (for instance, one may choose $\alpha = 0.204$ by proposition \ref{prop:lower_bound_2_independent_set}). Let $G = (V, E)$ a $n$-vertices 3-regular graph. Then with high probability on $G$, for all 2-independent set $V_0$ of $G$ such that $|V_0| \geq \alpha n$, the expected improvement of algorithm \ref{alg:local_maxcut_improvement} on a cut sampled from the $p = 1$ QAOA applied to $G$ is
    \begin{align}
        & \frac{3}{4}\left(1 + 2\sin(2\beta)\cos^2\gamma\sin\gamma + \sin^2\beta\cos^4\gamma\sin^2\gamma\right)|V_0| + o(n).
    \end{align}
\end{prop}
In fact, given the improvement can be expressed as a sum of local operators as shown in equation \ref{eq:qaoa_local_improvement_full}, concentration arguments apply and the improvement is close to its expected value with high probability (with respect to the probability distribution of the bitstrings sampled from QAOA).

The calculation above can be generalized to $d > 1$ by observing that given a $(2d - 1)$-independent set $V_0$ of size $\geq \alpha n$, for all but $o(n)$ $v_0 \in V_0$, the neighbourhood $B_G(v_0, d)$ is a tree. Therefore, it suffices to consider all possible $\{-1, 1\}$ assignments (up to tree isomorphism) of the leaves of the depth-$d$ tree $B_G(v_0, d)$. For each of this assignment, one then computes the expected number of satisfied edges in the tree for the state sampled by the QAOA and the maximum cut on the tree conditioned on the values of the leave vertices. The calculations on the QAOA state only involve expectation values of local observables and is therefore classically tractable (through tree tensor network simulations).

We carried out this procedure for $p \in \{1, 2, 3, 4\}$ and depth parameters $d \in \{1, 2, 3\}$ in the limit $n \to \infty$. The results are expressed on figure \ref{fig:local_update_improvement_infinite_size} as the increase of the fraction of satisfied edges. To evaluate the latter, we need an estimate for the size of the $d$-independent set $V_0$ as a function of $n$. We use both a ``pessimistic" estimate (corresponding to the lower bound established in \cite{beis_duckworth_zito_2007}) and an ``optimistic" estimate (upper bound from same reference), recapitulated in table. We used the optimal QAOA parameters reported in \cite{1909.02559}.
\begin{table}[!htbp]
    \centering
    \begin{subfigure}{0.48\textwidth}
        \begin{tabular}{c|c|c|c|c}
        \backslashbox[25mm]{$d$}{$p$} & 1 & 2 & 3 & 4\\
            \hline
            1 & 0.06584 & 0.04776 & 0.02788 & 0.01440\\
            \hline
            2 & 0.06398 & 0.04336 & 0.02586 & 0.01532\\
            \hline
            3 & 0.04269 & 0.02982 & 0.01839 & 0.01190
        \end{tabular}
        \caption{Lower bound}
    \end{subfigure}\\
    \begin{subfigure}{0.48\textwidth}
        \begin{tabular}{c|c|c|c|c}
    \backslashbox[25mm]{$d$}{$p$} & 1 & 2 & 3 & 4\\
        \hline
        1 & 0.06928 & 0.05025 & 0.02933 & 0.01515\\
        \hline
        2 & 0.09851 & 0.06676 & 0.03981 & 0.02359\\
        \hline
        3 & 0.09394 & 0.06561 & 0.04049 & 0.02619
    \end{tabular}
        \caption{Upper bound}
    \end{subfigure}
    \caption{Improvement in fraction of satisfied edges}
    \label{fig:local_update_improvement_infinite_size}
\end{table}

\begin{table}[!htbp]
	\centering{
		\begin{tabular}{c|c|c}
			$d$ & $\frac{|V_0|}{n}$\textbf{, lower bound} & $\frac{|V_0|}{n}$\textbf{, upper bound}\\
			\hline
			1 & 0.4328 & 0.4554\\
			\hline
			2 & 0.090322 & 0.139057\\
			\hline
			3 & 0.022635 & 0.049812
	\end{tabular}}
	\caption{Lower and upper bounds for independent set size}
	\label{table:local_improvement_is_sizes}
\end{table}
\section{Generalizing Bravi et al.'s argument}
\label{sec:bravyi_et_al_generalization}

\begin{figure}[!htbp]
	\centering
	\includegraphics[width=0.2\textwidth]{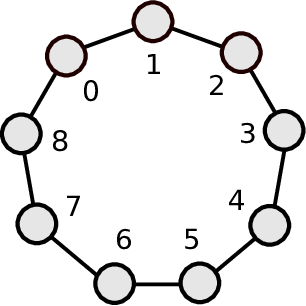}
	\caption{A 9-vertices ring graph}
	\label{fig:ring_graph}
\end{figure}

In a recent work \cite{1910.08980}, Bravyi et al. derived upper bounds on the performance on the MaxCut problem of a family of variational quantum algorithms generalizing QAOA. In this section, we give an alternative proof of \cite[theorem 2]{1910.08980} (proposition \ref{prop:bravyi_upper_bound_ring}); the statement is also more general. We then apply this result to two types of graphs: grid graphs and random regular graphs, extending the results of \cite{1910.08980} on ring graphs (figure \ref{fig:ring_graph}).

The following lemma will be our starting point.
\begin{lem}
	\label{lemma:bounding_ising_paths}
	Let $G = (V, E)$ a graph and let $\mathcal{P} = \left\{P_1, \ldots, P_p\right\}$ a set of paths included in $G$; given a path $P \in \mathcal{P}$, denote by $l(P)$ the length of $P$ (number of edges) and $P^1, \ldots, P^{l(P) + 1}$ the vertices of $P$. Let $\left(\sigma_e\right)_{e \in E} \in \{-1, 1\}^E$ a series of signs. Then
	\begin{align}
	\label{eq:bounding_ising_paths}
		|\mathcal{P}| - \sum_{P \in \mathcal{P}}Z_{P^1}Z_{P^{l(P) + 1}} & \preceq \sum_{e = \{e_0, e_1\} \in E}\left(1 + \sigma_eZ_{e_0}Z_{e_1}\right)\left|\left\{P \in \mathcal{P}:e \in P\right\}\right|.
	\end{align}
	\begin{proof}
    	It suffices to prove that for any path $P \in \mathcal{P}$
    	\begin{align}
    	\label{eq:path_lemma_main_inequality}
    	1 - Z_{P^1}Z_{P^{l(P) + 1}} & \preceq \sum_{1 \leq j \leq l(P)}\left(1 + \sigma_{\{P^j, P^{j + 1}\}}Z_{P^j}Z_{P^{j + 1}}\right).
    	\end{align}
    	The result will follow by summation over $P \in \mathcal{P}$. We first show:
    	\begin{align}
    	\label{eq:path_lemma_main_equality}
    	& 1 + (-1)^{l(P) + 1}\left(\prod_{e \in P}\sigma_e\right)Z_{P^1}Z_{P^{l(P) + 1}}\nonumber\\
    	& = \left(1 + \sigma_{\{P^1, P^2\}}Z_{P^1}Z_{P^2}\right) + \sum_{2 \leq j \leq l(P)}(-1)^{j - 1}\left(\prod_{1 \leq k < j}\sigma_{\{P^{k - 1}, P^k\}}\right)\nonumber\\
    	& \hspace{0.5\textwidth} Z_{P^1}Z_{P^j}\left(1 + \sigma_{\{P^j, P^{j + 1}\}}Z_{P^j}Z_{P^{j + 1}}\right).
    	\end{align}
    	To show this, we can prove by recursion on $n \in [2, l(P)]$ that
    	\begin{align*}
    	& 1 + (-1)^{n + 1}\left(\prod_{1 \leq  j \leq n}\sigma_{\{P^j, P^{j + 1}\}}\right)Z_{P^1}Z_{P^{n + 1}}\\
    	& = \left(1 + \sigma_{\{P^1, P^2\}}Z_{P^1}Z_{P^2}\right) + \sum_{2 \leq j \leq n}(-1)^{j - 1}\left(\prod_{1 \leq k < j}\sigma_{\{P^{k - 1}, P^k\}}\right)\\
    	& \hspace{0.6\textwidth} Z_{P^1}Z_{P^j}\left(1 + \sigma_{\{P^j, P^{j + 1}\}}Z_{P^j}Z_{P^{j + 1}}\right)
    	\end{align*}
    	For $n = 2$,
    	\begin{align*}
    	& 1 + (-1)^{n + 1}\left(\prod_{1 \leq j \leq n}\sigma_{\{P^j, P^{j + 1}\}}\right)Z_{P^1}Z_{P^{n + 1}}\\
    	& = 1 - \sigma_{\{P^1, P^2\}}\sigma_{\{P^2, P^3\}}Z_{P^1}Z_{P^3}\\
    	& = \left(1 + \sigma_{\{P^1, P^2\}}Z_{P^1}Z_{P^2}\right) + \left(-\sigma_{\{P^1, P^2\}}\right)Z_{P^1}Z_{P^2}\left(1 + \sigma_{\{P^2, P^3\}}Z_{P^2}Z_{P^3}\right)\\
    	& 
    	\end{align*}
    	which is the desired relation. Now, suppose the relation holds for $n \in [2, l(P))$. Then
    	\begin{align*}
    	& 1 + (-1)^{n + 2}\left(\prod_{1 \leq j \leq n + 1}\sigma_{\{P^j, P^{j + 1}\}}\right)Z_{P^1}Z_{P^{n + 2}}\\
    	& = 1 - \sigma_{\{P^{n + 1}, P^{n + 2}\}}Z_{P^{n + 1}}Z_{P^{n + 2}}(-1)^{n + 1}\left(\prod_{1 \leq j \leq n}\sigma_{\{P^j, P^{j + 1}\}}\right)Z_{P^1}Z_{P^{n + 1}}\\
    	& = 1 - \left(1 + \sigma_{\{P^{n + 1}, P^{n + 2}\}}Z_{P^{n + 1}}Z_{P^{n + 2}} - 1\right)(-1)^{n + 1}\left(\prod_{1 \leq j \leq n}\sigma_{\{P^j, P^{j + 1}\}}\right)Z_{P^1}Z_{P^{n + 1}}\\
    	& = 1 + (-1)^{n + 1}\left(\prod_{1 \leq j \leq n}\sigma_{\{P^j, P^{j + 1}\}}\right)Z_{P^1}Z_{P^{n + 1}}\\
    	& \hspace*{0.05\textwidth} + (-1)^n\left(\prod_{1 \leq j \leq n}\sigma_{\{P^j, P^{j + 1}\}}\right)Z_{P^1}Z_{P^{n + 1}}\left(1 + \sigma_{\{P^{n + 1}, P^{n + 2}\}}Z_{P^{n + 1}}Z_{P^{n + 2}}\right)\\
    	& = \left(1 + \sigma_{\{P^1, P^2\}}Z_{P^1}Z_{P^2}\right) + \sum_{2 \leq j \leq n}(-1)^{j - 1}\left(\prod_{1 \leq k < j}\sigma_{\{P^{k - 1}, P^k\}}\right)\\
    	& \hspace{0.55\textwidth} Z_{P^1}Z_{P^j}\left(1 + \sigma_{\{P^j, P^{j + 1}\}}Z_{P^j}Z_{P^{j + 1}}\right)\\
    	& \hspace*{0.05\textwidth} + (-1)^n\left(\prod_{1 \leq j \leq n}\sigma_{\{P^j, P^{j + 1}\}}\right)Z_{P^1}Z_{P^{n + 1}}\left(1 + \sigma_{\{P^{n + 1}, P^{n + 2}\}}Z_{P^{n + 1}}Z_{P^{n + 2}}\right)\\
    	& = \left(1 + \sigma_{\{P^1, P^2\}}Z_{P^1}Z_{P^2}\right) + \sum_{2 \leq j \leq n + 1}(-1)^{j - 1}\left(\prod_{1 \leq k < j}\sigma_{\{P^{k - 1}, P^k\}}\right)\\
    	& \hspace{0.6\textwidth} Z_{P^1}Z_{P^j}\left(1 + \sigma_{\{P^j, P^{j + 1}\}}Z_{P^j}Z_{P^{j + 1}}\right).
    	\end{align*}
    	This proves equation \ref{eq:path_lemma_main_equality}. Inequality \ref{eq:path_lemma_main_inequality} results from upper-bounding (in the operator sense) the r.h.s. of equation \ref{eq:path_lemma_main_equality}:
    	\begin{align*}
    	& \sum_{2 \leq j \leq l(P)}(-1)^{j - 1}\left(\prod_{1 \leq k < j}\sigma_{\{P^{k - 1}, P^k\}}\right)Z_{P^1}Z_{P^j}\left(1 + \sigma_{\{P^j, P^{j + 1}\}}Z_{P^j}Z_{P^{j + 1}}\right)\\
    	& = \sum_{2 \leq j \leq l(P)}\left[1 + (-1)^{j - 1}\left(\prod_{1 \leq k < j}\sigma_{\{P^{k - 1}, P^k\}}\right)Z_{P^1}Z_{P^j} - 1\right]\left(1 + \sigma_{\{P^j, P^{j + 1}\}}Z_{P^j}Z_{P^{j + 1}}\right)\\
    	& = \sum_{2 \leq j \leq l(P)}\left(1 + \sigma_{\{P^j, P^{j + 1}\}}Z_{P^j}Z_{P^{j + 1}}\right)\\
    	& \hspace*{0.05\textwidth} - \sum_{2 \leq j \leq l(P)}\left[1 + (-1)^j\left(\prod_{1 \leq k < j}\sigma_{\{P^{k - 1}, P^k\}}\right)Z_{P^1}Z_{P^j}\right]\left(1 + \sigma_{\{P^j, P^{j + 1}\}}Z_{P^j}Z_{P^{j + 1}}\right)\\
    	& \leq \sum_{2 \leq j \leq l(P)}\left(1 + \sigma_{\{P^j, P^{j + 1}\}}Z_{P^j}Z_{P^{j + 1}}\right).
    	\end{align*}
    	To obtain the last inequality, we used that the product of commuting positive operators is positive.
    \end{proof}
\end{lem}

This result may straightaway be applied to generalize Bravyi et al.'s upper bound \cite[Theorem 2]{1910.08980}. For that purpose, we start by recalling some definitions from \cite{1910.08980} (see section \ref{subsec:random_graphs} for graph-related notations):
\begin{defn}[$\mathbf{Z}_2$-symmetric states and circuits]
    A state $\ket{\psi} \in \mathbf{C}^{2^n}$ on $n$ qubits is called $\mathbf{Z}_2$-symmetric if $X^{\otimes n}\ket{\psi} = \ket{\psi}$. A quantum circuit $U \in \mathbf{U}(2^n)$ acting on $n$ qubits is called $\mathbf{Z}_2$-symmetric if $X^{\otimes n}UX^{\otimes n} = U$. A state will be called ``prepared by a $\mathbf{Z}_2$-symmetric circuit" if it can be obtained by applying a $\mathbf{Z}_2$-symmetric circuit to a $\mathbf{Z}_2$-symmetric product state.
\end{defn}

\begin{defn}
Let $G = (V, E)$ a graph. A circuit $U$ acting on qubits labelled by the vertices of $V$ is said to have range $R$ if for all $v \in V$, all single-qubit observable $\mathcal{O}_v$ supported on $v$, $U^{\dagger}\mathcal{O}_vU$ is supported on $B_G(v, R)$ (the $R$-neighbourhood of $v$ in $G$).
\end{defn}

We are now ready to state and prove our generalization of {\cite[Theorem 2]{1910.08980}}. The main difference with the original theorem is that the new result not only gives a lower bound on the expected energy of a ferromagnetic or antiferromagnetic Ising Hamiltonian, but applies to Ising Hamiltonians with arbitrary couplings in $\{-1, 1\}$. Incidentally, this removes the need for an extra step to relate the antiferromagnetic Ising Hamiltonian to the ferromagnetic one, which is required by the original approach of \cite{1910.08980} but does not generalize well beyond ring graphs.
\begin{prop}[Generalization of {\cite[Theorem 2]{1910.08980}}]
	\label{prop:bravyi_upper_bound_ring}
	Let $R \geq 1$ an integer, $k \geq 2$ an even integer, and $G$ a ring graph with $n = (2R + 1)k$ vertices. Then a state $\ket{\psi}$ prepared by a $\mathbf{Z}_2$-symmetric range-$R$ quantum circuit satisfies for all choices of signs $\left(\sigma_e\right)_{e \in E} \in \{-1, 1\}^E$:
	\begin{align}
		\braket{\psi|\sum_{e = \{e_0, e_1\} \in E}1 + \sigma_eZ_{e_0}Z_{e_1}|\psi} & \geq \frac{n}{2R + 1}.
	\end{align}
	In particular, for the MaxCut Hamiltonian $H_{\textnormal{MaxCut}} = \sum_{0 \leq i < k(2R + 1)}\frac{1 - Z_iZ_{i + 1}}{2}$ associated to the ring graph,
	\begin{align}
	\label{eq:maxcut_ring_ub}
		\braket{\psi|H_{\textnormal{MaxCut}}|\psi} & \leq n - \frac{n}{2(2R + 1)}.
	\end{align}
	\begin{proof}
		Following the notation of lemma \ref{lemma:bounding_ising_paths}, consider the $k = \frac{n}{2R + 1}$ edge-disjoint length-$(2R + 1)$ paths $P_0, \ldots, P_{k - 1}$ defined by $P_i^j := (2R + 1)i + j$ ($0 \leq j \leq 2R + 1$). Applying lemma \ref{lemma:bounding_ising_paths} then gives
		\begin{align*}
			\braket{\psi|\frac{n}{2R + 1} - \sum_{0 \leq i < k}Z_{(2R + 1)i}Z_{(2R + 1)i + 2R + 1}|\psi} & \leq \sum_{e = \{e_0, e_1\} \in E}\braket{\psi|1 + \sigma_eZ_{e_0}Z_{e_1}|\psi}
		\end{align*}
		Now, by the range-$R$ assumptions, $\braket{\psi|Z_{(2R + 1)i}Z_{(2R + 1)i + 2R + 1}|\psi} = 0$, giving the result.
		Finally, to upper bound the expectation of $H_{\textnormal{MaxCut}}$, it suffices to lower-bound the expectation of $\sum_{e = \{e_0, e_1\} \in E}\frac{1 + Z_{e_0}Z_{e_1}}{2}$ using the result just established.
	\end{proof}
\end{prop}

\begin{figure}[!tbp]
	\centering
	\includegraphics[width=0.3\textwidth]{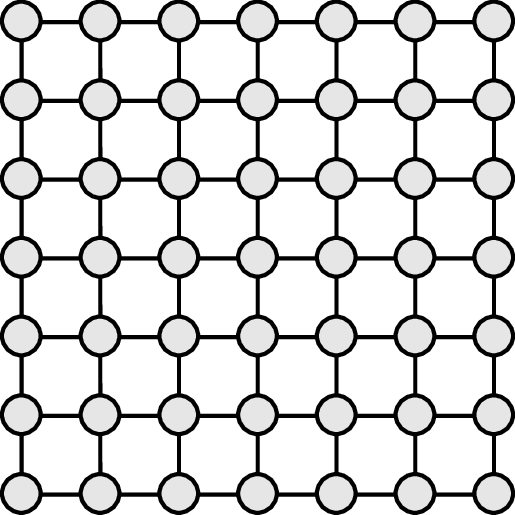}
	\caption{A 2D grid graph}
	\label{fig:2d_grid_graph}
\end{figure}

We now generalize this upper bound to grid graphs. In general, a grid graph is a graph whose vertices and edges are defined by a lattice in $\mathbf{R}^d$. Here, we will restrict to a square lattice. An example of such grid graph in $d = 2$ dimensions is given on figure \ref{fig:2d_grid_graph}.
The following proposition applies to a two-dimensional grid graph:
\begin{prop}
	Let $R \geq 1$ an integer and $G = (V, E)$ a 2D grid graph (with periodic boundary conditions) of dimensions $(2R + 1)k \times (2R + 1)k$, $k \geq 1$ integer. Then a state $\ket{\psi}$ prepared by a $\mathbf{Z}_2$-symmetric range-$R$ quantum circuit satisfies for all choices of signs $\left(\sigma_e\right)_{e \in E} \in \{-1, 1\}^E$:
	\begin{align}
		\braket{\psi|\sum_{e = \{e_0, e_1\} \in E}\left(1 + \sigma_eZ_{e_0}Z_{e_1}\right)|\psi} & \geq \frac{2n}{2R + 1}
	\end{align}
	where $n = (2R + 1)^2k^2$ is the number of vertices in $G$. In particular, for the MaxCut Hamiltonian\footnote{Note that the grid graph described here has $2n$ edges and that MaxCut is completely satisfiable for this graph.} $H_{\textrm{MaxCut}}$ of $G$, this implies
	\begin{align}
	\label{eq:maxcut_2d_grid_ub}
		\braket{\psi|H_{\textnormal{MaxCut}}|\psi} & \leq 2n - \frac{n}{2R + 1}.
	\end{align}
	\begin{proof}
		One may pack exactly $\underbrace{k}_{\substack{\textrm{horizontal paths}\\\textrm{per row}}} \times \underbrace{k(2R + 1)}_{\textrm{rows}} + \underbrace{k}_{\substack{\textrm{vertical paths}\\\textrm{per column}}} \times \underbrace{k(2R + 1)}_{\textrm{columns}} = 2k^2(2R + 1) = \frac{2n}{2R + 1}$ horizonzal and vertical edge-disjoint paths of length $2R + 1$ in $G$; an example is given on figure \ref{fig:grid_graph_paths} for $R = 1$ (length $3$ paths).
		\begin{figure}[!htbp]
			\centering
			\includegraphics[width=0.7\textwidth]{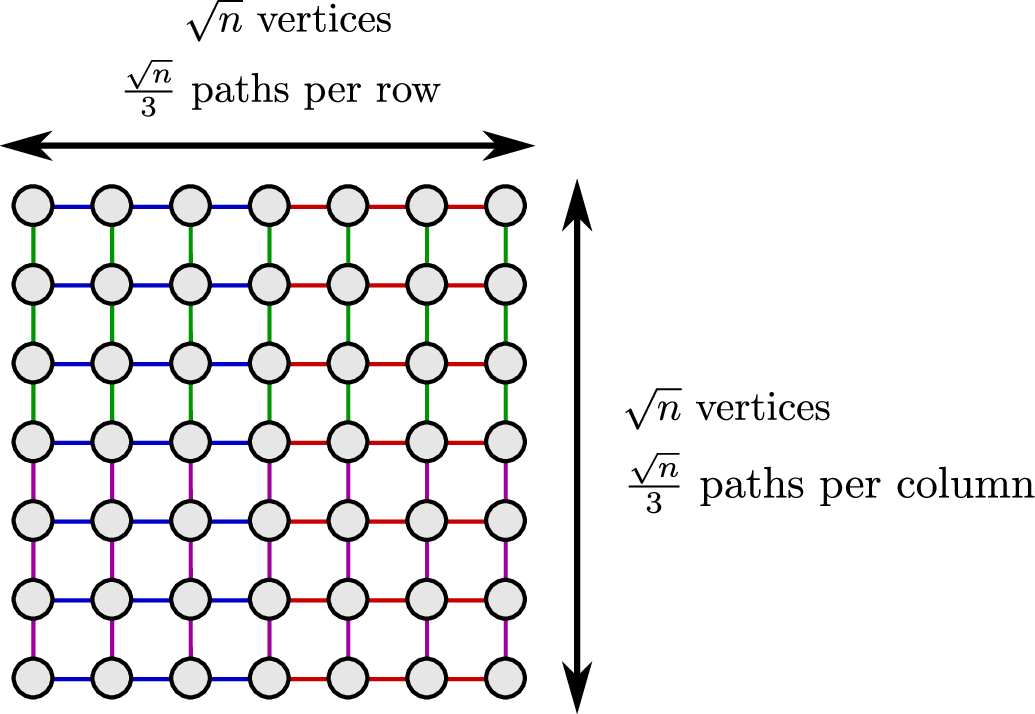}
			\caption{Packing length-3 paths (colored either red, green, blue or purple) in a 2D grid graph}
			\label{fig:grid_graph_paths}
		\end{figure}
		By the range-$R$ property of the circuit, the vertices at the extremities of each path are uncorrelated. Therefore, invoking lemma \ref{lemma:bounding_ising_paths} and taking the expectation of the operator inequality there on state $\ket{\psi}$ gives
		\begin{align*}
			\frac{2n}{2R + 1} & \leq \sum_{e = \{e_0, e_1\} \in E}\braket{\psi|1 + \sigma_eZ_{e_0}Z_{e_1}|\psi}.
		\end{align*}
	\end{proof}
\end{prop}

Note that the lower-bound $\frac{2n}{2R + 1}$ for a two-dimensional grid graph is twice the one for a ring graph (which is a one-dimensional grid graph). In fact, by reviewing the last proof, it is not hard to see that this extra factor comes from the number of dimensions. The previous proposition then generalizes immediately:

\begin{prop}
	Let $R \geq 1$ an integer and $G = (V, E)$ be a $d$-dimensional grid graph (with periodic boundary conditions) of dimensions $(2R + 1)k \times \ldots \times (2R + 1)k$, $k \geq 1$ integer. Then a state $\ket{\psi}$ prepared by a $\mathbf{Z}_2$-symmetric range-$R$ quantum circuit satisfies for all choices of signs $\left(\sigma_e\right)_{e \in E} \in \{-1, 1\}^E$:
	\begin{align}
	\braket{\psi|\sum_{e = \{e_0, e_1\} \in E}\left(1 + \sigma_eZ_{e_0}Z_{e_1}\right)|\psi} & \geq \frac{dn}{2R + 1}
	\end{align}
	where $n = (2R + 1)^2k^2$ is the number of vertices in $G$. In particular, for the MaxCut Hamiltonian\footnote{Note that the grid graph described here has $2n$ edges and that MaxCut is completely satisfiable for this graph.} $H_{\textnormal{MaxCut}}$ of $G$, this implies
	\begin{align}
	\label{eq:maxcut_dd_grid_ub}
		\braket{\psi|H_{\textnormal{MaxCut}}|\psi} & \leq dn - \frac{dn}{2R + 1}.
	\end{align}
\end{prop}

One may now wonder to what extent bounds \ref{eq:maxcut_ring_ub}, \ref{eq:maxcut_2d_grid_ub}, \ref{eq:maxcut_dd_grid_ub} are tight. For \ref{eq:maxcut_ring_ub}, which applies to a ring graph, \cite[lemma B.2]{1910.08980} shows a converse bound. A similar result can proved for two-dimensional graph:
\begin{prop}
	\label{prop:bravyi_lower_bound_2d_grid_graph}
	For a 2D grid graph $G$ with periodic boundary conditions of size $k(2R^2 + 2R + 1) \times k(2R^2 + 2R + 1)$, with $k \geq 2$ an even integer, exists a $\mathbf{Z}_2$-symmetric range-$R$ circuit such that
	\begin{align}
		\braket{\psi|H_{\textrm{MaxCut}}|\psi} & = 2n - \frac{2R + 1}{2R^2 + 2R + 1}n.
	\end{align}
	\begin{proof}
	Let us fix $k, R$ in the rest of the proof. We follow the strategy of proof of \cite[Lemma B.2]{1910.08980}, which consists to pack the graph with GHZ states ``cells" as illustrated on figure \ref{fig:grid_graph_packed_ghz_state}.
	\begin{figure}[!htbp]
		\centering
		\includegraphics[width=0.4\textwidth]{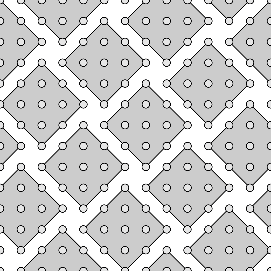}
		\caption{A $13 \times 13$ vertices grid graph (periodic boundary conditions) with packed GHZ states. Qubits forming a single GHZ state are delimited by a gray cell.}
		\label{fig:grid_graph_packed_ghz_state}
	\end{figure}
	For this packing to be exactly possible, each dimension of the lattice has to be a multiple of $2R^2 + 2R + 1$, hence the hypothesis of the proposition. A single cell comprises $2R^2 + 2R + 1$ qubits ---the case $R = 2$ is shown on figure \ref{fig:grid_graph_single_ghz_state}.
	\begin{figure}[!htbp]
		\centering
		\includegraphics[width = 0.4\textwidth]{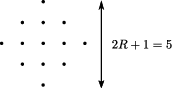}
		\caption{A GHZ state cell}
		\label{fig:grid_graph_single_ghz_state}
	\end{figure}
	Now, a GHZ state can be prepared in a cell starting with all qubits in state $\ket{0}$ and applying the circuit on figure \ref{fig:grid_graph_single_ghz_state_initial_circuit}.
	\begin{figure}
		\centering
		\includegraphics[width=\textwidth]{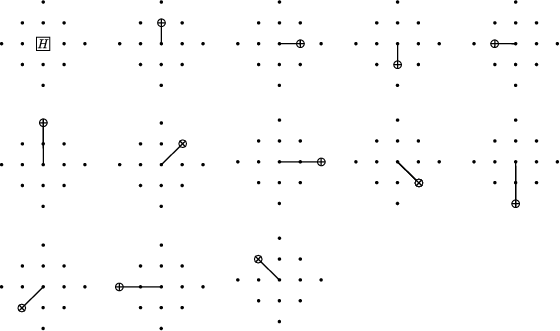}
		\caption{Circuit preparing a GHZ on the cell qubits.}
		\label{fig:grid_graph_single_ghz_state_initial_circuit}
	\end{figure}
	This circuit can be converted to a $\mathbf{Z}_2$-symmetric one (with range $R$) acting on the $\mathbf{Z}_2$-symmetric state $\ket{+}^{\otimes (2R^2 + 2R + 1)}$ (instead of $\ket{0}^{\otimes (2R^2 + 2R + 1)}$) using the same circuit identities as in the proof of \cite[Lemma B.1]{1910.08980}. The next step consists to apply to each GHZ cell a series of alternated bitflips as represented on figure \ref{subfig:alternated_bit_flip_pattern}. After performing these bit flips, every edge $\{i, j\}$ connecting vertices in the same GHZ cell will be satisfied ---meaning the quantum state $\ket{\psi}$ will satisfy $Z_iZ_j\ket{\psi} = -\ket{\psi}$ (illustration figure \ref{subfig:alternated_bit_flip_satisfied edges}).
	\begin{figure}
		\centering
		\begin{subfigure}{0.45\textwidth}
			\centering
			\includegraphics[width=0.6\textwidth]{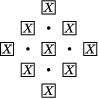}
			\caption{Alternated bit flip pattern}
			\label{subfig:alternated_bit_flip_pattern}
		\end{subfigure}
		\begin{subfigure}{0.45\textwidth}
			\centering
			\includegraphics[width=0.6\textwidth]{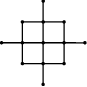}
			\caption{Resulting satisfied edges}
			\label{subfig:alternated_bit_flip_satisfied edges}
		\end{subfigure}
		\caption{Applying bit flips to GHZ cell}
	\end{figure}
	However, every edge $\{i, j\}$ connecting vertices not in the same GHZ cell will verify $\braket{\psi|Z_iZ_j|\psi} = 0$. For each GHZ cell, there are $4R^2$ edges belonging to the first category; since there are $\frac{n}{2R^2 + 2R + 1}$ GHZ cells, there are $\frac{4R^2}{2R^2 + 2R + 1}n$ edges belonging to the first category. There are then $2n - \frac{4R^2}{2R^2 + 2R + 1}n = \frac{4R + 2}{2R^2 + 2R + 1}n$ edges in the second category. This counting yields
	\begin{align*}
		\braket{\psi|H_{\textrm{MaxCut}}|\psi} & = \sum_{\textrm{edges }\{i, j\}}\braket{\psi|\frac{1 - Z_iZ_j}{2}|\psi}\\
		& = \frac{4R^2}{2R^2 + 2R + 1}n \times 1 + \frac{4R + 2}{2R^2 + 2R + 1}n \times \frac{1}{2}\\
		& = \frac{4R^2 + 2R + 1}{2R^2 + 2R + 1}n\\
		& = 2n - \frac{2R + 1}{2R^2 + 2R + 1}n.
	\end{align*}
\end{proof}
\end{prop}

We now apply lemma \ref{lemma:bounding_ising_paths} to random $d$-regular graphs. In previous cases: ring and grid graphs, the graphs were bipartite and the method bounded the cut achievable by the algorithm away from $|E|$, the number of edges in a maximum cut. This stands in contrast to random regular graph, as such a graph is, with high probability, not bipartite (see e.g. \cite{1503.03923} and remarks after proposition \ref{prop:bravyi_upper_bound_regular}). Unfortunately, Bravyi et al.'s method can only bound the achievable cut away from $|E|$ in this case; this means that the upper bound becomes trivial for large enough $R$, but may still be relevant for low $R$. The bound is precisely stated here:
\begin{prop}
	\label{prop:bravyi_upper_bound_regular}
	Let $d \geq 3$ and $G = (V, E)$ be a random $d$-regular graph. For any $\varepsilon > 0$, with probability $\geq 1 - \varepsilon$, for all choices of signs $\left(\sigma_e\right)_{e \in E} \in \{-1, 1\}^E$,
	\begin{align}
		(1 - \varepsilon)\frac{dn}{2R + 1} & \leq \sum_{e = \{e_0, e_1\} \in E}\braket{\psi|1 + \sigma_eZ_{e_0}Z_{e_1}|\psi}
	\end{align}
	holds for $n$ large enough for all state $\ket{\psi}$ prepared by a $\mathbf{Z}_2$-symmetric range-$R$ circuit. In particular, for the MaxCut Hamiltonian $H_{\textnormal{MaxCut}}$ associated to $G$,
	\begin{align}
		\label{eq:bravyi_bound_random_regular_graph}
		\braket{\psi|H_{\textnormal{MaxCut}}|\psi} & \leq \frac{nd}{2} - (1 - \varepsilon)\frac{nd}{2(2R + 1)}.
	\end{align}
	\begin{proof}
		Contrary to the proof for ring and grid graphs, we will choose here a set of paths $\mathcal{P}$ that overlap. Let us then consider the set of directed paths starting at any vertex of $G$ and having length $2R + 1$. (The directedness means that two paths consisting of the same edges but walked in opposite order are regarded as distinct.) Using the tree neighbourhood property of random regular graphs stated in proposition \ref{prop:tree_neighbourhood}, for large enough $n$, with probability $1 - \varepsilon$ at least $(1 - \varepsilon)n$ vertices in $G$ have trees as $(4R + 1)$-neighbourhoods. For each such vertex, there are $d(d - 1)^{2R}$ distinct paths starting from this vertex. Therefore, these vertices give rise to at least $\frac{1}{2}(1 - \varepsilon)nd(d - 1)^{2R}$ distinct paths, as illustrated on figure \ref{fig:regular_graph_paths_from_vertex}.
		\begin{figure}[!htbp]
			\centering
			\includegraphics[width=0.4\textwidth]{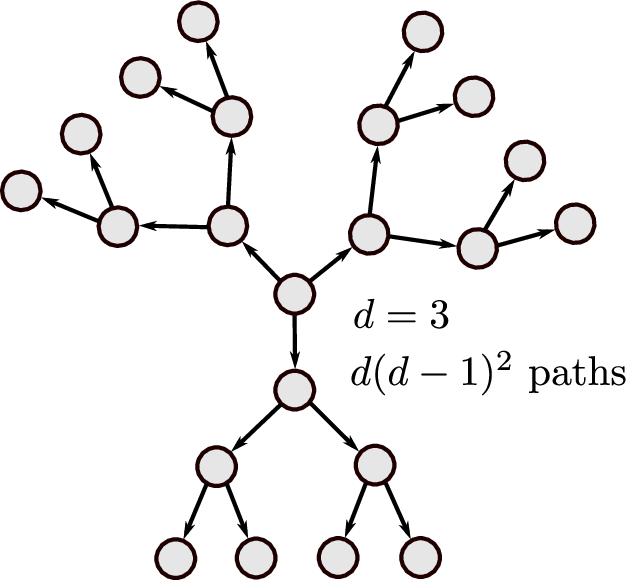}
			\caption{Directed paths arising from the central vertex for $d = 3$ and $R = 1$. There are 12 such paths, corresponding to the leaves of the tree.}
			\label{fig:regular_graph_paths_from_vertex}
		\end{figure}
		Each of these paths has their extremities $(2R + 1)$ apart in the graph ---otherwise, the $(4R + 1)$-neighbourhood of the vertex from which this path starts would not be a tree. This lower-bounds the left-hand-side of equation \ref{eq:bounding_ising_paths} from lemma \ref{lemma:bounding_ising_paths} by $\frac{1}{2}(1 - \varepsilon)nd(d - 1)^{2R}$. To upper-bound the right-hand side, it remains to upper-bound the number of paths an edge may appear in. Let then fix an edge $e \in E$. $e$ may then be the first edge of at most $(d - 1)^{2R}$ paths, the second edge of the same number of paths, and so on until the $(2R + 1)^{\textrm{th}}$ edge. This gives the upper-bound $(2R + 1)(d - 1)^{2R}\sum_{e = \{e_0, e_1\} \in E}\left(1 + \sigma_eZ_{e_0}Z_{e_1}\right)$. We then finally obtain:
		\begin{align*}
			(1 - \varepsilon)\frac{dn}{2R + 1} & \leq \sum_{e = \{e_0, e_1\}}\braket{\psi|1 + \sigma_eZ_{e_0}Z_{e_1}|\psi}.
		\end{align*}
	\end{proof}
\end{prop}

The bound \ref{eq:bravyi_bound_random_regular_graph} on MaxCut is unfortunately less satisfying than in the case of ring and grid graphs. Indeed, for these graphs, MaxCut was completely satisfiable and the right-hand side of the inequality was always smaller than the maximum cut. On the contrary, MaxCut is usually not completely satisfiable for a random regular graph. For instance, \cite{1503.03923} shows that with high probability, the MaxCut of a random $d$-regular graph $G = (V, E)$ includes $|E|\left(\frac{1}{2} + \frac{P_*}{\sqrt{d}}(1 + o_d(1))\right)$ edges, where $P_* \approx 0.7632\ldots$ is the Parisi constant, which is a constant factor below $|E|$. However, bound \ref{eq:bravyi_bound_random_regular_graph} merely guarantees $\braket{\psi|H_{\textnormal{MaxCut}}|\psi} \leq |E| - (1 - \varepsilon)\frac{|E|}{2R + 1}$ and therefore becomes trivial for large enough $R$.

In particular, it is hopeless to construct a circuit achieving the bound. However, one may still ask about the existence of a circuit similar to the ones constructed in \cite[lemma B.2]{1910.08980} or proposition \ref{prop:bravyi_lower_bound_2d_grid_graph} achieving some lower bound. Qualitatively, such a circuit is $\mathbf{Z}_2$-symmetric, has range $R$ and perfectly anticorrelates some edges $\{i, j\}$: $\braket{\psi|Z_iZ_j|\psi} = -1$ while perfectly decorrelating others: $\braket{\psi|Z_iZ_j|\psi} = 0$. We now demonstrate that for a random regular graph, a cut sampled from a circuit verifying these assumptions would necessarily have a poor approximation ratio. Therefore, maybe surprisingly, Bravyi et al.'s construction is inoperational in the case of random regular graphs. In fact, it will be interesting to start with a more general proposition:
\begin{prop}
\label{prop:upper_bound_sharp_maxcut_circuit}
	Let $G = (V, E)$ a graph with $n$ vertices. Assume that $\ket{\psi}$ is a $\mathbf{Z}_2$-symmetric state with range $R$ such that for all $e = \{e_0, e_1\}\in E$, $\braket{\psi|Z_{e_0}Z_{e_1}|\psi} \in \{0, -1\}$. Denote by $C_l$ the number of cycles in the graph with length $\leq l$.
	\begin{align}
	\braket{\psi|\sum_{e \in E}\frac{1 - Z_{e_0}Z_{e_1}}{2}|\psi} & \leq \frac{1}{2}\left(|E| + n + (4R + 1)C_{4R + 1}\right)
	\end{align}
	\begin{proof}
		Consider the edges $e$ such that $\braket{\psi|Z_{e_0}Z_{e_1}|\psi} = -1$; call these edges $E'$. Next, define $E''$ the set of edges in $E'$ which do not lie in a cycle of length $\leq 4R + 1$; therefore, $|E''| \geq |E'| - (4R + 1)C_{4R + 1}$. Now, we show that all paths with edges lying in $E''$ have length at most $2R$. Assume, for the sake of contradiction, that there exists a path of length $\geq 2R + 1$ with edges in $E''$ and denote by $v_0, \ldots, v_{2R + 1}$ the first $2R + 2$ vertices of this path. Then $v_0\ldots v_{2R + 1}$ is a shortest path between $v_0$ and $v_{2R + 1}$ ---otherwise, some edges $v_jv_{j + 1}$ would lie in a cycle of length $\leq 4R + 1$. But then,
		\begin{align*}
		1 & = \braket{\psi|1 + Z_{v_0}Z_{v_{2R + 1}}|\psi}\\
		& \leq \sum_{0 \leq j < 2R + 1}\braket{\psi|1 + Z_{v_j}Z_{v_{j + 1}}|\psi}\\
		& = 0,
		\end{align*}
		a contradiction. Therefore, all paths in $E''$ have length at most $2R$. Now, consider a connected component of the graph induced by the edge set $E''$. Such a component then has diameter at most $2R$. Besides, it has no cycle ---since such a cycle would be of length $\leq 4R$, contradicting the definition of $E''$. Therefore, the connected component must be a tree and its number of edges is then bounded by the number of vertices minus $1$. Summing over all connected components, one must then have $|E''| \leq n$, hence $|E'| \leq n + (4R + 1)C_{4R + 1}$. The bound follows by recalling that edges $\{i, j\}$ not in $E'$ satisfy $\braket{\psi|Z_iZ_j|\psi} = 0$.
	\end{proof}
\end{prop}
As previously hinted, this proposition implies that for a random $d$-regular graph, a circuit similar to the one proposed by Bravyi et al. produces a cut with at most $|E|\left(\frac{1}{2} + \frac{1}{d}\right)$ satisfied edges, which is bounded away from the typical value $|E|\left(\frac{1}{2} + \frac{0.7632\ldots}{\sqrt{d}}(1 + o_d(1))\right)$.
\begin{cor}
	Let $G = (V, E)$ a random $d$-regular graph with $n$ vertices. Assume that $\ket{\psi}$ is prepared by a $\mathbf{Z}_2$-symmetric range $R$ circuit such that for all $e = \{e_0, e_1\}\in E$, $\braket{\psi|Z_{e_0}Z_{e_1}|\psi} \in \{0, -1\}$. Then for all $\varepsilon > 0$, with probability exponentially close to $1$ as $n \to \infty$,
	\begin{align}
		\braket{\psi|\sum_{e = \{e_0, e_1\} \in E}\frac{1 - Z_{e_0}Z_{e_1}}{2}|\psi} & \leq |E|\left(\frac{1}{2} + \frac{1}{d} + \varepsilon\right)
	\end{align}
	for any state $\ket{\psi}$ prepared by a $\mathbf{Z}_2$-symmetric range-$R$ quantum circuit.
\begin{proof}
	Using for instance \cite[Theorem 1]{mckay_wormald_wysocka_2004}, for all $\varepsilon > 0$, a random $d$-regular graph $G$ has less than $\frac{\varepsilon}{4R + 1}n$ cycles of length less than $4R + 1$ with probability exponentially close to $1$ as $n \to \infty$. The result then follows from proposition \ref{prop:upper_bound_sharp_maxcut_circuit}.
\end{proof}
\end{cor}
Coming back to grid graphs, it is worth noting that proposition \ref{prop:upper_bound_sharp_maxcut_circuit} does not contradict the achievability result stated in proposition \ref{prop:bravyi_lower_bound_2d_grid_graph}. The reason is, the two-dimensional grid graph considered in the latter theorem has many short cycles compared to a random regular graph. For instance, for $R = 1$, $C_{4R + 1} = C_5 = n$, so proposition \ref{prop:upper_bound_sharp_maxcut_circuit} gives a loose $\frac{13}{2}n$ upper bound on the number of satisfied edges (while the grid graph only has $2n$ edges).

To conclude on the generalization of Bravyi et al.'s method, we finally show that proposition \ref{prop:bravyi_upper_bound_regular} can be applied to derive upper bounds on the \textit{overlap} of a typical cut sampled from a shallow circuit with a maximum cut. Contrary to previous upper bounds, these new bounds have the interest of being nontrivial both for grid graphs and random regular graphs. They will result from the following lemma:
\begin{lem}
	Let $G = (V, E)$ a graph. Assume that for some state $\ket{\psi}$ and constant $\alpha > 0$,
	\begin{align}
		\alpha|E| & \leq \sum_{e = \{e_0, e_1\} \in E}\braket{\psi|1 + \sigma_eZ_{e_0}Z_{e_1}|\psi}
	\end{align}
	for all choices of signs $\left(\sigma_e\right)_{e \in E} \in \{-1, 1\}^E$. Denote by $\textnormal{MaxCut}(G)$ the number of satisfied edges in a maximum cut of $G$. Then for all partition $E = E_1 \sqcup E_2$ of the edges of $G$,
	\begin{align}
		\braket{\psi|\sum_{e = \{e_0, e_1\}\in E_1}\frac{1 - Z_{e_0}Z_{e_1}}{2}|\psi} & \leq \frac{|E_1| + \textnormal{MaxCut}(G)}{2} - \frac{\alpha}{4}|E|.
	\end{align}
	\begin{proof}
		Fix a partition $E = E_1 \sqcup E_2$ of the edges of $G$. Define signs $\left(\sigma_e\right)_{e \in E}$ by letting $\sigma_e = 1$ if $e \in E_1$ and $\sigma_e = -1$ if $e \in E_2$. Then, by the assumption of the proposition,
		\begin{align*}
			\alpha|E| & \leq \braket{\psi|\sum_{e = \{e_0, e_1\} \in E_1}(1 + \sigma_eZ_{e_0}Z_{e_1})|\psi} + \braket{\psi|\sum_{e = \{e_0, e_1\} \in E_2}(1 - \sigma_eZ_{e_0}Z_{e_1})|\psi}.
		\end{align*}
		This implies successively:
		\begin{align*}
			& \alpha|E| \leq 2|E_1| - \braket{\psi|\sum_{e = \{e_0, e_1\} \in E_1}(1 - \sigma_eZ_{e_0}Z_{e_1})|\psi} + \braket{\psi|\sum_{e = \{e_0, e_1\} \in E_2}(1 - \sigma_eZ_{e_0}Z_{e_1})|\psi}\\
			& \alpha|E| - 2|E_1| + 2\braket{\psi|\sum_{e = \{e_0, e_1\} \in E_1}(1 - \sigma_eZ_{e_0}Z_{e_1})|\psi} \leq \braket{\psi|\sum_{e \in E}\left(1 - Z_{e_0}Z_{e_1}\right)|\psi}.
		\end{align*}
		But $\braket{\psi|\sum_{e \in E}\left(1 - Z_{e_0}Z_{e_1}\right)|\psi} \leq 2\textnormal{MaxCut}(G)$. The result follows.
	\end{proof}
\end{lem}

This lemma can right away be applied to a random regular graph:

\begin{prop}
	Let $G = (V, E)$ a random $d$-regular graph. Let $E_0$ the set of satisfied edges of a maximum cut. Then for all $\epsilon > 0$, with probability $\geq 1 - \varepsilon$ on the choice of $G$, for any state $\ket{\psi}$ prepared by a $\mathbf{Z}_2$-symmetric range-$R$ circuit, the overlap between $E_0$ and the satisfied edges of a cut sampled from $\ket{\psi}$ is bounded away from $|E_0|$:
	\begin{align}
		\braket{\psi|\sum_{e = \{e_0, e_1\} \in E_0}\frac{1 - Z_{e_0}Z_{e_1}}{2}|\psi} & \leq |E_0| - \frac{(1 - \varepsilon)}{2(2R + 1)}|E|.
	\end{align}
	\begin{proof}
		This results from the previous lemma, applied to $\alpha = \frac{2(1 - \varepsilon)}{2R + 1}$ (according to proposition \ref{prop:bravyi_upper_bound_regular}), $E_1 = E_0$ and $E_2 = E - E_0$.
	\end{proof}
\end{prop}

\section{Formulae for $p = 1$ QAOA}
\label{appendix:p1_qaoa}
\subsection{Ring graph}
Consider a vertex $0$ of a ring graph (with at least 3 vertices), having neighbours $1$ and $2$. $U_C(\gamma)^{\dagger}U_B(\beta)^{\dagger}Z_0U_B(\beta)U_C(\gamma)$ can be expressed as the sum of Pauli tensors represented on the next figure:
\begin{figure}[!htbp]
	\centering
	\includegraphics[width=0.7\columnwidth]{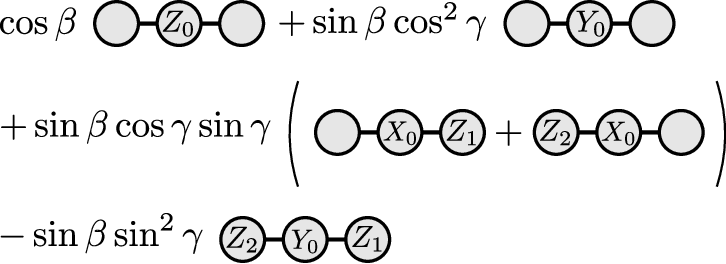}
	\caption{Terms of $U_C(\gamma)^{\dagger}U_B(\beta)^{\dagger}Z_0U_B(\beta)U_C(\gamma)$ for $p = 1$ QAOA on ring}
\end{figure}

\subsection{3-regular graph}
Consider a vertex $0$ of a 3-regular graph, having neighbours $1, 2, 3$. $U_C(\gamma)^{\dagger}U_B(\beta)^{\dagger}Z_0U_B(\beta)U_C(\gamma)$ can be expressed as the sum of Pauli tensors represented on the next figure:
\begin{figure}[!htbp]
	\centering
	\includegraphics[width=\columnwidth]{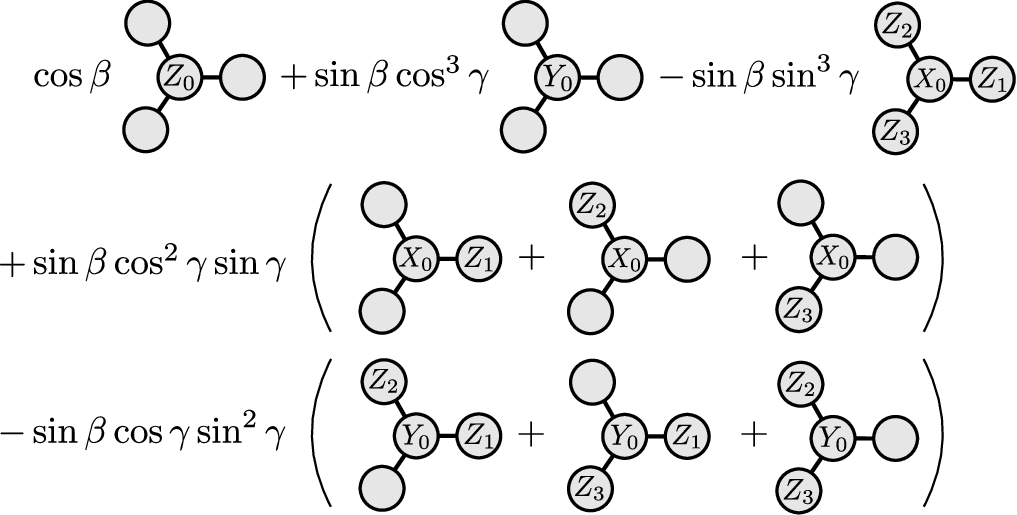}
	\caption{Terms of $U_C(\gamma)^{\dagger}U_B(\beta)^{\dagger}Z_0U_B(\beta)U_C(\gamma)$ for $p = 1$ QAOA on 3-regular graph}
\end{figure}

% ------------------------------------------------------------------------------

% ------------------------------------------------------------------------------
\end{document}